\documentclass[12pt]{article}
\usepackage{tikz}
\usetikzlibrary{decorations.pathreplacing,angles,quotes}
\usepackage{etex}
\usepackage[T1]{fontenc}
\usepackage{amssymb,amsmath,geometry}
\usepackage{lscape} 
\usepackage{mathtools}
\usepackage{graphicx}
\usepackage{fancyhdr}
\usepackage{fancybox}
\usepackage{shadow}
\usepackage{empheq}
\usepackage{titlesec}
\usepackage{titletoc}
\usepackage{soul, color}
\usepackage{float}
\usepackage{shorttoc}
\usepackage{xcolor}
\usepackage{fancybox}
\usepackage{natbib}

\usepackage{array}
\usepackage{tabularx}
\usepackage{booktabs}
\usepackage{rotating}
\usepackage{bigstrut}
\usepackage{natbib}
\usepackage{aeguill}

\usepackage{changepage}
\newenvironment{changemargin}[2]{\begin{list}{}{%
\setlength{\topsep}{0pt}%
\setlength{\leftmargin}{0pt}%
\setlength{\rightmargin}{0pt}%
\setlength{\listparindent}{\parindent}%
\setlength{\itemindent}{\parindent}%
\setlength{\parsep}{0pt plus 1pt}%
\addtolength{\leftmargin}{#1}%
\addtolength{\rightmargin}{#2}%
}\item }{\end{list}}

\usepackage{hyperref}
\def\sym#1{\ifmmode^{#1}\else\(^{#1}\)\fi}
\usepackage{caption}
\usepackage{blindtext}

\usepackage{caption}
\usepackage{subcaption}
\usepackage{mathtools}

\usepackage{graphicx}

\usepackage{enumerate}

\usepackage{hyperref}
\hypersetup{backref, colorlinks=true, citecolor=blue!50!black}
\usepackage{graphicx}

\renewcommand{\thetable}{\Roman{table}}
\renewcommand{\thefigure}{\Roman{figure}}

\usepackage{titlesec}
\usepackage{setspace}
\usepackage{indentfirst}

\makeatletter
  \newcommand\smalls{\@setfontsize\smalls{10.3pt}{6}}
\makeatother

\makeatletter
  \newcommand\footnotesizes{\@setfontsize\footnotesizes{9.6pt}{6}}
\makeatother

\newsavebox\tmpbox

\usepackage{caption}
\usepackage{subcaption}

\begin{document}

\title{Constitutions, Education and Gender Norms Change: Evidence from Colombia
\thanks{We thank seminar participants at several seminars, and  Violeta Martínez Calle for her invaluable efforts in searching for all the anecdotal evidence for the study. All errors or omissions are ours.}
}
\author{Hector Galindo-Silva\thanks{%
Department of Economics, Pontificia Universidad Javeriana, E-mail: galindoh@javeriana.edu.co
} \\
Pontificia Universidad Javeriana\\
\and Paula  Herrera-Id\'arraga \thanks{%
Department of Economics, Pontificia Universidad Javeriana, E-mail: pherrera@javeriana.edu.co} \\
Pontificia Universidad Javeriana
}
\date{First Draft: October, 2019 \\This Draft: August 2025}

\maketitle

\begin{abstract}
How do gender norms change? This paper provides evidence that exposure to mandatory high school courses on the 1991 Colombian Constitution—which incorporated principles of gender equality—significantly influenced attitudes toward gender roles. Using a difference-in-differences framework, we compare individuals who were exposed to these courses with those who were not. The results show that constitutional education increased support for gender equality, particularly among men. The effect is stronger when a younger woman is present in the household, suggesting a possible motivation to support the empowerment of younger female relatives. We also document important gender differences in how these shifts manifest within households: women exposed to the courses were more likely to reject the idea that men should be the head of the household, whereas men’s views on intra-household roles remained largely unchanged. This contrast points to persistent resistance to gender norm change within the private sphere, even as broader attitudes become more egalitarian. Taken together, the findings underscore the role of institutional and cultural change—through constitutional reform and civic education—in fostering more egalitarian gender norms, while also highlighting the complexity of such transformations.


\end{abstract}

\newpage

\section{Introduction}

Culture has become an increasingly prominent topic in the economics literature (for reviews, see \citealp{fernandez2011handbook}; \citealp{acemoglu2021culture}). Traditionally, scholars have conceptualized culture as a set of persistent values, norms, attitudes, and orientations that exert long-term influence on economic and political outcomes \citep[][]{GuisoSapienzaZingales2006, GuisoSapienzaZingales2009, FernandezFogli2006, FernandezFogli2009}. However, a growing body of research highlights the potential for rapid and discontinuous cultural change, as well as the responsiveness of culture to institutional and political developments \citep[][]{AlganCahuc2010, fernandez2013cultural, Bau2021, acemoglu2021culture, JaschkeSardoschauTabellini2022}.

\medskip

One domain in which culture is particularly salient is gender norms. Gender norms refer to socially constructed expectations about appropriate behaviors for men and women. These norms often sustain power asymmetries and perpetuate gender inequality. In many societies, traditional gender norms grant men disproportionate access to power and authority, while women are expected to take on caregiving and domestic responsibilities. Such norms constrain women’s opportunities and contribute to discriminatory outcomes.

\medskip

Although gender roles are often deeply entrenched and persistent \citep{alesina2013origins, FernandezFogli2009}, they are also subject to change \citep{Blauetalt2013, fernandez2013cultural}. Understanding the evolution of gender norms is therefore essential for addressing gender inequality and promoting inclusive development.

\medskip

This paper contributes to that understanding by examining the consequences of a major institutional reform in Colombia: the 1991 Constitution. The reform introduced a constitutional mandate for gender equality and was implemented alongside a requirement for mandatory high school instruction on the new constitution. We investigate whether exposure to this educational component affected individuals’ beliefs and attitudes toward gender equality.

\medskip

To estimate the effects of exposure, we implement a difference-in-differences strategy. We compare changes in outcomes across cohorts that, based on their age, were plausibly exposed to the mandatory constitutional curriculum in high school with those that were not either because they likely graduated before the reform or because they did not complete high school, even though they belonged to the same age group. Our identification strategy relies on the assumption that, in the absence of the reform, these groups would have followed similar trends. To strengthen the credibility of this assumption, we restrict the analysis to narrow age windows, improving comparability across treated and untreated cohorts.

\medskip

Our primary outcome measures general support for gender equality, focusing on beliefs about women's participation in and access to opportunities relative to men. We also examine perceptions of gender-based discrimination and attitudes toward gender roles within the household. Our analysis relies on two main data sources: the \emph{Political Culture Survey}, a nationally representative biennial survey, and the \emph{Time Use Survey} (ENUT),  conducted every five years. Both surveys are carried out by the National Administrative Department of Statistics (DANE).

\medskip

We find that exposure to the constitutional curriculum   increased support for gender equality, with more pronounced and statistically significant effects among men. Specifically, men exposed to courses on the 1991 Constitution are 2 to 3 percentage points more likely to express pro-equality views. The effect is concentrated among men who co-reside with females under the age of 18 (e.g., daughters, nieces, or granddaughters), for whom support increases by approximately 8 percentage points. This pattern is consistent with the argument by \citet{doepke2009women} that men’s support for women’s rights increases when they have a direct stake in the welfare and bargaining power of female family members.

  \medskip

We also explore whether these attitudinal changes plausibly translate into behavioral changes. In particular, we analyze self-reported experiences of discrimination. Our estimates suggest that women are less likely to report being treated unfairly, rejected, or mistreated forms of discrimination often shaped by patriarchal norms and male-dominated institutions.

  \medskip

Finally, we study gender-role attitudes in the household domain. We find gender differences in the effects: among women, exposure is associated with a greater likelihood of rejecting the view that men should be the head of the household. Among men, we detect no significant change on this dimension. This contrast highlights persistent resistance to gender norm change within the private sphere, even as broader attitudes become more egalitarian, underscoring the complexity of such transformations.

  \medskip

These results are robust to alternative model specifications, including different sets of fixed effects (cohort, survey year, region), additional controls (ethnicity), and various combinations (e.g., school-by-region and survey year-by-region fixed effects). They are also robust to alternative age windows. For all outcomes, we find no evidence of differential pre-trends prior to the reform, consistent with our main identification assumption. Moreover, for years following the reform—when students may have received longer exposure, and even non-graduates may have experienced partial treatment—we find no statistically significant effects, suggesting that the main impact occurred among those first exposed during high school.

\bigskip

This study makes several contributions to the existing literature. Beyond the previous work on the persistence and evolution of culture and its economic implications, we contribute to a growing body of research examining the enduring nature of gender roles and their links to economic and political outcomes. For instance, prior studies have explored how grammatical gender structures affect female labor force participation, access to credit, land ownership, and political representation \citep[][]{gay2013grammatical, galor2017geographical}. Religion has also been identified as a key determinant of gender attitudes \citep[][]{guiso2003people, bertocchi2011enfranchisement}.\footnote{Other historical shocks-such as wars or the transatlantic slave trade-have also disrupted the relative status of women by altering sex ratios and prompting women to assume roles traditionally held by men \citep[][]{teso2019long, goldin2013shocking, acemoglu2004women, fernandez2004mothers}.}

\medskip

We contribute to this literature by documenting a novel mechanism that shapes support for gender equality: the incorporation of gender equality provisions into national constitutions, reinforced through mandatory civic education. This institutional strategy, while not unique to Colombia, has become increasingly common since the late twentieth century.\footnote{For instance, \citet{unwomen2023gender} review constitutional reforms in 98 countries that amended or replaced their constitutions between 1978 and 2020. Since 2008, gender-related provisions have become increasingly common, addressing topics such as equality and non-discrimination, family and marriage, political participation, affirmative action, labor protections, and public services. In particular, 40 countries explicitly refer to equality of opportunity, and even more refer to equality of rights. Appendix Table \ref{ONU} presents a subset of Latin American countries that adopted gender equality provisions in their constitutions, along with the year of adoption and affected population. These reforms occurred at different moments, reflecting diverse institutional trajectories.} Our findings highlight the importance of these legal commitments, particularly when embedded in educational settings, and underscore the need to understand their broader social and cultural consequences.

\medskip

A complementary literature focuses on the role of economic incentives in shaping gender norms. A key contribution is \citet{doepke2009women}, who develop a theoretical model showing how the rising importance of human capital in the nineteenth century altered men’s incentives to support women’s rights, leading to institutional reforms in England and the United States. In their model, the expansion of women’s rights was driven by men’s increasing stake in their children’s welfare and education. Our paper provides empirical evidence for a related mechanism: institutional reform, when combined with civic education, can shift social norms by altering the perceived value of gender equality.

\medskip

In the Colombian context, \citet{iregui2020path} identify four major phases in the evolution of gender equality during the twentieth century. Despite gains in female educational attainment, men consistently enjoyed higher labor force participation, lower unemployment, and higher wages. Although women’s labor force participation rose steadily over time, it stabilized in the early twenty-first century, and large gender gaps persisted in wages and political representation. Building on this work, our study shifts attention from labor market outcomes to the role of institutional reform and civic education in shaping gender-related attitudes. In particular, we examine whether constitutional reforms can influence public opinion and promote social support for gender equality.

\medskip

In summary, this study analyzes a relatively recent institutional reform that formalized gender equality as a constitutional principle and introduced mandatory high school instruction on its content. While prior work has emphasized the importance of long-run historical processes, relatively little is known about the role of contemporary institutional reforms. By analyzing the effects of constitutional mandates and associated civic education, we shed light on the mechanisms through which institutional change can shape gender norms.

\medskip

The remainder of the paper is organized as follows. Section \ref{Background} provides an overview of the drafting and adoption of the 1991 Constitution, with a focus on the inclusion of women’s rights. Section \ref{DataandEmpiricalStrategy} describes the data and outlines the empirical strategy. Section \ref{MainResults} presents the main results. Section \ref{Section_robustness} provides additional evidence supporting our identification assumptions and includes a series of robustness checks. Section \ref{Conclusion} concludes.

\section{Background} \label{Background}

For much of the 20th century, Colombia adhered to conservative gender norms rooted in its long-standing political constitution of 1886. Although a major constitutional reform was enacted in 1936, and various legislative measures were introduced to expand women’s economic and social rights, the formal recognition of women’s rights as a constitutional principle only occurred with the drafting of a new political constitution in 1991 \citep{iregui2020path}.

\medskip

The 1991 National Constituent Assembly (NCA) played a pivotal role in this transformation. Convened specifically to draft a new constitution, the NCA sought to address Colombia’s persistent cycles of violence and armed conflict. The resulting Constitution of 1991 marked a fundamental shift by establishing a new social pact and laying the foundations for a ``social state under the rule of law,”  thereby promoting peaceful conflict resolution and expanding democratic representation.

\medskip

A central achievement of the 1991 Constitution was the recognition and protection of the rights of historically marginalized groups, including women-whose rights had been largely absent from the 1886 constitution. Earlier legal reforms had granted women access to higher education, control over marital assets, and the right to vote and hold public office. However, these changes occurred mostly through legislation rather than through constitutional guarantees. Despite these legal advances, gender-based discrimination and violence remained widespread.\footnote{\cite{iregui2020path} identify four phases in Colombia’s trajectory toward gender equality. During the “women left behind” phase (1905–1935), women faced limited educational opportunities, high fertility rates, and exclusion from the labor market and political life. The second stage, ``first steps toward empowerment” (1936–1965), included the 1936 reform and key legislative gains such as equal access to universities and labor protections, as well as women’s suffrage in 1954. Yet structural inequalities persisted, including low enrollment in higher education and limited labor force participation. The third stage (1966–1985), labeled “the rise of empowerment,” saw deeper reforms, including the legalization of divorce, equal rights in the civil code, and ratification of CEDAW. Fertility declined and female labor force participation rose. The final stage, ``still catching up” (1985–present), has featured sustained progress in education and some gains in political representation, but persistent labor market inequality and wage gaps remain.}

\medskip

Feminist movements played a crucial role in shaping the 1991 reform. Although women’s organizations did not present a unified candidate for the NCA-due in part to debates over dual affiliations and partisan alignments-they mounted a national campaign to promote a gender-inclusive constitutional agenda. As documented by \citet{quintero2006mujeres}, this campaign gathered roughly 15,000 signatures from cities across the country and submitted their proposals to the NCA Presidency. To amplify their message, activists disseminated press releases, ran radio spots-including a 15-day campaign on Caracol, a major national broadcaster-and organized public discussions. Appendix Figure~\ref{fig:constituyente}, Panel (a), displays a prominent newspaper ad from El Tiempo featuring their central slogan: “WITHOUT WOMEN'S RIGHTS, DEMOCRACY WON’T SUCCEED!”

\medskip

These efforts resonated with some members of the Constituent Assembly. Panel (b) of Appendix Figure~\ref{fig:constituyente} shows a statement titled Women’s Rights \citep{serpa_perry_verano:1991}, signed by delegates Horacio Serpa, Guillermo Perry, and Eduardo Verano de la Rosa. The statement emphasized the constitutional enshrinement of equality across sex, race, ethnicity, language, religion, and belief, and called for raising public awareness to combat discrimination. The NCA ultimately received over 150 formal proposals on women’s rights from both civil society and constituent delegates \citep{quintero2006mujeres}.

\medskip

The new Constitution included several provisions that explicitly address gender equality. Article 13 guarantees equal protection under the law and prohibits discrimination on the basis of sex, among other categories. Article 43 affirms the equal rights of women and men and mandates special state protection for pregnant women and female heads of household. Article 53 outlines fundamental labor rights, including specific protections for women, maternity, and minors. Article 40 guarantees women the right to participate in the formation and exercise of political power.\footnote{The relevant constitutional provisions are as follows: \textbf{Article 13:} “All individuals are born free and equal before the law, shall receive the same protection and treatment from the authorities, and shall enjoy the same rights, freedoms, and opportunities without any discrimination based on sex, race, national or family origin, language, religion, political or philosophical opinion.” \textbf{Article 43:} “Women and men have equal rights and opportunities. Women shall not be subjected to any form of discrimination. During pregnancy and after childbirth, they shall receive special assistance and protection from the State, and shall be entitled to food subsidies if unemployed or unprotected at that time. The State shall give special support to women who are heads of household.” \textbf{Article 40:} “Every citizen has the right to participate in the formation, exercise, and control of political power. To make this right effective, the authorities shall guarantee the adequate and effective participation of women in the decision-making levels of public administration.” \textbf{Article 53:} “Congress shall issue the Labor Statute. The corresponding law shall take into account at least the following fundamental minimum principles: equality of opportunity for workers; a vital and adjustable minimum wage, proportional to the quantity and quality of work; job stability; the inalienability of minimum benefits established by labor regulations; the ability to settle and conciliate rights that are uncertain or disputed; the most favorable condition to the worker in case of doubt in the application or interpretation of legal sources; primacy of reality over formalities established by the parties in labor relations; guarantee of social security, training, skill development, and necessary rest; \textbf{special protection for women}, maternity, and underage workers.”} Together, these provisions marked a fundamental shift: from viewing women as passive subjects in need of protection to recognizing them as rights-bearing individuals capable of exercising autonomy and fully participating in public life.

\medskip

An important feature of the 1991 Constitution's implementation, central to our empirical strategy, was the introduction of mandatory civic education on the Constitution in Colombian high schools. Law 107, enacted in January 1994, required the completion of fifty hours of Constitutional Studies as a prerequisite for graduation. This mandate was reinforced by Law 115, the General Law of Education (February 1994), which emphasized the inclusion of constitutional principles and civic instruction across all educational levels. Both laws were aligned with Article 41 of the Constitution, which requires the teaching of the Constitution and civic education in all educational institutions.

\medskip

The Ministry of Education played an active role in developing pedagogical materials to support this mandate. Our archival research uncovered multiple Ministry-led initiatives, including guides aimed at promoting gender equality and raising awareness of women’s rights. One early example is the 1994 booklet Combating Stereotyped Gender Roles (Appendix Figure~\ref{fig:booklet1}), which aimed to enhance students’ understanding of constitutional rights and democratic values. Panel B of Figure~\ref{fig:booklet1} highlights one objective: “to foster youth’s ability to identify gender roles that hinder the realization of both men’s and women’s potential.” Additional guides, such as a 1998 coeducation manual and a 1999 framework for non-discriminatory education, extended these efforts (Appendix Figure~\ref{fig:booklet2}). These materials explicitly referenced Articles 13 and 43 and encouraged educational development grounded in social equity.\footnote{It is difficult to assess the extent to which these materials were adopted across schools. Colombia’s education system comprises both public and private institutions; while public schools generally adhere to Ministry of Education guidelines, private schools enjoy considerable curricular autonomy under Article 27 of the Constitution, which guarantees academic freedom. As a result, curricular content varies substantially, and no centralized repository of syllabi exists. In the analysis that follows, we acknowledge the potential heterogeneity in implementation—particularly regarding the emphasis on gender equality in civic education—and discuss its implications for our identification strategy. In particular, although we cannot directly observe this heterogeneity, we note that most individuals in our sample likely attended public schools, and private institutions were nevertheless required to follow, at least in part, the Ministry’s directives.}



\section{Data and Empirical Strategy}\label{DataandEmpiricalStrategy}


\subsection{Data}\label{subdata}

Our primary data source is the \textit{Political Culture Survey}, conducted biennially since 2015 by Colombia’s National Administrative Department of Statistics (DANE). This nationally representative survey collects detailed information on citizens’ political and social attitudes across the country, covering key regions such as Bogotá and the Atlantic, Central, Eastern, and Pacific zones. It captures perceptions and behaviors related to civic life, enabling the analysis of attitudes toward gender equality and discrimination. The sample includes individuals aged 18 and older residing in private households throughout Colombia.

\medskip

As a complementary source, we use the \textit{Time Use Survey} (ENUT), also administered by DANE. Conducted every five years, ENUT surveys individuals aged 10 and older and provides nationally representative data on time allocation across paid work, unpaid domestic labor, caregiving, and personal activities. It also contains questions on intra-household gender dynamics, such as perceptions of household leadership, contributions to household income, and caregiving responsibilities. While ENUT offers valuable insights into household gender roles, it has two main limitations for our purposes. First, its temporal coverage is limited to three rounds: 2012–2013, 2016–2017, and 2020–2021. Second, because our empirical strategy-discussed in the next section-requires estimating individuals’ age in 1994 and relies on the exact survey year, the absence of precise fieldwork dates introduces measurement error in the age-based exposure variable. Moreover, ENUT focuses primarily on household-level attitudes rather than broader beliefs about gender equality.

\medskip

Our main outcomes are: (1) \textit{Support for gender equality}, measured by agreement with statements endorsing equal opportunities and participation for women and men (available in the 2019 and 2021 waves of the Political Culture Survey); (2) \textit{Perceptions of sex-based discrimination}, based on a question included only in the 2020 wave; and (3) \textit{Attitudes toward household gender roles}, constructed from the 2016–2017 and 2020–2021 ENUT rounds.
These measures allow for partial comparability across surveys and over time.

\medskip

Several data limitations may affect identification and interpretation. First, both surveys are repeated cross-sections and do not follow the same individuals over time. Second, the absence of month and day of interview may result in cohort misclassification when assigning age relative to 1994. Third, the outcomes are only available for 2016–2021, more than two decades after the treatment year (1994).  This gap complicates the identification of mechanisms, as attitudes in recent years may have been influenced by other events potentially correlated with the treatment. Nonetheless, finding an effect after such a long interval is consistent with a substantial and persistent attitudinal shift-originating at a critical stage of life-that was not offset by major intervening events in Colombia, such as the election of the country’s first female vice president in 2018 and the first female mayor of Bogotá in 2019.


\subsection{Empirical Strategy}\label{Sect_EmpiricalStrategy}

We study the effects of the implementation of Colombia’s 1991 Constitution. Our hypothesis is that the introduction of mandatory high school courses on the new Constitution generated a `cultural shock' that influenced gender-related attitudes. The staggered exposure to this curricular reform across birth cohorts creates a quasi-experimental setting that allows us to identify its causal impact on these attitudes.

\medskip

Our analysis focuses on individuals who were aged 16 or younger in 1994, the year the constitutional curriculum was introduced. We select this threshold because, in Colombia, students typically begin their last year of secondary school at age 16.\footnote{The Colombian education system consists of preschool, basic education (primary grades 1–5 and lower secondary grades 6–9), and upper secondary (grades 10–11), culminating in the high school diploma (\emph{bachiller}). Assuming students start school at age 5 and progress without repetition, they typically enter grade 11 at age 16 and graduate around age 17. To verify this, we examined Ministry of Education statistics from 2018 onward and found that the average graduation age is indeed close to 17, with a difference of approximately six months between public and private schools. Moreover, we find no compelling reason to believe this gap would have been significantly larger in earlier years.} In our survey data, this corresponds to individuals aged 39 or younger in 2017, 41 or younger in 2019, and 43 or younger in 2021. For those in this group who completed high school, at least one year-long course on the 1991 Constitution would have been mandatory. We refer to this group as the affected cohort. In contrast, individuals who were older than 16 in 1994 were not exposed to the curriculum and constitute the unaffected cohort. The use of 1994 as the treatment year is based on the legal mandate requiring all high schools to begin offering the course-at a minimum-to final-year students, making it unlikely that individuals who completed secondary school before 1994 received any exposure.\footnote{As noted in Section \ref{Background}, anecdotal evidence suggests that while the curriculum was formally introduced in 1994, its implementation became more robust over time due to improved textbooks, pedagogical innovations, and increased institutional support. As a result, later cohorts may have received more intensive exposure. Nevertheless, all individuals classified as treated (i.e., those aged 16 or younger in 1994) would have received at least some exposure to the reform. Those who completed high school likely encountered more sustained and comprehensive instruction. In Section \ref{Section_effectpost1994}, we explore heterogeneity across later cohorts.}

\medskip

Our empirical approach is based on a difference-in-differences (DiD) framework. Specifically, we estimate the following model:
\begin{equation}
\label{baselineDiD}
\begin{multlined}
Y_{ics}=\phi_{c}+\gamma_{s}+\beta_{1}(\text{HighSchool}_{ics})+\beta_{2}\big(1{\text{[Age $\leq 16$ in 1994}]_{c}}]\times 
\text{HighSchool}_{ics}\big)+\epsilon_{ics}
\end{multlined}
\end{equation}

\medskip

Here,   $Y_{ics}$ denotes the outcome of interest for individual  $i$, from birth cohort   $c$ and observed in survey year $s$. The specification includes cohort fixed effects ($\phi_c$) and survey year fixed effects ($\gamma_s$). The variable $\text{HighSchool}_{ics}$ is an indicator equal to one if the individual completed high school. The indicator $1{\text{[Age $\leq 16$ in 1994}]_{c}}$ identifies cohorts potentially exposed to the reform. The coefficient of interest, $\beta_{2}$, captures the differential effect of high school completion for those exposed to the constitutional curriculum.

\medskip

All models also include region fixed effects and, in some specifications, interactions between cohort and region, as well as between high school completion and region. We additionally control for individual-level characteristics such as gender and ethnicity. Standard errors are clustered at the region level. Given the small number of clusters, we report wild bootstrap confidence intervals following \citet{Cameronetalt2008}.\footnote{Regions are defined as follows: Caribbean (Atlántico, Bolívar, Cesar, Córdoba, La Guajira, Magdalena, Sucre); Central (Antioquia, Caldas, Caquetá, Huila, Quindío, Risaralda, Tolima); Eastern (Boyacá, Cundinamarca, Meta, Norte de Santander, Santander); Bogotá D.C.; and Pacific (Cauca, Chocó, Nariño, Valle del Cauca)-yielding six regions in total.}

\medskip

Our main identification assumption is that, in the absence of treatment, average outcomes for the treated and untreated groups would have evolved in parallel across cohorts. This assumption is essential for interpreting $\beta_2$ as a causal effect. It may be violated if changes in attitudes over time or across cohorts are systematically related to high school completion for reasons unrelated to the reform.

\medskip

One potential threat involves differential fertility trends. For example, if individuals without a high school degree had more children after 1994 due to unrelated shocks-and if family size affects gender attitudes (e.g., through raising daughters)-then post-treatment differences may reflect demographic shifts rather than the effects of the curriculum.

\medskip

We implement three strategies to assess the plausibility of our identification assumption. First, we restrict the sample to individuals just below and just above the age threshold of 16 in 1994. This focus on adjacent cohorts improves comparability and strengthens the case for parallel trends.\footnote{We estimate models across a range of age bandwidths. Our preferred specification uses a ±3-year window, which balances sufficient post-reform exposure with comparability across age groups.} Second, we estimate equation (\ref{baselineDiD}) using pre-1994 data. As shown in Section \ref{Section_robustness}, we find no evidence of systematic pre-treatment differences in outcomes. Third, we test whether our results are robust to controlling for household composition, proxied by the number of children. Results reported in Section \ref{Section_robustness} indicate that  our main estimates are unaffected.

\subsection{Additional Limitations and Potential Biases}

While the evidence supports our identifying assumption, several limitations and potential sources of bias remain due to data constraints and design features.

\medskip

First, our survey data do not report the month of interview, limiting our ability to construct precise measures of age at exposure and forcing us to rely on year-based definitions. Nonetheless, we believe annual data provide sufficient resolution to compare individuals within narrow age bands.

\medskip

Second, we do not observe the precise timing of high school completion. Because treatment assignment depends on whether individuals were enrolled in high school after the reform, this introduces measurement error that may attenuate our estimates. However, as we previously argued, in Colombia, high school completion typically occurs at standard ages, reducing the risk of severe misclassification.

\medskip

Third, variation in the implementation of the constitutional curriculum across schools may affect exposure. Public schools likely adhered more closely to the reform, but our data do not distinguish between public and private school graduates. Nonetheless, approximately 63\% of high school graduates from 1990 to 1999 attended public institutions \citep{ramirez2007educacion}, suggesting that the majority of students were plausibly exposed. To the extent that private schools lagged in implementation, our estimates may be biased toward zero.


\section{Main Results}\label{MainResults}

\subsection{Effect of Treatment on Attitudes Toward Gender Equality} \label{Section_effectin1994}

Table \ref{table_didgenderequality} presents our baseline estimates of equation (\ref{baselineDiD}), using an outcome from the \emph{Political Culture Survey} that captures support for gender equality in terms of women’s participation and opportunities.\footnote{The survey question, in Spanish, is: “¿Usted est\'a de acuerdo con que las mujeres participen en igualdad de condiciones y oportunidades?” (``Do you agree that women should participate on equal terms and have the same opportunities?"). This outcome is available only for 2019 and 2021.} Columns (2) and (5) show results without controls or fixed effects; columns (3) and (6) add fixed effects for cohort, region, and survey year. Columns (1), (4), and (7) further include controls for gender, ethnicity, and interactions between cohort and region, as well as cohort and high school fixed effects. The table reports estimates for the full sample (column (1)), women (columns (2)–(4)), and men (columns (5)–(7)). As discussed in Section \ref{Sect_EmpiricalStrategy}, we restrict the sample to individuals within a narrow age window (three years above and below the cutoff, i.e., $[-3,3]$) to strengthen identification. 

\medskip

Column (1) shows a positive, though imprecise, average effect. When disaggregating by gender, the coefficients remain positive but are larger and more precisely estimated for men (columns (5)–(7)). In magnitude, the estimates imply an increase of roughly 2–3 percentage points in men’s likelihood of expressing pro-gender-equality attitudes. These results suggest that civic education on the 1991 Colombian Constitution led to more progressive gender attitudes, especially among men.

\medskip

\begin{figure}[H]
             \caption{Effect of Exposure to Courses on the 1991 Constitution on Support for Gender Equality: Lags and Leads}
        \label{fig_didgenderequality_gender}
\begin{subfigure}{0.5\textwidth}
\caption{Women's Support for Gender Equality} \label{}
\includegraphics[width=\linewidth]{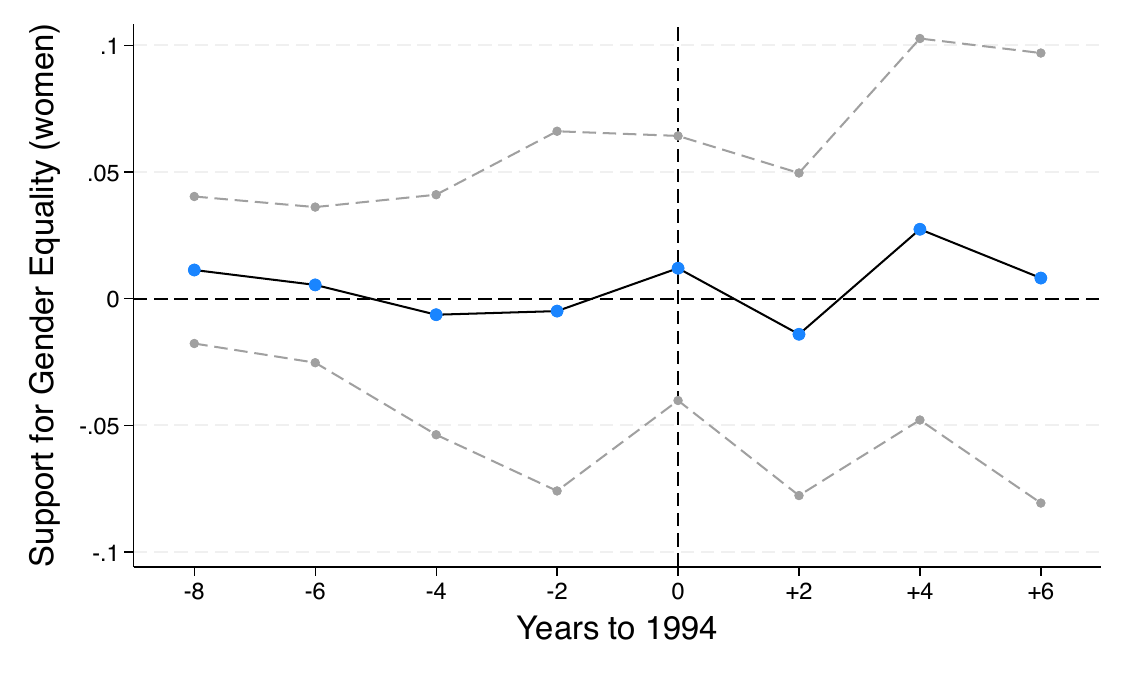}
\label{fig_didgenderequality_gender_a}
\end{subfigure}\hspace*{\fill}
\begin{subfigure}{0.5\textwidth}
\caption{Men's Support for Gender Equality} \label{}
\includegraphics[width=\linewidth]{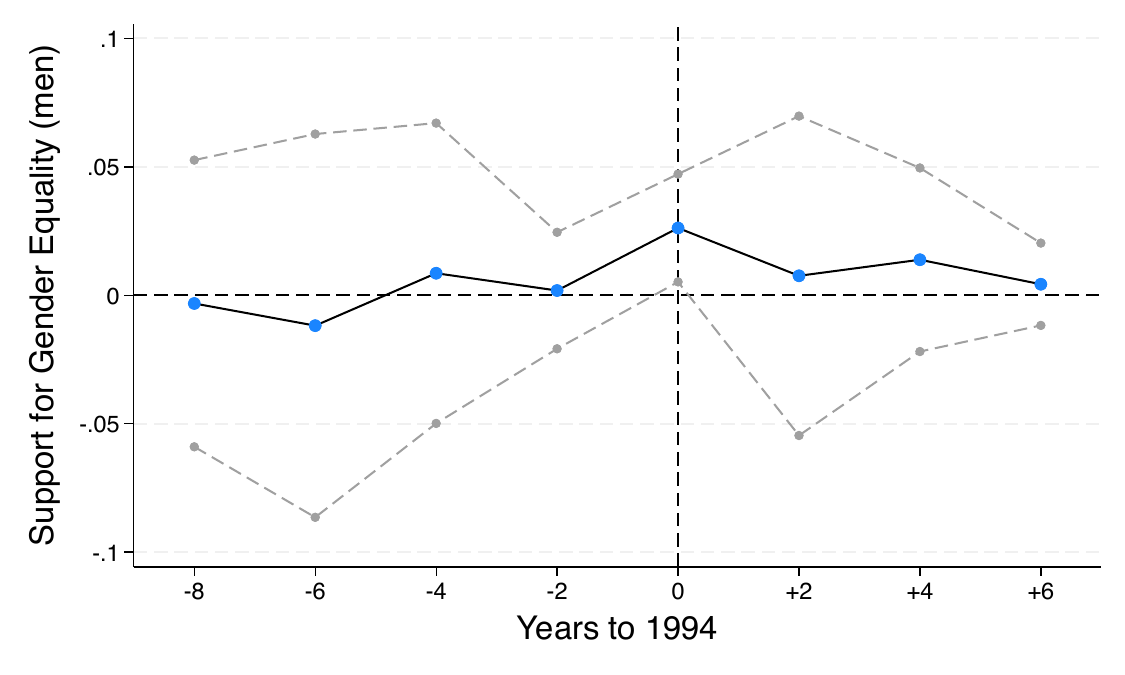}
\label{fig_didgenderequality_gender_b}
\end{subfigure}
     \begin{minipage}{16cm} \footnotesizes These figures show the estimates from Eq. (\ref{baselineDiD}) using an age window of [-3, 3], in a specification that includes cohort $\times$ region, high school $\times$ region, and survey year $\times$ region fixed effects, along with a set of ethnic group dummies. Dashed lines indicate 95\% confidence intervals.
\end{minipage}
\end{figure}

The results in Table \ref{table_didgenderequality} are robust to alternative specifications, including different combinations of controls and fixed effects. As detailed in Section \ref{Section_robustness}, they also hold under alternative age windows and, crucially, are consistent with the identifying assumption of parallel trends. In particular, we find no evidence of differential pre-trends across cohorts prior to 1994. Furthermore, we observe no comparable effects among individuals who completed high school after 1994, relative to similarly aged peers who did not. We return to this point in Subsection \ref{Section_effectpost1994}. These two additional sets of results are summarized in Figure \ref{fig_didgenderequality_gender}, which shows null effects before and after 1994, but a statistically significant increase for men in that year.

\bigskip

What might explain these results? We argue that they are driven by exposure to high school courses on the 1991 Constitution, which emphasized gender equality and promoted norms of equal rights for men and women. These messages, communicated by authority figures such as teachers, likely influenced students’ attitudes,  particularly among young men who may have been less predisposed to hold such views. As these individuals internalized such messages, they may have contributed to a broader shift in social norms surrounding gender equality.

\medskip

Although we cannot directly observe changes in prevailing norms, we present suggestive evidence consistent with this mechanism. Specifically, we explore: (i) whether effects are stronger among individuals more responsive to social norms; (ii) whether women’s reports of gender-based discrimination declined; and (iii) whether household attitudes toward gender roles shifted in ways consistent with broader attitudinal change. While exploratory, these analyses offer insight into the potential pathways through which the courses influenced attitudes.

\medskip

We begin by hypothesizing that the effect among men should be stronger for those who later had daughters. The intuition is that norms favoring gender equality may resonate more among men with altruistic preferences toward girls in their families. This is consistent with \citet{doepke2009women}, who argue that men with daughters are more likely to support gender equality to improve their daughters’ bargaining power. Related evidence in sociology and economics includes \citet{warner1991does}, who finds that having daughters shapes feminist views in Canada and the U.S. (except among American men), and \citet{washington2008female} and \citet{borrell2019mighty}, who show that U.S. and UK male politicians with daughters adopt more progressive stances on women’s issues.

\medskip

We test this hypothesis in Table \ref{table_didawareness_m_minorwhousehold} by interacting the treatment with the presence of young women (under 18) in men’s households-e.g., daughters, nieces, or granddaughters.\footnote{While the data do not allow us to identify specific relationships, our argument rests on the presence of a close personal connection, regardless of kinship.} We find that the treatment effect is larger for men living with young women (columns (1)–(4)) and negligible for those who do not (columns (5)–(8)). This provides suggestive evidence for one pathway through which civic education may have influenced men’s gender attitudes.

\medskip

 We next examine perceptions of gender-based discrimination, using women’s self-reported experiences. If male-dominated social norms are a primary driver of such discrimination, and if the reform shifted men’s attitudes toward greater gender equality, we would expect a decline in reported discrimination. This reduction should be most pronounced among treated women whose male peers were also treated, while women in the control group might experience smaller indirect effects through interactions with treated men but still face discrimination from untreated men in their cohorts.
 Table \ref{table_didsexdiscriminationy_gender} presents estimates of the likelihood of reporting gender-based discrimination in the past 12 months, using data from the \emph{Political Culture Survey}.\footnote{The survey question, in Spanish, is: “Durante los últimos 12 meses, ¿usted ha sentido que lo han discriminado [...] por alguno de los siguientes motivos: Sexo [...]?” (``In the past 12 months, have you felt discriminated against [...] based on sex?"). This outcome is available only for the 2021 wave.} We find negative and statistically significant effects for women (column (2)), consistent with a reduction in reported experiences of discrimination. As expected-and serving as a placebo test-we find no corresponding effects for men (column (3)).

\medskip

Finally, we analyze attitudes toward gender roles within the household using data from the ENUT survey. We focus on disagreement with three statements that directly reflect gender norms: (i) women are better suited for domestic tasks; (ii) the husband should make decisions about his wife’s life; and (iii) the household head should be male.\footnote{Respondents rated agreement on a four-point scale, from “strongly disagree” to “strongly agree.” We recode responses into a binary variable equal to 1 if the respondent answered “disagree” or “strongly disagree,” and 0 otherwise. The exact Spanish wording of the questions is: (i) “¿Usted está de acuerdo con que las mujeres son mejores para el trabajo doméstico que los hombres?”; (ii) “¿Usted está de acuerdo con que el esposo debe tomar las decisiones relacionadas con la vida de la esposa?”; and (iii) “¿Usted está de acuerdo con que la cabeza del hogar debe ser el hombre?”. These questions are available in the 2016-2017 and 2020-2021 ENUT waves.} As shown in Table \ref{table_did_domestic1}, we find no statistically significant effects for women or men on the first two statements (columns (1)–(4)). For the third statement,regarding male household headship, we find a positive effect for women and no effect for men (column (5)).

\medskip

How should we interpret these results? First, the observed effect is consistent with our broader hypothesis of cultural change. Women exposed to the 1991 Constitution curriculum are significantly more likely to reject the notion that men should lead the household, a belief reflecting explicit gender inequality. This norm may have been more susceptible to change, perhaps because it was less deeply entrenched by 1994. By contrast, the other two statements may reflect more persistent norms that were either widely accepted or already broadly rejected at the time of the reform, limiting the scope for change.\footnote{Although we cannot definitively explain the null effects for statements (i) and (ii), one speculative interpretation, offered with appropriate caution, is as follows. First, by 1994 (and still today), a large majority of Colombians strongly endorsed statement (i), i.e., the view that women are “naturally” better suited for domestic tasks- making this belief especially resistant to change. Second, by 1994, most Colombians-particularly women- already rejected statement (ii), which endorses overt male authority in household decision-making. Consequently, limited variation remained for either belief to shift further. In contrast, statement (iii) may have represented a more malleable norm, and thus was more responsive to the civic education intervention.}

\medskip

Second, and perhaps more importantly, these gender-differentiated effects, which contrast with those observed in Table \ref{table_didgenderequality}, highlight an important dimension of how cultural change around gender norms may unfold. While the intervention appears to have influenced men’s attitudes in the public sphere (e.g., regarding civic equality), it affected women’s views in the domestic sphere. In other words, women exposed to the curriculum were more likely to challenge traditional household hierarchies, whereas men may have adopted egalitarian views in public discourse while maintaining traditional norms in private settings. These findings underscore the particular challenge of shifting household-level gender norms, relative to more visible public attitudes. This interpretation aligns with \citet{doepke2009women}, who argue that men are more likely to support gender equality for other men’s wives than for their own.

\subsection{No Effects for Post-1994 Graduates} \label{Section_effectpost1994}

In the previous section, we focused on individuals who completed secondary school in 1994 or shortly thereafter-cohorts likely exposed to instruction on the 1991 Constitution-and compared their attitudes toward gender equality with those of similarly aged individuals who did not complete secondary education and were therefore unlikely to have received this instruction. A natural extension is to examine whether similar or stronger effects emerge among those who graduated in later years and may have been exposed to the revised curriculum for longer periods.

\medskip

On one hand, students graduating after 1994 -such as in 1996- were likely to receive civic education not only in their final year, but also in earlier grades, as the reform was gradually implemented throughout secondary school (see Section \ref{Background}). This would suggest more intensive treatment exposure.

\medskip

On the other hand, identifying untreated individuals becomes more difficult in post-1994 cohorts. Since we do not observe the specific grade at which students dropped out, we cannot determine with certainty whether those who did not complete high school avoided exposure to the new curriculum. Many may have received partial exposure, particularly if they left school during grades where the reform was already in place. This blurs the treatment–control distinction, complicating causal inference.

\medskip

\begin{figure}[H]
        \begin{center}
             \caption{Effect of Exposure to Courses on the 1991 Constitution on Women's Self-reported Perception of Gender Discrimination}
        \label{fig_didgenderequality_sexdiscr_women}
        \vspace{-0.3cm}
\includegraphics[width=100mm]{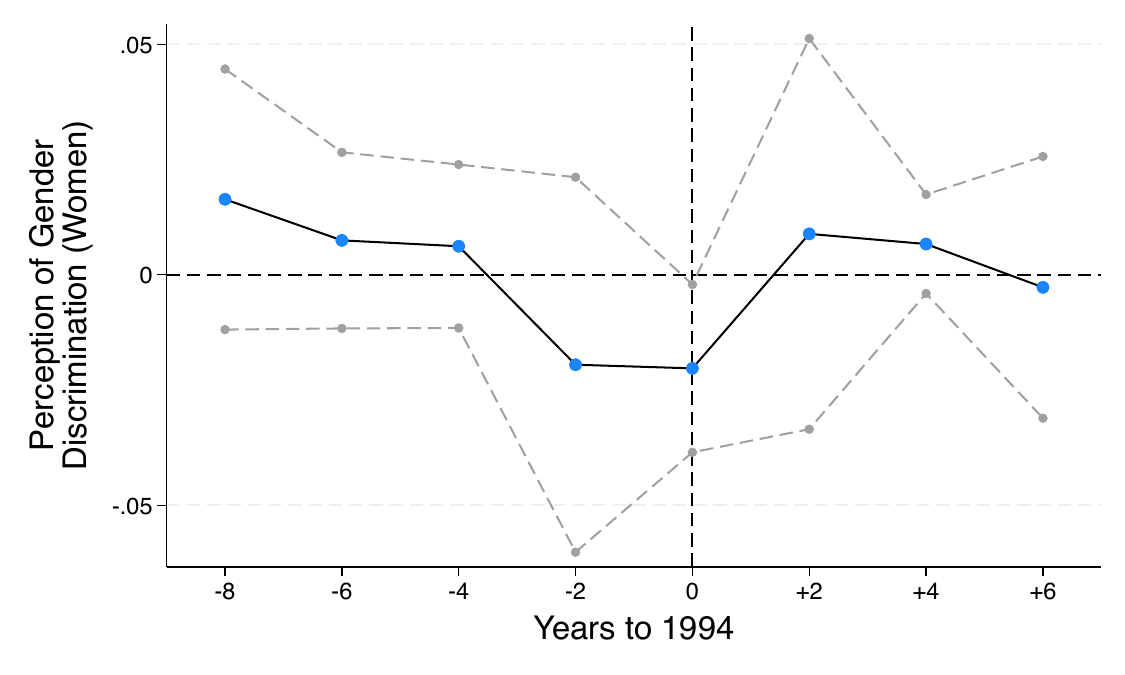}
     \begin{minipage}{12cm} \footnotesizes These figures show the estimates from Eq. (\ref{baselineDiD}), in a specification that includes cohort $\times$ region, high school $\times$ region, and survey year $\times$ region fixed effects, along with a set of ethnic group dummies. Dashed lines indicate 90\% and 95\% confidence intervals.
\end{minipage}
\end{center}
\end{figure}

\begin{figure}[htb]
\centering
\caption{Effect of Exposure to Courses on the 1991 Constitution on Women's Support for Gender Equality in the Domestic Sphere}
        \label{fig_did_domestic1}
        \vspace{-0.3cm}
\begin{subfigure}{0.32\linewidth}
    \includegraphics[width=5cm,height=5cm]{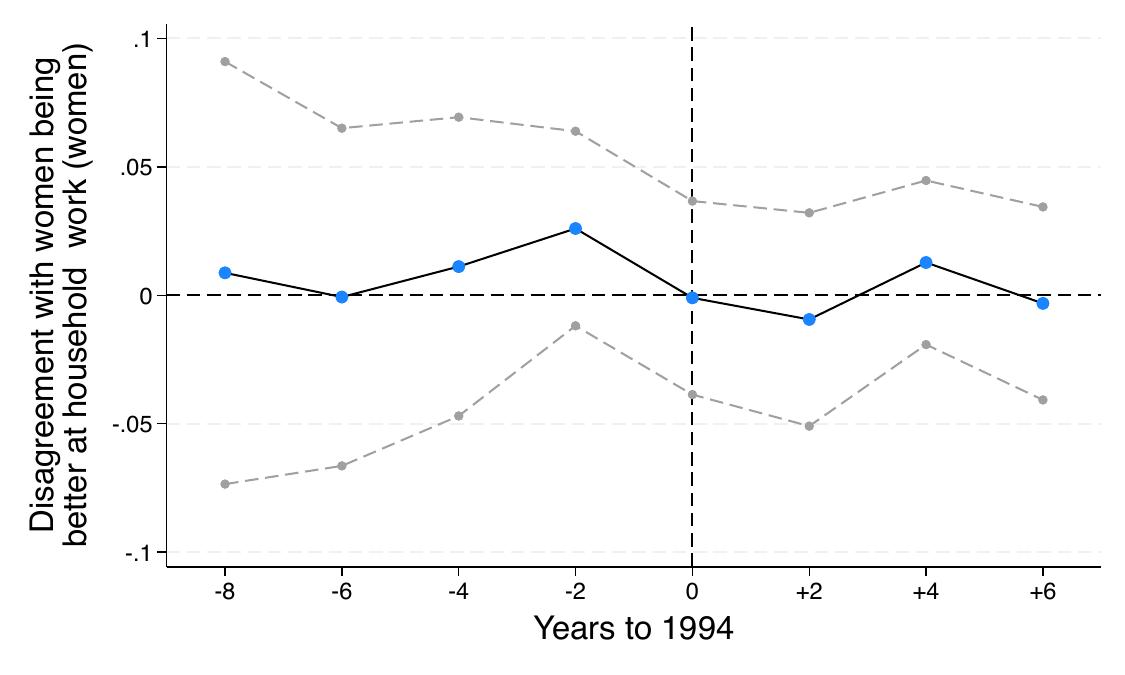}
    \label{fig_did_domestic1_a}
\end{subfigure}
\begin{subfigure}{0.32\linewidth}
    \includegraphics[width=5cm,height=5cm]{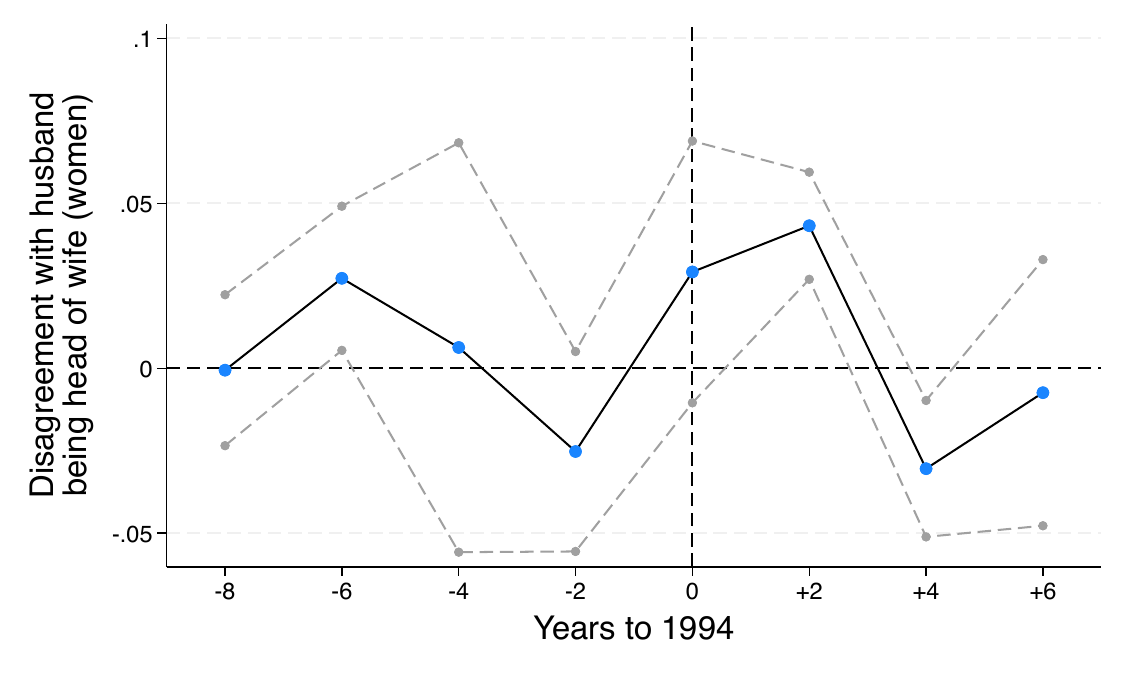}
    \label{fig_did_domestic1_b}
\end{subfigure}
\begin{subfigure}{0.32\linewidth}
    \includegraphics[width=5cm,height=5cm]{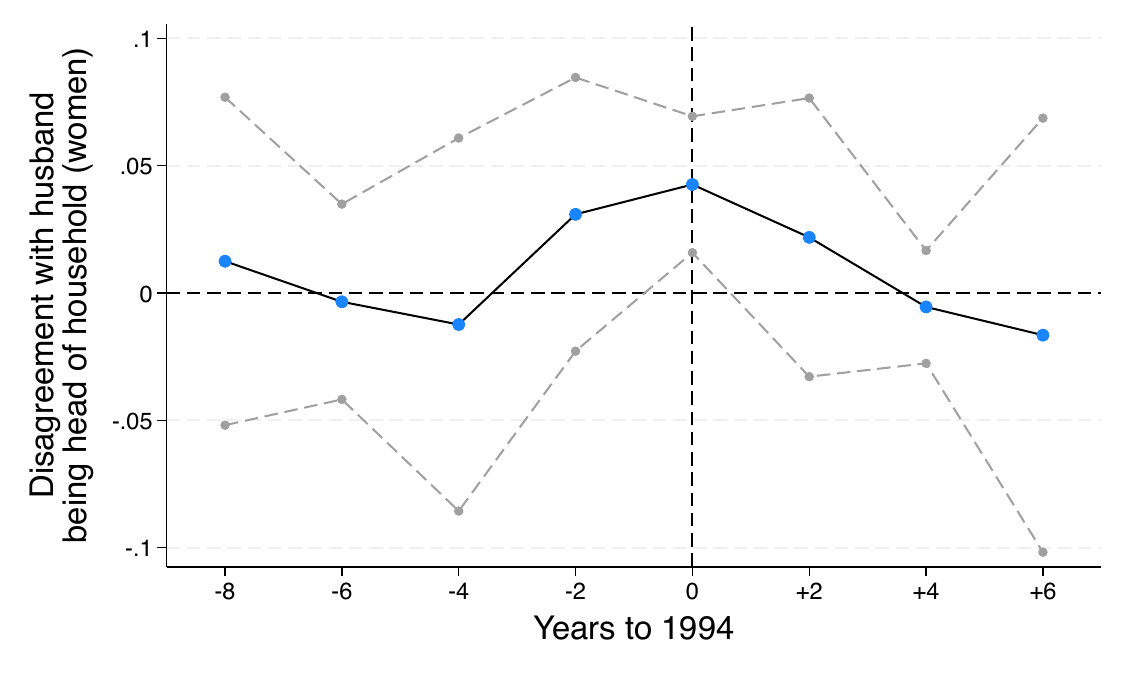}
    \label{fig_did_domestic1_c}
\end{subfigure}
     \begin{minipage}{15cm} \footnotesizes These figures show the estimates from Eq. (\ref{baselineDiD}) using an age window of [-3, 3], in a specification that includes cohort $\times$ region, high school $\times$ region, and survey year $\times$ region fixed effects, along with a set of ethnic group dummies. Dashed lines indicate 95\% confidence intervals.
\end{minipage}
\end{figure}

With this caveat in mind, we re-estimate our models using alternative graduation cutoffs-1996, 1998, 2000, 2002, and 2004-and focus on the same outcome variables. Results are presented in Table \ref{table_didgenderequalitypost1994} and in Tables \ref{table_didsexdiscrimpost1994} to \ref{table_diddomesticpost1994_op6}, some of which are previewed in Figures \ref{fig_didgenderequality_gender}, \ref{fig_didgenderequality_sexdiscr_women}, and \ref{fig_did_domestic1}.

\medskip

Across nearly all cohorts and outcomes, we find no statistically significant effects for either gender.\footnote{There are a few exceptions, limited to attitudes toward gender roles in the household. Most notably, for the 1996 cutoff, we observe statistically significant and opposing effects for men and women on the belief that “the husband should make decisions concerning his wife’s life” (Table \ref{table_diddomesticpost1994_op5}, columns (2) and (3)). Treated women show greater disagreement with this statement, while treated men show less. While this pattern may align with our general hypothesis-namely, increased opposition to patriarchal norms among women-it contrasts with the null effect in 1994 and our earlier interpretation that this belief was already broadly rejected. One possible, though speculative, explanation is that greater curricular exposure in later years enhanced women’s awareness and rejection of this norm, while men may have reacted defensively to perceived shifts in power. The fact that this effect is concentrated in a narrow post-reform window suggests caution in interpretation, as it may also reflect convergence between treatment and control groups over time. A second exception arises in Table \ref{table_diddomesticpost1994_op4}, column (1), for the 1998 cutoff. Here, we find a statistically significant effect on the pooled sample’s response to the statement that “women are better suited for domestic tasks.” However, the effect disappears when men and women are analyzed separately, suggesting it is small and detectable only in aggregate.} While these null results must be interpreted with caution, given the weaker treatment contrast in later years, they suggest that the effects are concentrated among students first exposed to the revised curriculum in high school and may reflect a broader and faster than anticipated expansion of the revised curriculum across lower secondary grades. If non-graduates were also exposed to civic education content, the difference between treatment and control groups would be attenuated, reducing our ability to detect effects.


\subsection{Alternative Explanations}

The results in Section~\ref{Section_effectin1994} could, in principle, be driven by mechanisms other than the cultural channel we emphasize. This section examines several theoretically plausible alternatives that, however, are less consistent with the empirical evidence.

\medskip

One possibility is that exposure to the 1991 Constitution curriculum shifted individuals’ political ideology—specifically, reducing alignment with right-wing conservatism, which is often associated with lower support for gender equality. This interpretation is consistent with one of the Constitution’s central features: the redefinition of Colombia as a social state under the rule of law.\footnote{The 1991 Constitution defined Colombia as a democratic, participatory, and pluralist state (Arts.~1–2) and introduced mechanisms for direct democracy, including plebiscites, referendums, popular consultations, legislative initiatives, and recall of elected officials (Arts.40, 103). It also strengthened political pluralism through party rights and minority protections (Arts.107–108).} We assess this hypothesis in Panel A of Table \ref{table_alternexplanationsreligionideologylabor}, which examines responses to the \emph{Political Culture Survey} question on self-reported ideological position along a left–right scale (1 to 10, where higher values indicate more right-leaning views).\footnote{The survey question, in Spanish, is: “Las personas cuando piensan en política utilizan los términos izquierda y derecha. En una escala de 1 a 10 donde 1 significa izquierda y 10 significa derecha, ¿dónde se ubicaría usted?” (“People, when thinking about politics, use the terms left and right. On a scale from 1 to 10, where 1 means left and 10 means right, where would you place yourself?”). This outcome is available in the 2019 and 2021 waves.} We find no statistically significant effects, suggesting this channel does not account for our results.

\medskip

A second possibility is that the curriculum reduced religiosity or increased openness to religious pluralism. In Colombia, religiosity is often associated with Catholicism, which traditionally promotes more conservative gender norms. In contrast, greater religious tolerance could reflect a broader openness to egalitarian values. This hypothesis is also consistent with the 1991 Constitution’s recognition of religious diversity, a topic likely addressed in civic instruction.\footnote{The 1991 Constitution replaced the 1886 confessional model with a secular state, guaranteeing freedom of conscience (Art.18) and equal protection for all faiths (Art.19).}  We assess this in Panels B and C of Table \ref{table_alternexplanationsreligionideologylabor}, which examine responses to the \emph{Political Culture Survey} questions on self-reported religiosity (Panel B) and religious intolerance (Panel C).\footnote{Religiosity is measured using a 1–5 scale of importance; the survey question, in Spanish, is: “¿Qué tan importante es la religión en su vida?” (``How important is religion in your life?'').  Religious intolerance is measured by whether the respondent would object to having neighbors of a different religion; the survey question, in Spanish, is:  ``De las siguientes personas, a quién no quisiera tener de vecino(a): Personas que profesan una religión diferente.'' (``Of the following people, who would you not want to have as a neighbor: People who practice a different religion."). Both outcomes are available in the 2019 and 2021 waves.} We find no significant effects on either measure, providing no support for this mechanism.

\medskip

A third hypothesis is that the curriculum improved women’s labor market outcomes—such as higher employment or earnings—and that such economic empowerment, in turn, shifted attitudes toward gender roles. We test this in Panel D of Table \ref{table_alternexplanationsreligionideologylabor}, which examines the \emph{Political Culture Survey} measure of labor force participation.\footnote{Labor force participation is based on the survey question, in Spanish: “¿En qué actividad ocupó la mayor parte del tiempo la semana pasada?” (“In what activity did you spend most of your time last week?”). Respondents are coded as 1 if they reported working or actively seeking work. This outcome is available in the 2019 and 2021 waves.} The estimates are statistically indistinguishable from zero, providing no evidence for this channel.\footnote{An earlier version of the paper, using wider age windows, found statistically significant increases in female labor force participation. Figure~\ref{fig_windowslaborforcepart_gender} reports gains of approximately five percentage points for women in the $[-6,6]$ to $[-8,8]$ age windows, with no corresponding effects for men. Similar but less precise estimates are reported in column~(2) of Panel~D. While these findings may suggest a link between civic education and women’s economic empowerment, other concurrent constitutional reforms could also have contributed. We therefore present this evidence cautiously.}

\medskip

While not exhaustive, this list likely captures the main alternative cultural mechanisms plausibly influenced by exposure to high school courses on the 1991 Constitution. Nonetheless, caution is warranted before ruling out other explanations. For instance, male students nearing the end of secondary school might have been influenced by other types of coursework—such as science, sociology, or history—or by classroom environments that facilitated greater interaction with female peers. Although plausible, we are not aware of curricular reforms of this nature around 1994, making this explanation less likely.


\section{Robustness Checks}\label{Section_robustness}

This section presents additional evidence in support of our identifying assumptions. While, as discussed in Section \ref{Sect_EmpiricalStrategy}, our empirical analysis is subject to several limitations, we believe the evidence provided here substantially strengthens the credibility of the findings reported in the preceding section.

\medskip

The first and most important set of robustness checks examines the absence of differential pre-trends across age groups in the years preceding 1994. For the outcome on overall support for gender equality, Table \ref{table_didbaseline_placebo23_boot} shows that this condition is satisfied for the $[-2,2]$ and $[-3,3]$ age windows in the years 1992, 1990, 1988, 1986, and 1984. Table \ref{table_didbaseline_placebo456_boot} in the appendix presents similar results for the same years using wider age windows. For our other main outcomes — perceptions of gender discrimination and disagreement with statements inconsistent with gender equality in the domestic sphere — Tables \ref{table_didsexdiscrim_placebo23_boot} to \ref{table_opinion6_placebo45_boot} in the appendix report analogous findings.\footnote{A small number of estimates in these tables are statistically different from zero. However, they are either only marginally significant, rely on wider age windows than our baseline specification (i.e., $[-3,3]$), or correspond to years well before the onset of treatment. In all cases, they do not reveal a pattern suggestive of a violation of the parallel trends assumption. Specifically, the only estimates that are statistically significant at conventional levels (i.e., at the 5\% level or below) represent less than 4\% of all estimates and occur either in years well before 1994 (see Tables \ref{table_didsexdiscrim_placebo23_boot} and \ref{table_didsexdiscrim_placebo45_boot}, columns (3), (4), and (6); Table \ref{table_opinion4_placebo23_boot}, column (1); Tables \ref{table_opinion5_placebo23_boot} and \ref{table_opinion5_placebo45_boot}, column (5); Table \ref{table_opinion6_placebo45_boot}, column (6)) or for outcomes not disaggregated by gender (see Table \ref{table_opinion4_placebo45_boot}, column (4)).}

\medskip

Taken together, these results underscore the specificity of the effect observed in 1994. They show no evidence of pre-existing differences in attitudes between treated and untreated individuals prior to the intervention, thereby providing critical support for the parallel trends assumption.

\medskip

The second set of robustness checks focuses on the age window used in our main analysis, defined as $[-3,3]$. As shown in Figure \ref{fig_didgenderequalityagewindows_gender} and \ref{table_baselineresults_agewindows}, the findings on attitudes toward gender equality are robust to alternative bandwidths. While Tables \ref{table_sexdiscrimination_agewindows} to \ref{table_domesticopinion6_agewindows} present analogous robustness analyses for our other key outcomes. These results remain consistent when using slightly wider age windows.

\medskip

A final set of robustness checks addresses the possibility that unobserved factors not accounted for in the baseline specification could influence our results. As discussed in Section \ref{Sect_EmpiricalStrategy}, one such factor is the disparity in the number of children between women with and without a high school education. To address this concern, we re-estimate our main specifications controlling for the number of children per household.\footnote{The number of children in a household may constitute a “bad control,” as it could itself be affected by the treatment. These results should therefore be interpreted with caution.} The results, presented in Tables \ref{table_supportgenderequality_robustnchildren} to \ref{table_opinion6_robustnchildren} in the Appendix, show that the coefficient on the interaction of interest remains virtually unchanged across all specifications, suggesting that our main findings are not driven by differences in household composition.


\section{ Conclusion}\label{Conclusion}
Our findings suggest that exposure to mandatory high school courses on the 1991 Colombian Constitution—particularly those highlighting gender equality principles—contributed to shaping more egalitarian attitudes toward gender roles and opportunities. We find that exposure to these courses led to more favorable views on gender equality, particularly among men, and was associated with a lower likelihood that women reported experiencing gender-based discrimination.

\medskip
We also document important gender differences in how these shifts manifest within the household. Women exposed to the courses were more likely to reject the idea that men should be the head of the household, whereas men’s views on intra-household roles remained largely unchanged. This contrast highlights an area of persistent resistance to gender norm change within the private sphere, even as broader attitudes become more egalitarian. However, men who co-reside with younger women are more likely to express progressive attitudes, suggesting that intergenerational considerations may play a role in shaping support for gender equality.

\medskip

Our findings underscore the complexity of changing gender norms. Institutional interventions, such as constitutional reform and civic education, can play a significant role in reshaping public attitudes and reducing perceived discrimination. However, household-level dynamics are more resistant to change and may require targeted strategies to shift deeply embedded power structures. These results align with theories emphasizing the importance of bargaining power within the household \citep{agarwal1997bargaining}, and point to the need for gender equality policies that address both public and private spheres.

\medskip

Furthermore, future research could explore the intergenerational effects of these cultural changes. While earlier generations may encounter greater obstacles in breaking gender stereotypes within the home, subsequent generations are expected to possess improved resources and perspectives to reassess and redefine traditional gender roles. Understanding the evolution of these cultural shifts and their impact on future generations is vital for fostering sustainable progress towards gender equality. By examining the interplay between cultural transformations, generational dynamics, and gender norms, we can gain valuable insights into the long-term implications of these changes and inform targeted interventions to accelerate advancements in gender equality.

\hbox {} \newpage

\begin{landscape}
\section*{Figures and Tables}

\begin{table}[H]
\vspace{1cm}
\begin{center}
\renewcommand{\arraystretch}{1}
\setlength{\tabcolsep}{3pt}
\caption {Effect of Exposure to Courses on the 1991 Constitution on Support for Gender Equality}  \label{table_didgenderequality}
\footnotesize
\vspace{-0.3cm}\centering  \begin{tabular}{lccccccc}
\hline\hline \addlinespace[0.15cm]
    & (1)& (2)& (3)& (4)& (5)& (6)& (7)\\\addlinespace[0.12cm]\cmidrule[0.2pt](l){2-8}\addlinespace[0.12cm]
      & \multicolumn{7}{c}{Dependent Variable: Support for Gender Equality}\\\addlinespace[0.1cm]\cmidrule[0.2pt](l){2-8}\addlinespace[0.1cm]
                      &\multicolumn{1}{c}{All}&\multicolumn{3}{c}{Women}&\multicolumn{3}{c}{Men}\\
                \addlinespace[0.1cm]\cmidrule[0.2pt](l){2-2} \cmidrule[0.2pt](l){3-5}\cmidrule[0.2pt](l){6-8}\addlinespace[0.12cm]
    (Age $\leq$ 16 in 1994) $\times$ (Highschool)&         0.021         &       0.011         &       0.012         &       0.012         &       0.024         &       0.030         &       0.026         \\
                    &       (0.013)         &     (0.020)         &     (0.019)         &     (0.019)         &     (0.004)\sym{***} &     (0.006)\sym{***} &     (0.008)\sym{**}  \\
                    &[-0.008 0.059]           &[-0.036 0.053]          &[-0.036 0.049]           &[-0.036 0.052]          &[0.013 0.035]\sym{***}            &[0.018 0.044]\sym{***}           &[0.010 0.063]\sym{***}                 \\
(Highschool)        &         0.084         &       0.022         &       0.021         &       0.178         &       0.022         &       0.023         &       0.001         \\
                    &      (0.006)\sym{***}&     (0.025)         &     (0.021)         &     (0.009)\sym{***}&     (0.019)         &     (0.018)         &     (0.004)         \\
                    &[-0.236 0.389]             &[-0.034 0.147]              &[-0.028 0.144]            &[-0.270 0.691]             &[-0.024 0.084]       &[-0.020 0.082]            &[-0.213 0.195]              \\
(Age $\leq$ 16 in 1994)&               &       0.002         &                &               &      -0.032         &         &               \\
                    &       &     (0.020)         &          &        &     (0.013)*         &            &     \\
                    &         &[-0.040 0.055]            &    &        &[-0.099 0.009]          &         &         \\
R-sq                &         0.021         &       0.002         &       0.012         &       0.019         &       0.003         &       0.016         &       0.033         \\
Observations        &          11816         &        6275         &        6275         &        6275         &        5541         &        5541         &        5541         \\

    \addlinespace[0.15cm]\hline    \addlinespace[0.15cm]
\multicolumn{1}{l}{Cohort, region \& year fixed effects}     &  Y& N& Y& Y & N& Y& Y\\
\multicolumn{1}{l}{Interacted fixed effects}     & Y& N& N& Y & N& N& Y\\
\addlinespace[0.15cm]\hline\hline\addlinespace[0.15cm]
\multicolumn{8}{p{22.cm}}{\scriptsize{\textbf{Notes:} All columns report estimates from Equation (\ref{baselineDiD}) using an age window of [-3, 3]. The regression samples are drawn from the 2019 and 2021 waves of the Political Culture Survey conducted by DANE. The dependent variable in all columns is the response to the question (in Spanish): ``¿Usted est\'a de acuerdo con que las mujeres participen en igualdad de condiciones y oportunidades?'' (``Do you agree that women should participate on equal terms and have the same opportunities?''). The specifications in columns (2) and (5) include fixed effects for cohort, region, and survey year. Columns (1), (3), and (6) additionally include cohort-by-region, high school-by-region, and survey year-by-region fixed effects, as well as a set of ethnic group dummies. Robust standard errors, clustered at the region level, are reported in parentheses. Wild bootstrap confidence intervals are shown in square brackets.  *, **, and *** indicate statistical significance at the 10\%, 5\%, and 1\% levels, respectively.} }
\end{tabular}
\end{center}
\end{table}
\end{landscape}


\newpage
\begin{landscape}
\begin{table}[H]
\vspace{2cm}
\begin{center}
\renewcommand{\arraystretch}{1}
\setlength{\tabcolsep}{3pt}
\caption {Effect of Exposure to Courses on the 1991 Constitution on Men's Support for Gender Equality: Heterogeneous Effects by Presence of Younger Women in the Household}  \label{table_didawareness_m_minorwhousehold}
\vspace{-0.3cm}
\footnotesize
\centering  \begin{tabular}{lcccccccc}
\hline\hline \addlinespace[0.15cm]
    & (1)& (2)& (3)& (4)& (5)& (6)\\\addlinespace[0.12cm]\hline\addlinespace[0.15cm]
          & \multicolumn{6}{c}{Dependent Variable: Men's Support for gender equality }\\\addlinespace[0.1cm]\cmidrule[0.2pt](l){2-7}\addlinespace[0.1cm]
  & \multicolumn{3}{c}{At Least One Younger} & \multicolumn{3}{c}{No Younger}\\
    & \multicolumn{3}{c}{Women in Household} & \multicolumn{3}{c}{Women in Household}\\\addlinespace[0.1cm]\cmidrule[0.2pt](l){2-4}\cmidrule[0.2pt](l){5-7}\addlinespace[0.1cm]
    (Age $\leq$ 16 in 1994) $\times$ (Highschool)&       0.080         &       0.086         &       0.084         &      -0.018         &      -0.014         &      -0.025         \\
                    &     (0.008)\sym{***}&     (0.008)\sym{***}&     (0.010)\sym{**} &     (0.007)         &     (0.010)         &     (0.013)         \\
                    &[0.043 0.102]\sym{***}            &[0.060 0.106]\sym{***}                &[0.052 0.131]\sym{***}          &[-0.031 0.027]\sym{*}          &[-0.037 0.017]           &[-0.056 0.005]                          \\
(Highschool)        &      -0.017         &      -0.015         &       0.007         &       0.051         &       0.050         &      -0.010         \\
                    &     (0.012)         &     (0.012)         &     (0.005)         &     (0.026)         &     (0.025)         &     (0.007)         \\
                    &[-0.058 0.021]             &[-0.056 0.022]                &[-0.200 0.355]           &[0.002 0.127]\sym{***}         &[-0.023 0.127]\sym{*}            &[-0.386 0.424]                       \\
(Age $\leq$ 16 in 1994)&      -0.055         &               &             &      -0.012         &              &               \\
                    &     (0.010)\sym{**} &             &     &     (0.017)         &          &           \\
                    &[-0.146 -0.030]\sym{***}           &          &     &[-0.044 0.046]        &      &            \\
R-sq                &       0.005         &       0.017         &       0.039         &       0.004         &       0.020         &       0.042         \\
Observations        &        2388         &        2388         &        2388         &        3153         &        3153         &        3153         \\

    \addlinespace[0.15cm]\hline    \addlinespace[0.15cm]
\multicolumn{1}{l}{Cohort, region \& year fixed effects}     & N& Y& Y& N& Y& Y \\
\multicolumn{1}{l}{Interacted fixed effects}     & N& N& Y& N& N& Y \\
\addlinespace[0.15cm]\hline\hline\addlinespace[0.15cm]
\multicolumn{7}{p{21cm}}{\scriptsize{\textbf{Notes:} All columns report estimates from Equation (\ref{baselineDiD}) using an age window of [-3, 3]. The regression samples are drawn from the 2019 and 2021 waves of the Political Culture Survey conducted by DANE. The dependent variable in all columns is the response to the question (in Spanish): ``¿Usted est\'a de acuerdo con que las mujeres participen en igualdad de condiciones y oportunidades?'' (``Do you agree that women should participate on equal terms and have the same opportunities?''). The specifications in columns (2) and (5) include fixed effects for cohort, region, and survey year. Columns (1), (3), and (6) additionally include cohort-by-region, high school-by-region, and survey year-by-region fixed effects, as well as a set of ethnic group dummies. Robust standard errors, clustered at the region level, are reported in parentheses. Wild bootstrap confidence intervals are shown in square brackets.  *, **, and *** indicate statistical significance at the 10\%, 5\%, and 1\% levels, respectively.} }
\end{tabular}
\end{center}
\end{table}
\end{landscape}


\begin{figure}[H]
        \begin{center}
             \caption{Effect of Exposure to to Courses on the 1991 Constitution on Support for Gender Equality by Presence of Younger Women in the Household: Lags and Leads}
    \label{fig_didgenderequality_men_minorswhousehold}
\includegraphics[width=100mm]{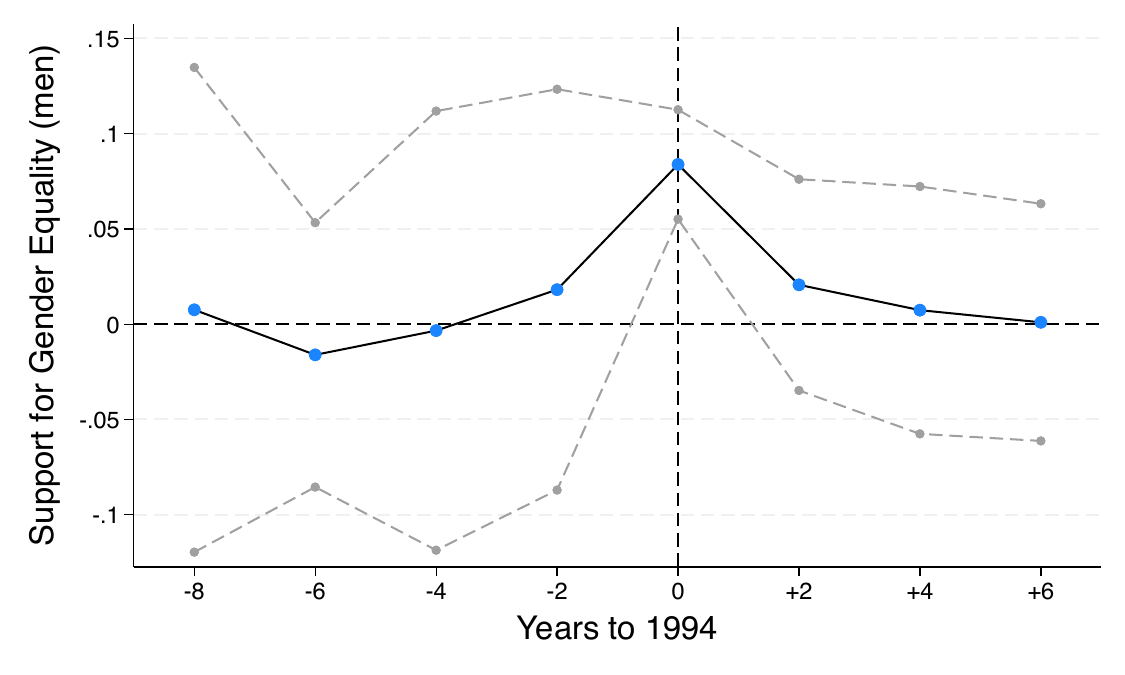}
     \begin{minipage}{12cm} \footnotesizes These figures show the estimates from Eq. (\ref{baselineDiD}), in a specification that includes cohort $\times$ region, high school $\times$ region, and survey year $\times$ region fixed effects, along with a set of ethnic group dummies. Dashed lines indicate 90\% and 95\% confidence intervals.
\end{minipage}
\end{center}
\end{figure}

\newpage
\begin{table}[h!]
\vspace{3cm}
\begin{center}
{
\renewcommand{\arraystretch}{1}
\setlength{\tabcolsep}{10pt}
\caption {Effect of Exposure to Courses on the 1991 Constitution on Self-Reported Perceptions of Gender Discrimination}  \label{table_didsexdiscriminationy_gender}
\vspace{-0.3cm}
\footnotesize
\centering  \begin{tabular}{lccccccccc}
\hline\hline \addlinespace[0.15cm]
    & (1)& (2)& (3)\\\addlinespace[0.12cm]\cmidrule[0.2pt](l){2-4}\addlinespace[0.12cm]
      & \multicolumn{3}{c}{Dep. var.: Perception of Gender Discrimination}\\\addlinespace[0.1cm]\cmidrule[0.2pt](l){2-4}\addlinespace[0.1cm]
                      &\multicolumn{1}{c}{All}&\multicolumn{1}{c}{Women}&\multicolumn{1}{c}{Men}\\
                \addlinespace[0.1cm]\cmidrule[0.2pt](l){2-2} \cmidrule[0.2pt](l){3-3}\cmidrule[0.2pt](l){4-4}\addlinespace[0.12cm]
    (Age $\leq$ 16 in 1994) $\times$ (Highschool)&      -0.005         &      -0.020         &       0.012         \\
                    &     (0.003)         &     (0.007)\sym{*}  &     (0.010)         \\
                    &[-0.033 0.003]         &[-0.059 -0.005]***        &[-0.000 0.044]         \\
(Highschool)        &      -0.002         &      -0.002         &      -0.003         \\
                    &     (0.001)         &     (0.004)         &     (0.005)         \\
                    &[-0.105 0.269]        &[-0.169 0.362]        &[-0.474 0.172]         \\
R-sq                &       0.013         &       0.020         &       0.013         \\
Observations        &        6555         &        3495         &        3060         \\

    \addlinespace[0.15cm]\hline    \addlinespace[0.15cm]
\multicolumn{1}{l}{Cohort, region \& year fixed effects}     &  Y& Y& Y \\
\multicolumn{1}{l}{Interacted fixed effects}    &  Y& Y& Y \\
\addlinespace[0.15cm]\hline\hline\addlinespace[0.15cm]
\multicolumn{4}{p{14.5cm}}{\scriptsize{\textbf{Notes:} All columns report estimates from Equation (\ref{baselineDiD}) using an age window of [-3, 3]. The regression samples are drawn from 2021 wave of the Political Culture Survey conducted by DANE. The dependent variable in all columns is the response to the question (in Spanish): ``Durante los \'ultimos 12 meses, usted ha sentido que lo han discriminado [...] por alguno de los siguientes motivos: Sexo [...]'' (``In the past 12 months, have you felt discriminated against [...] based on sex?''). All specifications include cohort-by-region, high school-by-region, and survey year-by-region fixed effects, as well as a set of ethnic group dummy variables. Robust standard errors, clustered at the region level, are reported in parentheses. Wild bootstrap confidence intervals are shown in square brackets.  *, **, and *** indicate statistical significance at the 10\%, 5\%, and 1\% levels, respectively. } }
\end{tabular}
}
\end{center}
\end{table}


\begin{landscape}
\begin{table}[H]
\vspace{3cm}
\begin{center}
{
\renewcommand{\arraystretch}{1}
\setlength{\tabcolsep}{2pt}
\caption {Effect of Exposure to Courses on the 1991 Constitution on Support for Gender Equality in the Domestic Sphere}  \label{table_did_domestic1}
\vspace{-0.3cm}
\small
\centering  \begin{tabular}{lcccccc}
\hline\hline \addlinespace[0.15cm]
    & (1)& (2)& (3)& (4)& (5)& (6)\\\addlinespace[0.12cm]\hline\addlinespace[0.15cm]
      &\multicolumn{6}{c}{Dependent Variable: Disagreement with Statement}\\\cmidrule[0.2pt](l){2-7}\addlinespace[0.1cm]
         &\multicolumn{2}{c}{Women are better suited}&\multicolumn{2}{c}{The husband should make}&\multicolumn{2}{c}{The head of the household}\\\addlinespace[0.1cm]
                                                    
                    &\multicolumn{2}{c}{for domestic tasks}&\multicolumn{2}{c}{decisions concerning his wife’s life}&\multicolumn{2}{c}{should be the man}\\\cmidrule[0.2pt](l){2-3}\cmidrule[0.2pt](l){4-5}\cmidrule[0.2pt](l){6-7}\addlinespace[0.1cm]
                   &Women&Men& Women&Men& Women&Men\\\addlinespace[0.1cm]\cmidrule[0.2pt](l){2-2} \cmidrule[0.2pt](l){3-3}\cmidrule[0.2pt](l){4-4}\cmidrule[0.2pt](l){5-5}\cmidrule[0.2pt](l){6-6}\cmidrule[0.2pt](l){7-7}\addlinespace[0.12cm]                 (Age $\leq$ 16 in 1994) $\times$ (Highschool)&      -0.003         &      -0.010         &       0.028         &      -0.018         &       0.042         &      -0.007         \\
                    &     (0.016)         &     (0.009)         &     (0.016)         &     (0.012)         &     (0.011)\sym{**}  &     (0.013)         \\
                    &[-0.046 0.041]           &[-0.030 0.019]           &[-0.019 0.067]          &[-0.053 0.004]             &[0.011 0.076]\sym{***}               &[-0.033 0.026]         \\
(Highschool)        &       0.174         &       0.164         &       0.073         &       0.126         &       0.132         &       0.132         \\
                    &     (0.008)\sym{***}&     (0.005)\sym{***}&     (0.008)\sym{***}&     (0.007)\sym{***}&     (0.007)\sym{***}&     (0.009)\sym{***}\\
                    &[0.012 0.234]\sym{***}            &[0.107 0.215]\sym{***}                &[-0.009 0.088]\sym{*}           &[0.096 0.202]\sym{***}            &[0.070 0.254]\sym{***}             &[0.019 0.212]\sym{**}          \\
R-sq                &       0.050         &       0.056         &       0.611         &       0.478         &       0.067         &       0.098         \\
Observations        &       13933         &       12128         &       13963         &       12187         &       14131         &       12261         \\

\addlinespace[0.15cm]\hline\hline\addlinespace[0.15cm]
\multicolumn{7}{p{22cm}}{\scriptsize{\textbf{Notes:} All columns report estimates from Equation (\ref{baselineDiD}) using an age window of [-3, 3]. The regression samples are drawn from the 2016-2017 and 2020-2021 waves of the ENUT Survey, conducted by DANE. The dependent variable in all columns is a binary indicator equal to 1 if the respondent expresses disagreement with the corresponding statement. This variable is a recoding of the original survey question, which asks respondents to indicate their level of agreement with the statement on a 1-to-4 scale. The original (pre-recoding) question used in columns (1) and (2) is: “¿Usted está de acuerdo con que las mujeres son mejores para el trabajo doméstico que los hombres?” (“Do you agree that women are better suited for domestic tasks than men?”). Columns (3) and (4) use: “¿Usted está de acuerdo con que el esposo debe tomar las decisiones relacionadas con la vida de la esposa?” (“Do you agree that the husband should make decisions concerning his wife’s life?”). Columns (5) and (6) use: “¿Usted está de acuerdo con que la cabeza del hogar debe ser el hombre?” (“Do you agree that the head of the household should be the man?”). All specifications include cohort-by-region, high school-by-region, and survey year-by-region fixed effects, as well as a set of ethnic group dummy variables. Robust standard errors, clustered at the region level, are reported in parentheses. Wild bootstrap confidence intervals are shown in square brackets. *, **, and *** indicate statistical significance at the 10\%, 5\%, and 1\% levels, respectively.} }
\end{tabular}
}
\end{center}
\end{table}
\end{landscape}

\begin{landscape}
\begin{table}[H]
\vspace{1cm}
\begin{center}
{
\renewcommand{\arraystretch}{0.6}
\setlength{\tabcolsep}{5pt}
\caption {Effect on Support for Gender Equality ( Years Prior to 1994 and Windows [-2, 2] and [-3, 3])}  \label{table_didbaseline_placebo23_boot}
\vspace{-0.3cm}
\scriptsize
\centering  \begin{tabular}{lcccccc}
\hline\hline \addlinespace[0.15cm]
    & (1)& (2)& (3)  & (4)& (5)& (6) \\\addlinespace[0.12cm]\cmidrule[0.2pt](l){2-7}\addlinespace[0.12cm]
& \multicolumn{6}{c}{Dependent Variable: Support for Gender Equality}\\\addlinespace[0.15cm]\cmidrule[0.2pt](l){2-7} \addlinespace[0.10cm]   
      & \multicolumn{3}{c}{Age window [-2,2]} & \multicolumn{3}{c}{Age window [-3,3]}\\\addlinespace[0.1cm]\cmidrule[0.2pt](l){2-4}\cmidrule[0.2pt](l){5-7}
                                 &\multicolumn{1}{c}{All}&\multicolumn{1}{c}{Women}&\multicolumn{1}{c}{Men}&\multicolumn{1}{c}{All}&\multicolumn{1}{c}{Women}&\multicolumn{1}{c}{Men}\\\addlinespace[0.1cm]\cmidrule[0.2pt](l){2-2} \cmidrule[0.2pt](l){3-3}\cmidrule[0.2pt](l){4-4}\cmidrule[0.2pt](l){5-5}\cmidrule[0.2pt](l){6-6}\cmidrule[0.2pt](l){7-7}\addlinespace[0.12cm]

        (Age $\leq$ 16 in 1992) $\times$ (Highschool)&      -0.005         &      -0.021         &       0.013         &      -0.003         &      -0.006         &       0.002         \\
                    &     (0.019)         &     (0.035)         &     (0.020)         &     (0.013)         &     (0.026)         &     (0.008)         \\
                    &[-0.050 0.188]         &[-0.077 0.341]         &[-0.039 0.053]         &[-0.028 0.151]          &[-0.045 0.273]         &[-0.021 0.057]         \\
R-sq                &       0.020         &       0.018         &       0.031         &       0.021         &       0.021         &       0.030         \\
Observations        &        7685         &        4082         &        3603         &       10945         &        5848         &        5097         \\
\addlinespace[0.1cm]\cmidrule[0.2pt](l){2-2} \cmidrule[0.2pt](l){3-3}\cmidrule[0.2pt](l){4-4}\cmidrule[0.2pt](l){5-5}\cmidrule[0.2pt](l){6-6}\cmidrule[0.2pt](l){7-7}\addlinespace[0.12cm]    
        (Age $\leq$ 16 in 1990) $\times$ (Highschool)&       0.003         &      -0.000         &       0.009         &       0.000         &      -0.006         &       0.009         \\
                    &     (0.007)         &     (0.015)         &     (0.012)         &     (0.008)         &     (0.017)         &     (0.021)         \\
                    &[-0.061 0.026]           &[-0.093 0.038]               &[-0.026 0.043]        &[-0.020 0.032]        &[-0.056 0.045]      &[-0.049 0.067]         \\
R-sq                &       0.016         &       0.022         &       0.019         &       0.019         &       0.021         &       0.025         \\
Observations        &        7582         &        4020         &        3562         &       10660         &        5615         &        5045         \\
\addlinespace[0.1cm]\cmidrule[0.2pt](l){2-2} \cmidrule[0.2pt](l){3-3}\cmidrule[0.2pt](l){4-4}\cmidrule[0.2pt](l){5-5}\cmidrule[0.2pt](l){6-6}\cmidrule[0.2pt](l){7-7}\addlinespace[0.12cm]
        (Age $\leq$ 16 in 1988) $\times$ (Highschool)&      -0.004         &      -0.012         &       0.005         &      -0.003         &       0.006         &      -0.012         \\
                    &     (0.015)         &     (0.019)         &     (0.030)         &     (0.016)         &     (0.010)         &     (0.027)         \\
                    &[-0.049 0.050]            &[-0.052 0.116]           &[-0.087 0.093]          &[-0.049 0.038]         &[-0.006 0.044]        &[-0.125 0.057]         \\
R-sq                &       0.017         &       0.022         &       0.020         &       0.015         &       0.020         &       0.019         \\
Observations        &        7542         &        4009         &        3533         &       10705         &        5664         &        5041         \\
\addlinespace[0.1cm]\cmidrule[0.2pt](l){2-2} \cmidrule[0.2pt](l){3-3}\cmidrule[0.2pt](l){4-4}\cmidrule[0.2pt](l){5-5}\cmidrule[0.2pt](l){6-6}\cmidrule[0.2pt](l){7-7}\addlinespace[0.12cm]
        (Age $\leq$ 16 in 1986) $\times$ (Highschool)&      -0.015         &       0.001         &      -0.033         &       0.004         &       0.012         &      -0.004         \\
                    &     (0.008)         &     (0.023)         &     (0.021)         &     (0.004)         &     (0.011)         &     (0.020)         \\
                    &[-0.073 -0.002]          &[-0.100 0.044]          &[-0.058 0.043]            &[-0.019 0.016]          &[-0.019 0.028]         &[-0.044 0.059]         \\
R-sq                &       0.014         &       0.019         &       0.017         &       0.015         &       0.019         &       0.019         \\
Observations        &        7640         &        4063         &        3577         &       10574         &        5649         &        4925         \\
\addlinespace[0.1cm]\cmidrule[0.2pt](l){2-2} \cmidrule[0.2pt](l){3-3}\cmidrule[0.2pt](l){4-4}\cmidrule[0.2pt](l){5-5}\cmidrule[0.2pt](l){6-6}\cmidrule[0.2pt](l){7-7}\addlinespace[0.12cm]
        (Age $\leq$ 16 in 1984) $\times$ (Highschool)&       0.015         &       0.010         &       0.023         &       0.016         &       0.011         &       0.021         \\
                    &     (0.020)         &     (0.025)         &     (0.015)         &     (0.018)         &     (0.022)         &     (0.022)         \\
                    &[-0.024 0.068]            &[-0.035 0.070]         &[-0.013 0.111]          &[-0.022 0.074]           &[-0.036 0.061]          &[-0.044 0.090]         \\
R-sq                &       0.017         &       0.020         &       0.023         &       0.017         &       0.021         &       0.022         \\
Observations        &        7544         &        4080         &        3464         &       10697         &        5730         &        4967         \\

\addlinespace[0.15cm]\hline\hline\addlinespace[0.15cm]
\multicolumn{7}{p{17cm}}{\scriptsize{\textbf{Notes:} All columns report estimates from Equation (\ref{baselineDiD}), but instead of using the indicator for being at most 16 years old in 1994, we replace it with an indicator for being that age in the specific year indicated in each row. Specifications in columns (1)–(3) use an age window of $[-2, 2]$, while those in columns (4)–(6) use $[-3, 3]$.  The regression samples are drawn from the 2019 and 2021 waves of the Political Culture Survey conducted by DANE. The dependent variable in all columns is the response to the question (in Spanish): ``¿Usted est\'a de acuerdo con que las mujeres participen en igualdad de condiciones y oportunidades?'' (``Do you agree that women should participate on equal terms and have the same opportunities?''). All the specifications include cohort-by-region, high school-by-region, and survey year-by-region fixed effects, as well as a set of ethnic group dummies. Robust standard errors, clustered at the region level, are reported in parentheses. Wild bootstrap confidence intervals are shown in square brackets.  *, **, and *** indicate statistical significance at the 10\%, 5\%, and 1\% levels, respectively.} }
\end{tabular}
}
\end{center}
\end{table}
\end{landscape}

\begin{table}[H]
\begin{center}
{
\renewcommand{\arraystretch}{0.8}
\setlength{\tabcolsep}{8pt}
\caption {Effect of Exposure to 1991 Constitution Courses on Support for Gender Equality After 1994}  \label{table_didgenderequalitypost1994}
\vspace{-0.3cm}
\small
\centering  \begin{tabular}{lccc}
\hline\hline \addlinespace[0.15cm]
    & (1)& (2)& (3)\\\addlinespace[0.12cm]\cmidrule[0.2pt](l){2-4}\addlinespace[0.12cm]
      & \multicolumn{3}{c}{Dependent Variable: Support for Gender Equality}\\\addlinespace[0.1cm]\cmidrule[0.2pt](l){2-4}\addlinespace[0.1cm]
                      &\multicolumn{1}{c}{All}&\multicolumn{1}{c}{Women}&\multicolumn{1}{c}{Men}\\
                \addlinespace[0.1cm]\cmidrule[0.2pt](l){1-1} \cmidrule[0.2pt](l){2-2} \cmidrule[0.2pt](l){3-3}\cmidrule[0.2pt](l){4-4}\addlinespace[0.05cm]
                
(Age $\leq$ 16 in 1996) $\times$ (Highschool)&      -0.003   &      -0.014   &       0.008   \\
                    &     (0.007)   &     (0.023)   &     (0.022)   \\
                    &[-0.017 0.014]     &[-0.059 0.067]          &[-0.040 0.056]   \\
R-sq                &       0.024   &       0.025   &       0.032   \\
Observations        &       12328   &        6608   &        5720   \\
\addlinespace[0.1cm]\cmidrule[0.2pt](l){1-1} \cmidrule[0.2pt](l){2-2} \cmidrule[0.2pt](l){3-3}\cmidrule[0.2pt](l){4-4}\addlinespace[0.05cm]
(Age $\leq$ 16 in 1998) $\times$ (Highschool)&       0.020   &       0.028   &       0.014   \\
                    &     (0.011)   &     (0.028)   &     (0.013)   \\
                    &[-0.004 0.037]         &[-0.030 0.101]       &[-0.014 0.055]   \\
R-sq                &       0.021   &       0.027   &       0.023   \\
Observations        &       12905   &        6976   &        5929   \\
\addlinespace[0.1cm]\cmidrule[0.2pt](l){1-1} \cmidrule[0.2pt](l){2-2} \cmidrule[0.2pt](l){3-3}\cmidrule[0.2pt](l){4-4}\addlinespace[0.05cm]
(Age $\leq$ 16 in 2000) $\times$ (Highschool)&       0.006   &       0.007   &       0.003   \\
                    &     (0.014)   &     (0.031)   &     (0.006)   \\
                    &[-0.092 0.051]       &[-0.183 0.104]       &[-0.011 0.024]   \\
R-sq                &       0.020   &       0.029   &       0.019   \\
Observations        &       12678   &        6934   &        5744   \\
\addlinespace[0.1cm]\cmidrule[0.2pt](l){1-1} \cmidrule[0.2pt](l){2-2} \cmidrule[0.2pt](l){3-3}\cmidrule[0.2pt](l){4-4}\addlinespace[0.05cm]
(Age $\leq$ 16 in 2002) $\times$ (Highschool)&      -0.027   &      -0.032   &      -0.022   \\
                    &     (0.011)*  &     (0.025)   &     (0.017)   \\
                    &[-0.062 -0.008]         &[-0.094 0.008]           &[-0.064 0.029]   \\
R-sq                &       0.021   &       0.027   &       0.019   \\
Observations        &       12535   &        6818   &        5717   \\
\addlinespace[0.1cm]\cmidrule[0.2pt](l){1-1} \cmidrule[0.2pt](l){2-2} \cmidrule[0.2pt](l){3-3}\cmidrule[0.2pt](l){4-4}\addlinespace[0.05cm]
                
(Age $\leq$ 16 in 2004) $\times$ (Highschool)&      -0.019   &      -0.002   &      -0.034   \\
                    &     (0.010)   &     (0.013)   &     (0.013)*  \\
                    &[-0.045 0.003]      &[-0.109 0.022]       &[-0.073 0.048]   \\
R-sq                &       0.018   &       0.023   &       0.017   \\
Observations        &       12699   &        6851   &        5848   \\

\addlinespace[0.15cm]\hline\hline\addlinespace[0.15cm]
\multicolumn{4}{p{15cm}}{\scriptsize{\textbf{Notes:} All columns report estimates from Equation (\ref{baselineDiD}), but instead of using the indicator for being at most 16 years old in 1994, we replace it with an indicator for being that age in the specific year indicated in each row. All the specifications use an age window of $[-3, 3]$.  The regression samples are drawn from the 2019 and 2021 waves of the Political Culture Survey conducted by DANE. The dependent variable in all columns is the response to the question (in Spanish): ``¿Usted est\'a de acuerdo con que las mujeres participen en igualdad de condiciones y oportunidades?'' (``Do you agree that women should participate on equal terms and have the same opportunities?''). All the specifications include cohort-by-region, high school-by-region, and survey year-by-region fixed effects, as well as a set of ethnic group dummies. Robust standard errors, clustered at the region level, are reported in parentheses. Wild bootstrap confidence intervals are shown in square brackets.  *, **, and *** indicate statistical significance at the 10\%, 5\%, and 1\% levels, respectively.} }
\end{tabular}
}
\end{center}
\end{table}

\begin{table}[H]
\begin{center}
{
\renewcommand{\arraystretch}{0.8}
\setlength{\tabcolsep}{10pt}
\caption {Effect of Exposure to 1991 Constitution Courses on Political Ideology, Religious Tolerance and Labor Market Participation}  \label{table_alternexplanationsreligionideologylabor}
\vspace{-0.3cm}
\small
\centering  \begin{tabular}{lccc}
\hline\hline \addlinespace[0.15cm]
    & (1)& (2)& (3)\\\addlinespace[0.12cm]\cmidrule[0.2pt](l){2-4}\addlinespace[0.12cm]
                      &\multicolumn{1}{c}{All}&\multicolumn{1}{c}{Women}&\multicolumn{1}{c}{Men}\\\addlinespace[0.10cm]\hline
                \addlinespace[0.15cm]
      \multicolumn{1}{l}{\emph{\underline{Panel A}:  }}      & \multicolumn{3}{c}{Dep. Var: Right-Wing Ideology}\\\addlinespace[0.1cm]\cmidrule[0.2pt](l){2-4}\addlinespace[0.1cm]               
(Age $\leq$ 16 in 1994) $\times$ (Highschool)&      -0.086   &      -0.061   &      -0.134   \\
                    &     (0.050)   &     (0.047)   &     (0.118)   \\
                    &[-0.262 0.083]     &[-0.213 0.403]          &[-0.502 0.058]   \\
R-sq                &       0.039   &       0.038   &       0.048   \\
Observations        &        9494   &        4952   &        4542   \\

\addlinespace[0.15cm]\hline\addlinespace[0.15cm]
      \multicolumn{1}{l}{\emph{\underline{Panel B}:  }}      & \multicolumn{3}{c}{Dep. Var: Religiosity}\\\addlinespace[0.1cm]\cmidrule[0.2pt](l){2-4}\addlinespace[0.1cm]               
(Age $\leq$ 16 in 1994) $\times$ (Highschool)&       0.041   &       0.036   &       0.044   \\
                    &     (0.037)   &     (0.057)   &     (0.046)   \\
                    &[-0.038 0.126]         &[-0.147 0.348]          &[-0.411 0.177]   \\
R-sq                &       0.056   &       0.054   &       0.043   \\
Observations        &       12071   &        6418   &        5653   \\

\addlinespace[0.15cm]\hline\addlinespace[0.15cm]
      \multicolumn{1}{l}{\emph{\underline{Panel C}:  }}      & \multicolumn{3}{c}{Dep. Var: Religious Intolerance}\\\addlinespace[0.1cm]\cmidrule[0.2pt](l){2-4}\addlinespace[0.1cm]               
(Age $\leq$ 16 in 1994) $\times$ (Highschool)&      -0.002   &      -0.003   &       0.000   \\
                    &     (0.003)   &     (0.007)   &     (0.002)   \\
                    &[-0.012 0.003]        &[-0.043 0.010]      &[-0.003 0.012]   \\
R-sq                &       0.011   &       0.013   &       0.017   \\
Observations        &       12147   &        6455   &        5692   \\

\addlinespace[0.15cm]\hline\addlinespace[0.15cm]
      \multicolumn{1}{l}{\emph{\underline{Panel D}:  }}      & \multicolumn{3}{c}{Dep. Var: Labor Market Participation}\\\addlinespace[0.1cm]\cmidrule[0.2pt](l){2-4}\addlinespace[0.1cm]               
(Age $\leq$ 16 in 1994) $\times$ (Highschool)&       0.027   &       0.038   &       0.011   \\
                    &     (0.018)   &     (0.038)   &     (0.011)   \\
                    &[-0.034 0.069]     &[-0.093 0.112]       &[-0.021 0.041]   \\
R-sq                &       0.210   &       0.094   &       0.014   \\
Observations        &       12098   &        6425   &        5673   \\

\addlinespace[0.15cm]\hline\hline\addlinespace[0.15cm]
\multicolumn{4}{p{15cm}}{\scriptsize{\textbf{Notes:} All columns report estimates from Equation (\ref{baselineDiD}) using an age window of [-3, 3]. The regression samples are drawn from the 2019 and 2021 waves of the Political Culture Survey conducted by DANE. The dependent variable in Panel A measures the respondent’s political ideology, with higher values indicating a more right-wing position. The exact question is: ``Las personas cuando piensan en pol\'itica utilizan los t\'erminos izquierda y derecha. En una escala de 1 a 10 donde 1 significa izquierda y 10 significa derecha ¿d\'onde se ubicar\'ia usted?'' (``People often use the terms left and right when thinking about politics. On a scale from 1 to 10, where 1 means left and 10 means right, where would you place yourself?'').  The dependent variable in Panel B measures the importance of religion in the respondent’s life. The exact question is: ``En una escala de 1 a 5, donde 1 significa nada importante y 5 muy importante, qu\'e tan importantes son los siguientes grupos de personas o elementos en su vida: La religi\'on.'' (“On a scale from 1 to 5, where 1 means not at all important and 5 means very important, how important are the following groups or elements in your life: Religion?”).  The dependent variable in Panel C measures tolerance toward people of a different religion. The exact question is: ``De las siguientes personas, ¿a qui\'en no quisiera tener de vecino(a)?: Personas que profesan una religi\'on diferente.'' (“Which of the following types of people would you not want to have as a neighbor? People who profess a religion different from yours.”).  The dependent variable in Panel D captures labor market participation. The exact question is: ¿En qu\'e actividad ocup\'o <...> la mayor parte del tiempo la semana pasada?'' (: “What activity occupied most of your time last week?”).  A binary variable is coded as 1 if the respondent reported either working or looking for work.  All specifications include cohort-by-region, high school-by-region, and survey year-by-region fixed effects, as well as a set of ethnic group dummy variables. Robust standard errors, clustered at the region level, are reported in parentheses. Wild bootstrap confidence intervals are shown in square brackets.} }
\end{tabular}
}
\end{center}
\end{table}

\bibliographystyle{ecca}
\bibliography{bib}



\newpage
\hbox {} 
\appendix
\section{Appendix}
\setcounter{table}{0}
\setcounter{figure}{0}
\renewcommand{\thefigure}{\Alph{section}\arabic{figure}}
\renewcommand{\thetable}{\Alph{section}\arabic{table}}

\subsection{Additional Figures and Tables}

\begin{table}[h!]
\centering
\caption{Gender Equality Provisions in Latin American Constitutions}
\vspace{-0.3cm}
\label{ONU}
\begin{tabular}{@{}lclc@{}}
\hline\hline \addlinespace[0.15cm]
Country & Year & Article & Female Population \\ 
 & Introduced &  &  Affected (millions) \\ \cmidrule[0.2pt](l){1-1}\cmidrule[0.2pt](l){2-2}\cmidrule[0.2pt](l){3-3}\cmidrule[0.2pt](l){4-4}
Uruguay     & 1967             & Art. 8    & $\sim$1.4  \\
Mexico      & 1974             & Art. 4    & $\sim$28.7 \\
Chile       & 1980             & Art. 19   & $\sim$5.8 \\
Brazil      & 1988             & Art. 5    & $\sim$72.1  \\
Colombia    & 1991             & Art. 13   & $\sim$16.6 \\
Peru        & 1993             & Art. 2    & $\sim$11.7  \\
Argentina   & 1994             & Art. 16   & $\sim$17.6  \\
Ecuador     & 2008             & Art. 11   & $\sim$7.3  \\ \addlinespace[0.1cm]\hline\hline\addlinespace[0.1cm]
\multicolumn{4}{p{11cm}}{\scriptsize \textbf{Notes:} Constitutional texts can be consulted at \url{https://www.constituteproject.org/}. Female population data for the year each provision was introduced is available at the World Bank: \url{https://datos.bancomundial.org}.}
\end{tabular} 
\end{table}

\begin{figure}[h!]
 	\caption{True democracy cannot exist without gender equality}
    \label{fig:constituyente}
	\centering	
\begin{subfigure}[t]{0.25\textwidth}
	\centering
	\includegraphics[width=\textwidth]{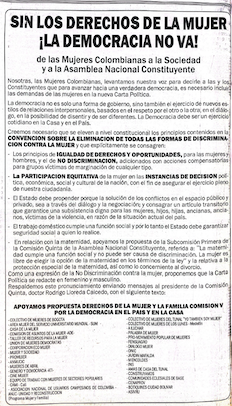}
	\caption{ El Tiempo, April 28th}
\end{subfigure}
\begin{subfigure}[t]{0.3\textwidth}
	\centering
	\includegraphics[width=\textwidth]{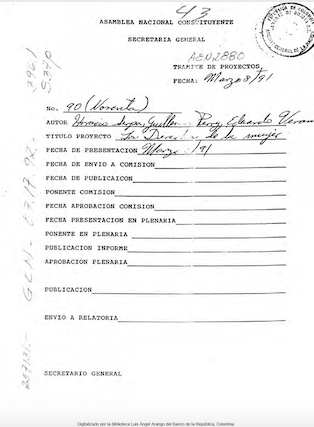}
	\caption{Constituyente, March 8 1991}
\end{subfigure}\\
\begin{minipage}{14cm} \scriptsize 
The left side features a publication in the newspaper El Tiempo on April 28th, endorsed by members of 35 organizations, directed towards society and the National Constituent Assembly. The publication emphasizes the main slogan of the Movement during that period: ``WITHOUT WOMEN'S RIGHTS, DEMOCRACY WON'T SUCCEED!". On the right side, there is the Statement titled ``Women's Rights" (Constituyente, 1991) signed by constituents Horacio Serpa, Guillermo Perry, and Eduardo Verano de la Rosa.
\end{minipage}
\end{figure}

\newpage

\begin{figure}[h!]
	\centering
 	\caption{Booklet titled ``Combating
Stereotyped Gender Roles” published in 1994 by the Ministry of National Education of Colombia}
 \begin{subfigure}[t]{0.3\textwidth}
	\centering
	\includegraphics[width=\textwidth]{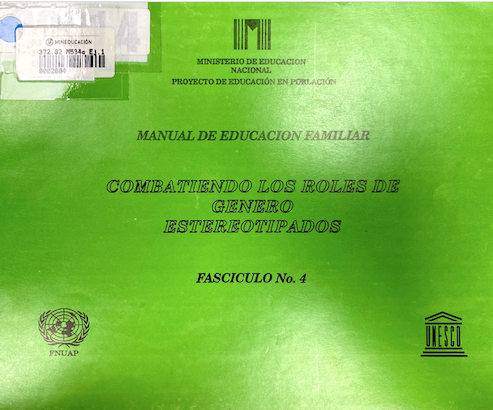}
\end{subfigure}
\begin{subfigure}[t]{0.3\textwidth}
	\centering
	\includegraphics[width=\textwidth]{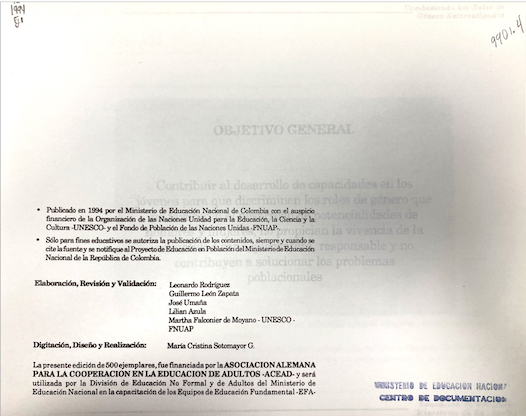}
\end{subfigure}
\begin{subfigure}[t]{0.3\textwidth}
	\centering
	\includegraphics[width=\textwidth]{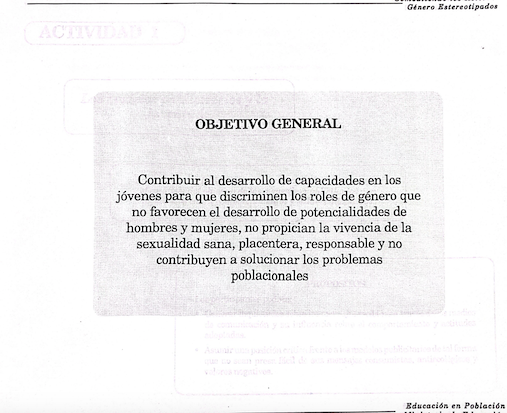}
\end{subfigure}
		\label{fig:booklet1}
\end{figure}


\begin{figure}[H]
	\centering
 	\caption{Booklet titled ``Coeducation guide" published in 1999 by the Ministry of National Education of Colombia}
 \begin{subfigure}[t]{0.3\textwidth}
	\centering
	\includegraphics[width=\textwidth]{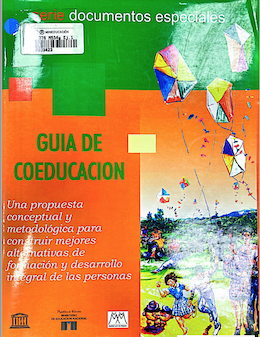}
\end{subfigure}
\begin{subfigure}[t]{0.3\textwidth}
	\centering
	\includegraphics[width=\textwidth]{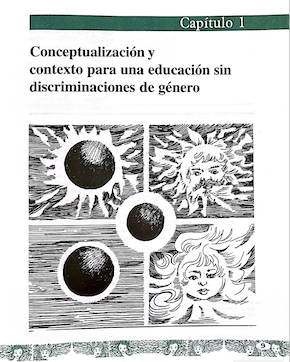}
\end{subfigure}
\begin{subfigure}[t]{0.3\textwidth}
	\centering
	\includegraphics[width=\textwidth]{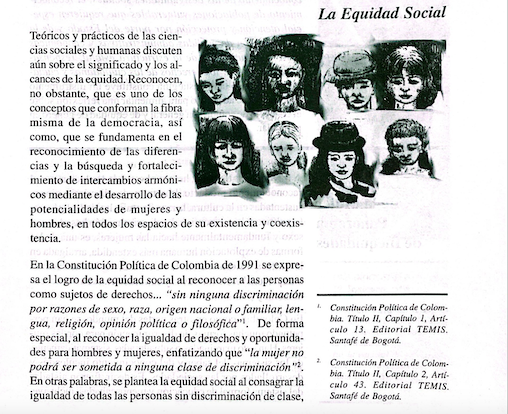}
\end{subfigure}
		\label{fig:booklet2}
\end{figure}

\newpage

\begin{table}[h!]
\begin{center}
{
\renewcommand{\arraystretch}{0.5}
\setlength{\tabcolsep}{10pt}
\caption {Effect on Perception of Gender Discrimination (Years After 1994)}  \label{table_didsexdiscrimpost1994}
\vspace{-0.3cm}
\footnotesize
\centering  \begin{tabular}{lccc}
\hline\hline \addlinespace[0.15cm]
    & (1)& (2)& (3)\\\addlinespace[0.12cm]\cmidrule[0.2pt](l){2-4}\addlinespace[0.05cm]
      & \multicolumn{3}{c}{Dep. Var.: Perception of Gender Discrimination}\\\addlinespace[0.1cm]\cmidrule[0.2pt](l){2-4}\addlinespace[0.1cm]
                      &\multicolumn{1}{c}{All}&\multicolumn{1}{c}{Women}&\multicolumn{1}{c}{Men}\\
                \addlinespace[0.1cm]\cmidrule[0.2pt](l){1-1} \cmidrule[0.2pt](l){2-2} \cmidrule[0.2pt](l){3-3}\cmidrule[0.2pt](l){4-4}\addlinespace[0.05cm]
                
(Age $\leq$ 16 in 1996) $\times$ (Highschool)&       0.007   &       0.009   &       0.004   \\
                    &     (0.010)   &     (0.015)   &     (0.005)   \\
                    &[-0.033 0.043]       &[-0.101 0.062]       &[-0.005 0.021]   \\
R-sq                &       0.019   &       0.022   &       0.043   \\
Observations        &        6755   &        3650   &        3105   \\
\addlinespace[0.1cm]\cmidrule[0.2pt](l){1-1} \cmidrule[0.2pt](l){2-2} \cmidrule[0.2pt](l){3-3}\cmidrule[0.2pt](l){4-4}\addlinespace[0.05cm]
(Age $\leq$ 16 in 1998) $\times$ (Highschool)&       0.002   &       0.007   &      -0.004   \\
                    &     (0.002)   &     (0.004)   &     (0.002)*  \\
                    &[-0.002 0.046]     &[-0.006 0.086]         &[-0.007 0.003]   \\
R-sq                &       0.015   &       0.017   &       0.032   \\
Observations        &        7163   &        3908   &        3255   \\
\addlinespace[0.1cm]\cmidrule[0.2pt](l){1-1} \cmidrule[0.2pt](l){2-2} \cmidrule[0.2pt](l){3-3}\cmidrule[0.2pt](l){4-4}\addlinespace[0.05cm]
(Age $\leq$ 16 in 2000) $\times$ (Highschool)&       0.000   &      -0.003   &       0.004   \\
                    &     (0.004)   &     (0.010)   &     (0.006)   \\
                    &[-0.046 0.006]         &[-0.117 0.014]       &[-0.006 0.024]   \\
R-sq                &       0.014   &       0.014   &       0.033   \\
Observations        &        7032   &        3864   &        3168   \\

\addlinespace[0.15cm]\hline\hline\addlinespace[0.15cm]
\multicolumn{4}{p{14.5cm}}{\scriptsize{\textbf{Notes:} All columns report estimates from Equation (\ref{baselineDiD}), but instead of using the indicator for being at most 16 years old in 1994, we replace it with an indicator for being that age in the specific year indicated in each row. All the specifications use an age window of $[-3, 3]$.  The regression samples are drawn from the 2019 and 2021 waves of the Political Culture Survey conducted by DANE. The dependent variable in all columns is the response to the question (in Spanish): ``Durante los \'ultimos 12 meses, usted ha sentido que lo han discriminado [...] por alguno de los siguientes motivos: Sexo [...]'' (``In the past 12 months, have you felt discriminated against [...] based on sex?''). All the specifications include cohort-by-region, high school-by-region, and survey year-by-region fixed effects, as well as a set of ethnic group dummies. Robust standard errors, clustered at the region level, are reported in parentheses. Wild bootstrap confidence intervals are shown in square brackets.  *, **, and *** indicate statistical significance at the 10\%, 5\%, and 1\% levels, respectively.} }
\end{tabular}
}
\end{center}
\end{table}


\begin{table}[H]
\begin{center}
{
\renewcommand{\arraystretch}{0.5}
\setlength{\tabcolsep}{10pt}
\caption {Effect of Treatment on Disagreement with the Statement that Women are Better Suited for Domestic Tasks than Men (Years After 1994)}  \label{table_diddomesticpost1994_op4}
\vspace{-0.3cm}
\footnotesize 
\centering  \begin{tabular}{lccc}
\hline\hline \addlinespace[0.05cm]
    & (1)& (2)& (3)\\\addlinespace[0.05cm]\cmidrule[0.2pt](l){2-4}\addlinespace[0.05cm]
  & \multicolumn{3}{c}{Dep. Var: Disagree with}\\\addlinespace[0.10cm]
   & \multicolumn{3}{c}{``women are better suited for domestic tasks''}\\    \addlinespace[0.1cm]\cmidrule[0.2pt](l){2-4}\addlinespace[0.1cm]
                      &\multicolumn{1}{c}{All}&\multicolumn{1}{c}{Women}&\multicolumn{1}{c}{Men}\\
                \addlinespace[0.1cm]\cmidrule[0.2pt](l){1-1} \cmidrule[0.2pt](l){2-2} \cmidrule[0.2pt](l){3-3}\cmidrule[0.2pt](l){4-4}\addlinespace[0.05cm]
                
(Age $\leq$ 16 in 1996) $\times$ (Highschool)&       0.001   &      -0.007   &       0.008   \\
                    &     (0.015)   &     (0.015)   &     (0.019)   \\
                    &[-0.039 0.041]     &[-0.050 0.083]    &[-0.034 0.047]   \\
R-sq                &       0.047   &       0.045   &       0.053   \\
Observations        &       27133   &       14535   &       12598   \\
\addlinespace[0.05cm]\cmidrule[0.2pt](l){1-1} \cmidrule[0.2pt](l){2-2} \cmidrule[0.2pt](l){3-3}\cmidrule[0.2pt](l){4-4}\addlinespace[0.05cm]
(Age $\leq$ 16 in 1998) $\times$ (Highschool)&       0.014   &       0.013   &       0.017   \\
                    &     (0.005)** &     (0.013)   &     (0.014)   \\
                    &[-0.009 0.022]**    &[-0.019 0.057]       &[-0.057 0.043]   \\
R-sq                &       0.048   &       0.046   &       0.054   \\
Observations        &       28217   &       15219   &       12998   \\
\addlinespace[0.05cm]\cmidrule[0.2pt](l){1-1} \cmidrule[0.2pt](l){2-2} \cmidrule[0.2pt](l){3-3}\cmidrule[0.2pt](l){4-4}\addlinespace[0.05cm]
(Age $\leq$ 16 in 2000) $\times$ (Highschool)&      -0.003   &      -0.003   &      -0.000   \\
                    &     (0.008)   &     (0.014)   &     (0.018)   \\
                    &[-0.042 0.030]         &[-0.059 0.034]      &[-0.034 0.068]   \\
R-sq                &       0.044   &       0.042   &       0.049   \\
Observations        &       28640   &       15468   &       13172   \\

\addlinespace[0.10cm]\hline\hline\addlinespace[0.10cm]
\multicolumn{4}{p{14.4cm}}{\scriptsize{\textbf{Notes:} All columns report estimates from Equation (\ref{baselineDiD}), but instead of using the indicator for being at most 16 years old in 1994, we replace it with an indicator for being that age in the specific year indicated in each row. All the specifications use an age window of $[-3, 3]$.  The regression samples are drawn from the 2017–2016 and 2020-2021 waves of the ENUT Survey, conducted by DANE. The dependent variable in all columns is a binary indicator equal to 1 if the respondent expresses disagreement with the following statement: “¿Las mujeres son mejores para el trabajo doméstico que los hombres?” (“Women are better suited for domestic tasks than men?”). All the specifications include cohort-by-region, high school-by-region, and survey year-by-region fixed effects, as well as a set of ethnic group dummies. Robust standard errors, clustered at the region level, are reported in parentheses. Wild bootstrap confidence intervals are shown in square brackets.  *, **, and *** indicate statistical significance at the 10\%, 5\%, and 1\% levels, respectively.} }
\end{tabular}
}
\end{center}
\end{table}


\begin{table}[H]
\begin{center}
{
\renewcommand{\arraystretch}{0.5}
\setlength{\tabcolsep}{10pt}
\caption {Effect of Treatment on Disagreement with ``the Husband Should Make Decisions Concerning his Wife’s Life'' (Years After 1994)}  \label{table_diddomesticpost1994_op5}
\vspace{-0.3cm}
\footnotesize 
\centering  \begin{tabular}{lccc}
\hline\hline \addlinespace[0.05cm]
    & (1)& (2)& (3)\\\addlinespace[0.05cm]\cmidrule[0.2pt](l){2-4}\addlinespace[0.05cm]
  & \multicolumn{3}{c}{Dep. Var: Disagreement with }\\\addlinespace[0.10cm]
   & \multicolumn{3}{c}{``husbands should make decisions concerning wifes’ life''}\\    \addlinespace[0.1cm]\cmidrule[0.2pt](l){2-4}\addlinespace[0.1cm]
                      &\multicolumn{1}{c}{All}&\multicolumn{1}{c}{Women}&\multicolumn{1}{c}{Men}\\
                \addlinespace[0.1cm]\cmidrule[0.2pt](l){1-1} \cmidrule[0.2pt](l){2-2} \cmidrule[0.2pt](l){3-3}\cmidrule[0.2pt](l){4-4}\addlinespace[0.05cm]
                
(Age $\leq$ 16 in 1996) $\times$ (Highschool)&       0.001   &       0.044   &      -0.038   \\
                    &     (0.010)   &     (0.007)***&     (0.014)** \\
                    &[-0.016 0.037]        &[0.025 0.066]***      &[-0.061 0.036]**   \\
R-sq                &       0.558   &       0.626   &       0.490   \\
Observations        &       27192   &       14546   &       12646   \\
\addlinespace[0.1cm]\cmidrule[0.2pt](l){1-1} \cmidrule[0.2pt](l){2-2} \cmidrule[0.2pt](l){3-3}\cmidrule[0.2pt](l){4-4}\addlinespace[0.05cm]
(Age $\leq$ 16 in 1998) $\times$ (Highschool)&      -0.023   &      -0.030   &      -0.016   \\
                    &     (0.010)*  &     (0.008)** &     (0.015)   \\
                    &[-0.040 0.011]           &[-0.053 -0.011]       &[-0.039 0.039]   \\
R-sq                &       0.578   &       0.643   &       0.511   \\
Observations        &       28294   &       15231   &       13063   \\
\addlinespace[0.1cm] \cmidrule[0.2pt](l){1-1}\cmidrule[0.2pt](l){2-2} \cmidrule[0.2pt](l){3-3}\cmidrule[0.2pt](l){4-4}\addlinespace[0.05cm]
(Age $\leq$ 16 in 2000) $\times$ (Highschool)&      -0.008   &      -0.008   &      -0.009   \\
                    &     (0.007)   &     (0.016)   &     (0.009)   \\
                    &[-0.055 0.017]       &[-0.053 0.042]        &[-0.050 0.014]   \\
R-sq                &       0.588   &       0.657   &       0.516   \\
Observations        &       28715   &       15468   &       13247   \\

\addlinespace[0.10cm]\hline\hline\addlinespace[0.10cm]
\multicolumn{4}{p{14.4cm}}{\scriptsize{\textbf{Notes:} All columns report estimates from Equation (\ref{baselineDiD}), but instead of using the indicator for being at most 16 years old in 1994, we replace it with an indicator for being that age in the specific year indicated in each row. All the specifications use an age window of $[-3, 3]$.  The regression samples are drawn from the 2016-2017 and 2020-2021 waves of the ENUT Survey, conducted by DANE. The dependent variable in all columns is a binary indicator equal to 1 if the respondent expresses disagreement with the following statement:  “El esposo debe tomar las decisiones relacionadas con la vida de la esposa?” (“The husband should make decisions concerning his wife’s life?”). All the specifications include cohort-by-region, high school-by-region, and survey year-by-region fixed effects, as well as a set of ethnic group dummies. Robust standard errors, clustered at the region level, are reported in parentheses. Wild bootstrap confidence intervals are shown in square brackets.  *, **, and *** indicate statistical significance at the 10\%, 5\%, and 1\% levels, respectively.} }
\end{tabular}
}
\end{center}
\end{table}

\newpage

\begin{table}[H]
\begin{center}
{
\renewcommand{\arraystretch}{0.5}
\setlength{\tabcolsep}{10pt}
\caption {Effect of Treatment on Disagreement with  ``The Head of the Household Should be the Man'' (Years After 1994)}  \label{table_diddomesticpost1994_op6}
\vspace{-0.3cm}
\footnotesize 
\centering  \begin{tabular}{lccc}
\hline\hline \addlinespace[0.05cm]
    & (1)& (2)& (3)\\\addlinespace[0.05cm]\cmidrule[0.2pt](l){2-4}\addlinespace[0.05cm]
                      &\multicolumn{1}{c}{All}&\multicolumn{1}{c}{Women}&\multicolumn{1}{c}{Men}\\\addlinespace[0.10cm]\hline
                \addlinespace[0.10cm]
    \multicolumn{1}{l}{\emph{\underline{Panel C}:  }}    & \multicolumn{3}{c}{Dep. Var: Disagreement with}\\\addlinespace[0.10cm]
   & \multicolumn{3}{c}{``the head of the household should be the man''}\\    
          
      \addlinespace[0.10cm]\cmidrule[0.2pt](l){1-1}\cmidrule[0.2pt](l){2-4}\addlinespace[0.05cm]  
                
(Age $\leq$ 16 in 1996) $\times$ (Highschool)&       0.005   &       0.022   &      -0.007   \\
                    &     (0.018)   &     (0.021)   &     (0.018)   \\
                    &[-0.034 0.053]         &[-0.041 0.071]         &[-0.036 0.046]   \\
R-sq                &       0.094   &       0.066   &       0.097   \\
Observations        &       27436   &       14721   &       12715   \\
\addlinespace[0.05cm]\cmidrule[0.2pt](l){1-1} \cmidrule[0.2pt](l){2-2} \cmidrule[0.2pt](l){3-3}\cmidrule[0.2pt](l){4-4}\addlinespace[0.05cm]
(Age $\leq$ 16 in 1998) $\times$ (Highschool)&       0.002   &      -0.006   &       0.007   \\
                    &     (0.008)   &     (0.009)   &     (0.013)   \\
                    &[-0.024 0.024]        &[-0.056 0.010]        &[-0.026 0.036]   \\
R-sq                &       0.094   &       0.067   &       0.098   \\
Observations        &       28551   &       15433   &       13118   \\
\addlinespace[0.05cm]\cmidrule[0.2pt](l){1-1} \cmidrule[0.2pt](l){2-2} \cmidrule[0.2pt](l){3-3}\cmidrule[0.2pt](l){4-4}\addlinespace[0.05cm]
(Age $\leq$ 16 in 2000) $\times$ (Highschool)&       0.004   &      -0.018   &       0.023   \\
                    &     (0.025)   &     (0.032)   &     (0.030)   \\
                    &[-0.058 0.079]           &[-0.164 0.088]     &[-0.044 0.214]   \\
R-sq                &       0.091   &       0.067   &       0.089   \\
Observations        &       29000   &       15698   &       13302   \\

\addlinespace[0.05cm]\hline\hline\addlinespace[0.05cm]
\multicolumn{4}{p{14.4cm}}{\scriptsize{\textbf{Notes:} All columns report estimates from Equation (\ref{baselineDiD}), but instead of using the indicator for being at most 16 years old in 1994, we replace it with an indicator for being that age in the specific year indicated in each row. All the specifications use an age window of $[-3, 3]$.  The regression samples are drawn from the 2017–2016 and 2020-2021 waves of the ENUT Survey, conducted by DANE. The dependent variable in all columns is a binary indicator equal to 1 if the respondent expresses disagreement with the following statement:   “¿La cabeza del hogar debe ser el hombre?” (“The head of the household should be the man?”). All the specifications include cohort-by-region, high school-by-region, and survey year-by-region fixed effects, as well as a set of ethnic group dummies. Robust standard errors, clustered at the region level, are reported in parentheses. Wild bootstrap confidence intervals are shown in square brackets.  *, **, and *** indicate statistical significance at the 10\%, 5\%, and 1\% levels, respectively.} }
\end{tabular}
}
\end{center}
\end{table}

\hbox{}
\begin{figure}[H]
             \caption{Effect of Exposure to Courses on the 1991 Constitution on Labor Force Participation: Heterogeneous Effects by Gender}
        \label{fig_windowslaborforcepart_gender}
\vspace{-0.3cm}
\begin{subfigure}{0.5\textwidth}
\caption{Women's Labor Force Participation} \label{fig_didbaseline_genderF}
\includegraphics[width=\linewidth]{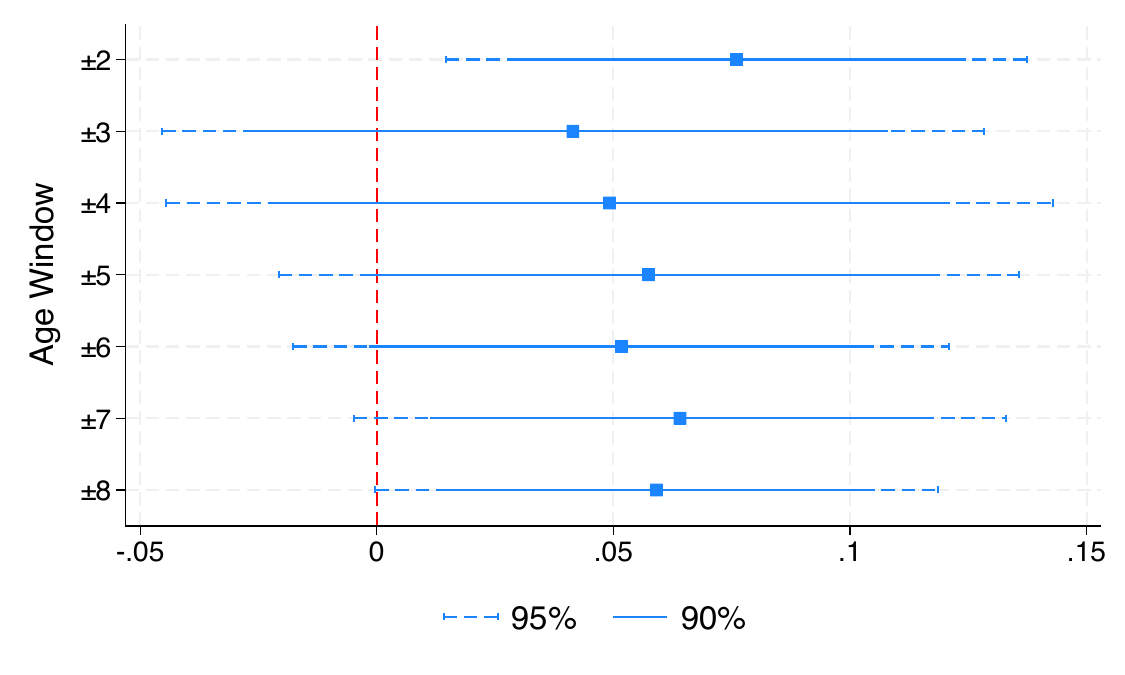}
\end{subfigure}\hspace*{\fill}
\begin{subfigure}{0.5\textwidth}
\caption{Men's Labor Force Participation} \label{fig_didbaseline_genderM}
\includegraphics[width=\linewidth]{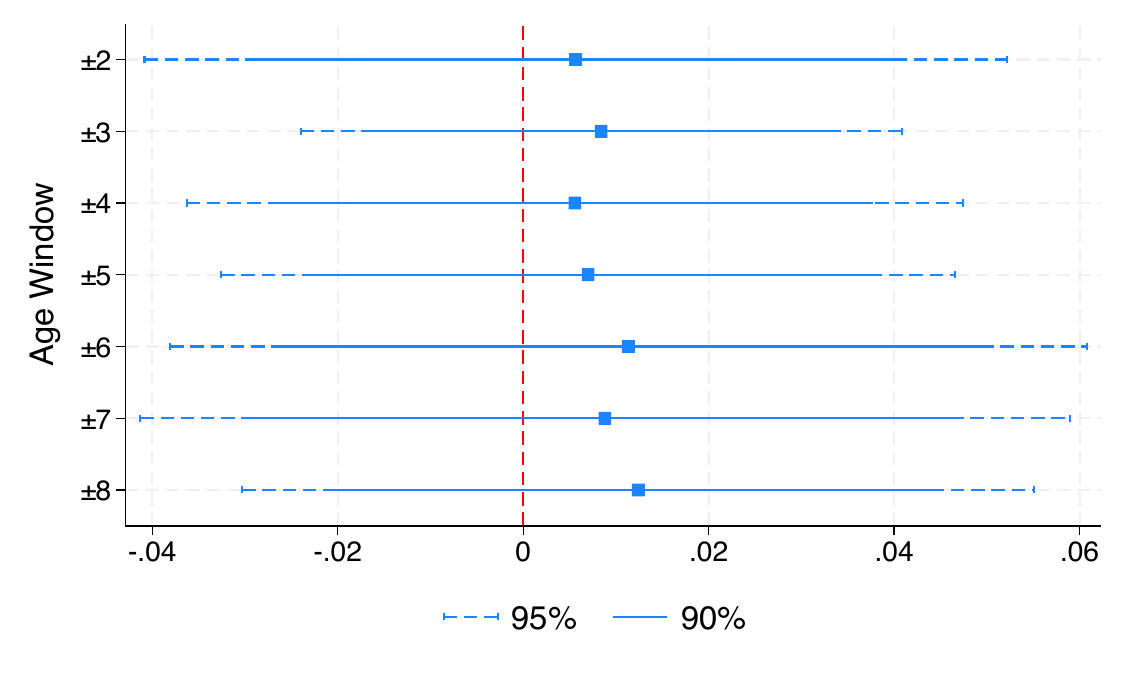}
\end{subfigure}
     \begin{minipage}{16cm} \footnotesizes These figures show the estimates from Eq. (\ref{baselineDiD}), in a specification that includes cohort $\times$ region, high school $\times$ region, and survey year $\times$ region fixed effects, along with a set of ethnic group dummies. 
     \end{minipage}
\end{figure}

\begin{landscape}
\begin{table}[H]
\vspace{1cm}
\begin{center}
{
\renewcommand{\arraystretch}{0.6}
\setlength{\tabcolsep}{0pt}
\caption {Effect on Support for Gender Equality (Years Prior to  1994, Windows [-4,4], [-5,5] and [-6,6])}  \label{table_didbaseline_placebo456_boot}
\vspace{-0.3cm}
\scriptsize
 \begin{tabular}{lccccccccc}
\hline\hline \addlinespace[0.15cm]
    & (1)& (2)& (3)  & (4)& (5)& (6) & (7)& (8)& (9) \\\addlinespace[0.12cm]\cmidrule[0.2pt](l){2-10}\addlinespace[0.12cm]
& \multicolumn{9}{c}{Dependent Variable: Support for Gender Equality}\\\addlinespace[0.15cm]\cmidrule[0.2pt](l){2-10} \addlinespace[0.10cm]   
      & \multicolumn{3}{c}{Age window [-4,4]} & \multicolumn{3}{c}{Age window [-5,5]}& \multicolumn{3}{c}{Age window [-6,6]}\\\addlinespace[0.1cm]\cmidrule[0.2pt](l){2-4}\cmidrule[0.2pt](l){5-7}\cmidrule[0.2pt](l){8-10}

&\multicolumn{1}{c}{all}&\multicolumn{1}{c}{women}&\multicolumn{1}{c}{men}&\multicolumn{1}{c}{all}&\multicolumn{1}{c}{women}&\multicolumn{1}{c}{men}\\\addlinespace[0.1cm]\cmidrule[0.2pt](l){2-2} \cmidrule[0.2pt](l){3-3}\cmidrule[0.2pt](l){4-4}\cmidrule[0.2pt](l){5-5}\cmidrule[0.2pt](l){6-6}\cmidrule[0.2pt](l){7-7}\cmidrule[0.2pt](l){8-8}\cmidrule[0.2pt](l){9-9}\cmidrule[0.2pt](l){10-10}\addlinespace[0.12cm]

        (Age $\leq$ 16 in 1992) $\times$ (Highschool)&       0.008   &      -0.007   &       0.024   &       0.005   &      -0.007   &       0.018   &       0.005   &      -0.011   &       0.023   \\
                    &     (0.013)   &     (0.020)   &     (0.010)*  &     (0.012)   &     (0.019)   &     (0.009)   &     (0.015)   &     (0.022)   &     (0.013)   \\
                    &[-0.024 0.087]             &[-0.047 0.175]          &[-0.004 0.053]        &[-0.021 0.101]         &[-0.058 0.194]    &[-0.004 0.046]       &[-0.018 0.133]     &[-0.047 0.240]      &[-0.008 0.067]   \\
R-sq                &       0.019   &       0.019   &       0.028   &       0.019   &       0.019   &       0.027   &       0.018   &       0.020   &       0.024   \\
Observations        &       14358   &        7594   &        6764   &       17795   &        9425   &        8370   &       21198   &       11309   &        9889   \\
\addlinespace[0.1cm]\cmidrule[0.2pt](l){2-2} \cmidrule[0.2pt](l){3-3}\cmidrule[0.2pt](l){4-4}\cmidrule[0.2pt](l){5-5}\cmidrule[0.2pt](l){6-6}\cmidrule[0.2pt](l){7-7}\cmidrule[0.2pt](l){8-8}\cmidrule[0.2pt](l){9-9}\cmidrule[0.2pt](l){10-10}\addlinespace[0.12cm]    
        (Age $\leq$ 16 in 1990) $\times$ (Highschool)&       0.006   &      -0.006   &       0.021   &       0.009   &       0.008   &       0.010   &       0.007   &      -0.001   &       0.016   \\
                    &     (0.010)   &     (0.014)   &     (0.022)   &     (0.013)   &     (0.009)   &     (0.024)   &     (0.015)   &     (0.008)   &     (0.025)   \\
                    &[-0.014 0.038]        &[-0.030 0.064]       &[-0.054 0.079]       &[-0.019 0.052]        &[-0.007 0.055]      &[-0.061 0.081]      &[-0.022 0.059]     &[-0.018 0.045]    &[-0.042 0.100]   \\
R-sq                &       0.018   &       0.019   &       0.025   &       0.018   &       0.019   &       0.023   &       0.017   &       0.017   &       0.024   \\
Observations        &       13763   &        7315   &        6448   &       17123   &        9105   &        8018   &       20517   &       10885   &        9632   \\
\addlinespace[0.1cm]\cmidrule[0.2pt](l){2-2} \cmidrule[0.2pt](l){3-3}\cmidrule[0.2pt](l){4-4}\cmidrule[0.2pt](l){5-5}\cmidrule[0.2pt](l){6-6}\cmidrule[0.2pt](l){7-7}\cmidrule[0.2pt](l){8-8}\cmidrule[0.2pt](l){9-9}\cmidrule[0.2pt](l){10-10}\addlinespace[0.12cm]
        (Age $\leq$ 16 in 1988) $\times$ (Highschool)&      -0.010   &      -0.003   &      -0.015   &      -0.001   &      -0.003   &       0.003   &       0.001   &      -0.001   &       0.006   \\
                    &     (0.014)   &     (0.011)   &     (0.022)   &     (0.015)   &     (0.011)   &     (0.025)   &     (0.012)   &     (0.016)   &     (0.018)   \\
                    &[-0.044 0.029]      &[-0.028 0.073]       &[-0.110 0.056]      &[-0.027 0.036]      &[-0.019 0.093]       &[-0.052 0.084]        &[-0.021 0.033]      &[-0.028 0.092]       &[-0.036 0.070]   \\
R-sq                &       0.014   &       0.019   &       0.018   &       0.016   &       0.018   &       0.023   &       0.017   &       0.019   &       0.023   \\
Observations        &       13741   &        7311   &        6430   &       16719   &        8890   &        7829   &       19744   &       10536   &        9208   \\
\addlinespace[0.1cm]\cmidrule[0.2pt](l){2-2} \cmidrule[0.2pt](l){3-3}\cmidrule[0.2pt](l){4-4}\cmidrule[0.2pt](l){5-5}\cmidrule[0.2pt](l){6-6}\cmidrule[0.2pt](l){7-7}\cmidrule[0.2pt](l){8-8}\cmidrule[0.2pt](l){9-9}\cmidrule[0.2pt](l){10-10}\addlinespace[0.12cm]
        (Age $\leq$ 16 in 1986) $\times$ (Highschool)&       0.002   &       0.004   &      -0.001   &       0.005   &       0.011   &      -0.002   &       0.006   &       0.012   &      -0.001   \\
                    &     (0.006)   &     (0.007)   &     (0.017)   &     (0.008)   &     (0.009)   &     (0.011)   &     (0.006)   &     (0.008)   &     (0.008)   \\
                    &[-0.026 0.023]      &[-0.046 0.021]    &[-0.030 0.052]      &[-0.033 0.029]    &[-0.056 0.036]      &[-0.018 0.036]    &[-0.007 0.019]    &[-0.001 0.033]     &[-0.018 0.021]   \\
R-sq                &       0.016   &       0.020   &       0.019   &       0.016   &       0.020   &       0.019   &       0.016   &       0.021   &       0.019   \\
Observations        &       13523   &        7230   &        6293   &       16705   &        8909   &        7796   &       19696   &       10529   &        9167   \\
\addlinespace[0.1cm]\cmidrule[0.2pt](l){2-2} \cmidrule[0.2pt](l){3-3}\cmidrule[0.2pt](l){4-4}\cmidrule[0.2pt](l){5-5}\cmidrule[0.2pt](l){6-6}\cmidrule[0.2pt](l){7-7}\cmidrule[0.2pt](l){8-8}\cmidrule[0.2pt](l){9-9}\cmidrule[0.2pt](l){10-10}\addlinespace[0.12cm]
(Age $\leq$ 16 in 1984) $\times$ (Highschool)&       0.015   &       0.015   &       0.015   &       0.013   &       0.014   &       0.010   &       0.013   &       0.014   &       0.011   \\
                    &     (0.014)   &     (0.018)   &     (0.015)   &     (0.009)   &     (0.016)   &     (0.011)   &     (0.010)   &     (0.014)   &     (0.011)   \\
                    &[-0.019 0.069]       &[-0.017 0.065]      &[-0.036 0.081]    &[-0.010 0.056]       &[-0.010 0.064]       &[-0.026 0.048]      &[-0.014 0.042]     &[-0.009 0.062]     &[-0.031 0.044]   \\
R-sq                &       0.017   &       0.021   &       0.019   &       0.018   &       0.023   &       0.020   &       0.017   &       0.021   &       0.020   \\
Observations        &       13595   &        7281   &        6314   &       16609   &        8924   &        7685   &       19383   &       10411   &        8972   \\

\addlinespace[0.15cm]\hline\hline\addlinespace[0.15cm]
\multicolumn{10}{p{20cm}}{\scriptsize{\textbf{Notes:} All columns report estimates from Equation (\ref{baselineDiD}), but instead of using the indicator for being at most 16 years old in 1994, we replace it with an indicator for being that age in the specific year indicated in each row. Specifications in columns (1)–(3) use an age window of $[-4, 4]$, those in columns (4)–(6) use $[-5, 5]$ and those in columns (7)–(9) use $[-6, 6]$.  The regression samples are drawn from the 2019 and 2021 waves of the Political Culture Survey conducted by DANE. The dependent variable in all columns is the response to the question (in Spanish): ``¿Usted est\'a de acuerdo con que las mujeres participen en igualdad de condiciones y oportunidades?'' (``Do you agree that women should participate on equal terms and have the same opportunities?''). All the specifications include cohort-by-region, high school-by-region, and survey year-by-region fixed effects, as well as a set of ethnic group dummies. Robust standard errors, clustered at the region level, are reported in parentheses. Wild bootstrap confidence intervals are shown in square brackets.  *, **, and *** indicate statistical significance at the 10\%, 5\%, and 1\% levels, respectively.} }
\end{tabular}
}
\end{center}
\end{table}
\end{landscape}

\begin{landscape}
\begin{table}[H]
\vspace{0cm}
\begin{center}
{
\renewcommand{\arraystretch}{0.6}
\setlength{\tabcolsep}{3pt}
\caption {Effect on Perception of Gender Discrimination (Years Prior to  1994, Windows [-2,2] and [-3,3])}  \label{table_didsexdiscrim_placebo23_boot}
\vspace{-0.3cm}
\scriptsize
\centering  \begin{tabular}{lcccccc}
\hline\hline \addlinespace[0.15cm]
    & (1)& (2)& (3)  & (4)& (5)& (6) \\\addlinespace[0.12cm]\cmidrule[0.2pt](l){2-7}\addlinespace[0.12cm]
& \multicolumn{6}{c}{Dependent Variable: Perception of Gender Discrimination}\\\addlinespace[0.15cm]\cmidrule[0.2pt](l){2-7} \addlinespace[0.10cm]   
      & \multicolumn{3}{c}{age window [-2,2]} & \multicolumn{3}{c}{age window [-3,3]}\\\addlinespace[0.1cm]\cmidrule[0.2pt](l){2-4}\cmidrule[0.2pt](l){5-7}
                                 &\multicolumn{1}{c}{all}&\multicolumn{1}{c}{women}&\multicolumn{1}{c}{men}&\multicolumn{1}{c}{all}&\multicolumn{1}{c}{women}&\multicolumn{1}{c}{men}\\\addlinespace[0.1cm]\cmidrule[0.2pt](l){2-2} \cmidrule[0.2pt](l){3-3}\cmidrule[0.2pt](l){4-4}\cmidrule[0.2pt](l){5-5}\cmidrule[0.2pt](l){6-6}\cmidrule[0.2pt](l){7-7}\addlinespace[0.12cm]

    (Age $\leq$ 16 in 1992) $\times$ (Highschool)&      -0.013         &      -0.022         &       0.000         &      -0.012         &      -0.020         &      -0.001         \\
                    &     (0.011)         &     (0.018)         &     (0.007)         &     (0.010)         &     (0.015)         &     (0.005)         \\
                    &[-0.040 0.023]                &[-0.061 0.029]            &[-0.019 0.023]             &[-0.030 0.019]        &[-0.044 0.025]              &[-0.014 0.017]         \\
R-sq                &       0.015         &       0.020         &       0.053         &       0.016         &       0.024         &       0.033         \\
Observations        &        4214         &        2255         &        1959         &        6004         &        3220         &        2784         \\

    \addlinespace[0.1cm]\cmidrule[0.2pt](l){2-2} \cmidrule[0.2pt](l){3-3}\cmidrule[0.2pt](l){4-4}\cmidrule[0.2pt](l){5-5}\cmidrule[0.2pt](l){6-6}\cmidrule[0.2pt](l){7-7}\addlinespace[0.12cm]    
(Age $\leq$ 16 in 1990) $\times$ (Highschool)&       0.004         &       0.013         &      -0.003         &      -0.001         &       0.006         &      -0.006         \\
                    &     (0.004)         &     (0.008)         &     (0.007)         &     (0.004)         &     (0.006)         &     (0.005)         \\
                    &[-0.007 0.049]           &[0.002 0.092]            &[-0.013 0.014]        &[-0.015 0.025]       &[-0.006 0.052]            &[-0.018 0.005]         \\
R-sq                &       0.017         &       0.018         &       0.052         &       0.013         &       0.018         &       0.036         \\
Observations        &        4168         &        2193         &        1975         &        5857         &        3081         &        2776         \\

        \addlinespace[0.1cm]\cmidrule[0.2pt](l){2-2} \cmidrule[0.2pt](l){3-3}\cmidrule[0.2pt](l){4-4}\cmidrule[0.2pt](l){5-5}\cmidrule[0.2pt](l){6-6}\cmidrule[0.2pt](l){7-7}\addlinespace[0.12cm]
(Age $\leq$ 16 in 1988) $\times$ (Highschool)&      -0.001         &      -0.001         &       0.000         &       0.005         &       0.007         &       0.002         \\
                    &     (0.003)         &     (0.005)         &     (0.004)         &     (0.004)         &     (0.007)         &     (0.003)         \\
                    &[-0.027 0.005]                 &[-0.047 0.008]              &[-0.012 0.012]           &[-0.017 0.012]           &[-0.025 0.018]                   &[-0.009 0.008]         \\
R-sq                &       0.016         &       0.018         &       0.042         &       0.015         &       0.020         &       0.033         \\
Observations        &        4102         &        2154         &        1948         &        5824         &        3059         &       2765         \\

\addlinespace[0.1cm]\cmidrule[0.2pt](l){2-2} \cmidrule[0.2pt](l){3-3}\cmidrule[0.2pt](l){4-4}\cmidrule[0.2pt](l){5-5}\cmidrule[0.2pt](l){6-6}\cmidrule[0.2pt](l){7-7}\addlinespace[0.12cm]
(Age $\leq$ 16 in 1986) $\times$ (Highschool)&       0.016         &       0.012         &       0.018         &       0.018         &       0.016         &       0.018         \\
                    &     (0.009)         &     (0.010)         &     (0.010)         &     (0.009)         &     (0.010)         &     (0.010)         \\
                    &[-0.018 0.032]            &[-0.019 0.033]          &[-0.021 0.037]        &[-0.002 0.035]            &[-0.012 0.044]               &[-0.012 0.040]         \\
R-sq                &       0.010         &       0.015         &       0.028         &       0.012         &       0.016         &       0.029         \\
Observations        &        4132         &        2158         &        1974         &        5721         &        3030         &       2691         \\

\addlinespace[0.1cm]\cmidrule[0.2pt](l){2-2} \cmidrule[0.2pt](l){3-3}\cmidrule[0.2pt](l){4-4}\cmidrule[0.2pt](l){5-5}\cmidrule[0.2pt](l){6-6}\cmidrule[0.2pt](l){7-7}\addlinespace[0.12cm]
(Age $\leq$ 16 in 1984) $\times$ (Highschool)&      -0.000         &       0.000         &      -0.001         &       0.006         &       0.007         &       0.006         \\
                    &     (0.004)         &     (0.006)         &     (0.001)         &     (0.004)         &     (0.006)         &     (0.002)\sym{*}  \\
                    &[-0.013 0.031]            &[-0.020 0.031]           &[-0.006 0.013]          &[0.002 0.018]***                &[-0.003 0.024]            &[0.000 0.011]***         \\
R-sq                &       0.014         &       0.019         &       0.036         &       0.012         &       0.018         &       0.027         \\
Observations        &        3964         &        2145         &        1819         &        5713         &        3058         &       2655         \\
         
\addlinespace[0.15cm]\hline\hline\addlinespace[0.15cm]
\multicolumn{7}{p{17cm}}{\scriptsize{\textbf{Notes:} All columns report estimates from Equation (\ref{baselineDiD}),  but instead of using the indicator for being at most 16 years old in 1994, we replace it with an indicator for being that age in the specific year indicated in each row. Specifications in columns (1)–(3) use an age window of $[-2, 2]$, while those in columns (4)–(6) use $[-3, 3]$. The regression samples are drawn from the 2019 and 2021 waves of the Political Culture Survey conducted by DANE. The dependent variable in all columns is the response to the question (in Spanish): ``Durante los \'ultimos 12 meses, usted ha sentido que lo han discriminado [...] por alguno de los siguientes motivos: Sexo [...]'' (``In the past 12 months, have you felt discriminated against [...] based on sex?''). The specifications in columns (2) and (5) include fixed effects for cohort, region, and survey year. Columns (1), (3), and (6) additionally include cohort-by-region, high school-by-region, and survey year-by-region fixed effects, as well as a set of ethnic group dummies. Robust standard errors, clustered at the region level, are reported in parentheses. Wild bootstrap confidence intervals are shown in square brackets.  *, **, and *** indicate statistical significance at the 10\%, 5\%, and 1\% levels, respectively.} }
\end{tabular}
}
\end{center}
\end{table}
\end{landscape}

\begin{landscape}
\begin{table}[H]
\vspace{0cm}
\begin{center}
\renewcommand{\arraystretch}{0.6}
\setlength{\tabcolsep}{5pt}
\caption {Effect on Perception of Gender Discrimination  (Years Prior to  1994, Windows [-4,4] and [-5,5])}  \label{table_didsexdiscrim_placebo45_boot}
\vspace{-0.3cm}
\scriptsize
\centering  \begin{tabular}{lcccccc}
\hline\hline \addlinespace[0.15cm]
    & (1)& (2)& (3)  & (4)& (5)& (6) \\\addlinespace[0.12cm]\cmidrule[0.2pt](l){2-7}\addlinespace[0.12cm]
& \multicolumn{6}{c}{Dependent Variable: Perception of Gender Discrimination}\\\addlinespace[0.15cm]\cmidrule[0.2pt](l){2-7} \addlinespace[0.10cm]   
      & \multicolumn{3}{c}{age window [-4,4]} & \multicolumn{3}{c}{age window [-5,5]}\\\addlinespace[0.1cm]\cmidrule[0.2pt](l){2-4}\cmidrule[0.2pt](l){5-7}
                                 &\multicolumn{1}{c}{all}&\multicolumn{1}{c}{women}&\multicolumn{1}{c}{men}&\multicolumn{1}{c}{all}&\multicolumn{1}{c}{women}&\multicolumn{1}{c}{men}\\\addlinespace[0.1cm]\cmidrule[0.2pt](l){2-2} \cmidrule[0.2pt](l){3-3}\cmidrule[0.2pt](l){4-4}\cmidrule[0.2pt](l){5-5}\cmidrule[0.2pt](l){6-6}\cmidrule[0.2pt](l){7-7}\addlinespace[0.12cm]

(Age $\leq$ 16 in 1992) $\times$ (Highschool)&      -0.009   &      -0.016   &       0.001   &      -0.009   &      -0.016   &       0.001   \\
                    &     (0.007)   &     (0.011)   &     (0.003)   &     (0.006)   &     (0.011)   &     (0.002)   \\
                    &[-0.021 0.011]     &[-0.038 0.010]     &[-0.005 0.013]    &[-0.021 0.009]   &[-0.038 0.010]      &[-0.004 0.010]   \\
R-sq                &       0.014   &       0.021   &       0.020   &       0.013   &       0.019   &       0.018   \\
Observations        &        7845   &        4157   &        3688   &        9830   &        5212   &        4618   \\

    \addlinespace[0.1cm]\cmidrule[0.2pt](l){2-2} \cmidrule[0.2pt](l){3-3}\cmidrule[0.2pt](l){4-4}\cmidrule[0.2pt](l){5-5}\cmidrule[0.2pt](l){6-6}\cmidrule[0.2pt](l){7-7}\addlinespace[0.12cm]    
(Age $\leq$ 16 in 1990) $\times$ (Highschool)&      -0.001   &       0.003   &      -0.003   &      -0.000   &       0.002   &      -0.000   \\
                    &     (0.004)   &     (0.006)   &     (0.006)   &     (0.004)   &     (0.007)   &     (0.004)   \\
                    &[-0.017 0.013]   &[-0.009 0.028]   &[-0.021 0.009]   &[-0.015 0.016]   &[-0.018 0.037]   &[-0.011 0.007]   \\
R-sq                &       0.012   &       0.018   &       0.024   &       0.012   &       0.019   &       0.024   \\
Observations        &        7545   &        3992   &        3553   &        9335   &        4947   &        4388   \\

        \addlinespace[0.1cm]\cmidrule[0.2pt](l){2-2} \cmidrule[0.2pt](l){3-3}\cmidrule[0.2pt](l){4-4}\cmidrule[0.2pt](l){5-5}\cmidrule[0.2pt](l){6-6}\cmidrule[0.2pt](l){7-7}\addlinespace[0.12cm]
(Age $\leq$ 16 in 1988) $\times$ (Highschool)&       0.008   &       0.009   &       0.006   &       0.010   &       0.013   &       0.006   \\
                    &     (0.004)   &     (0.005)   &     (0.007)   &     (0.004)*  &     (0.006)*  &     (0.008)   \\
                    &[-0.017 0.016]    &[-0.024 0.019]    &[-0.007 0.023]    &[-0.004 0.019]*    &[-0.003 0.027]*  &[-0.008 0.027]   \\
R-sq                &       0.012   &       0.016   &       0.031   &       0.011   &       0.016   &       0.031   \\
Observations        &        7516   &        3958   &        3558   &        9074   &        4811   &        4263   \\

\addlinespace[0.1cm]\cmidrule[0.2pt](l){2-2} \cmidrule[0.2pt](l){3-3}\cmidrule[0.2pt](l){4-4}\cmidrule[0.2pt](l){5-5}\cmidrule[0.2pt](l){6-6}\cmidrule[0.2pt](l){7-7}\addlinespace[0.12cm]
(Age $\leq$ 16 in 1986) $\times$ (Highschool)&       0.015   &       0.013   &       0.016   &       0.016   &       0.016   &       0.016   \\
                    &     (0.007)   &     (0.007)   &     (0.009)   &     (0.007)*  &     (0.008)   &     (0.008)*  \\
                    &[0.000 0.030]    &[-0.004 0.035]  &[-0.009 0.034]     &[0.001 0.031]      &[-0.004 0.040]   &[-0.005 0.033]*   \\
R-sq                &       0.015   &       0.018   &       0.040   &       0.013   &       0.019   &       0.030   \\
Observations        &        7239   &        3862   &        3377   &        8990   &        4783   &        4207   \\

\addlinespace[0.1cm]\cmidrule[0.2pt](l){2-2} \cmidrule[0.2pt](l){3-3}\cmidrule[0.2pt](l){4-4}\cmidrule[0.2pt](l){5-5}\cmidrule[0.2pt](l){6-6}\cmidrule[0.2pt](l){7-7}\addlinespace[0.12cm]
(Age $\leq$ 16 in 1984) $\times$ (Highschool)&       0.002   &      -0.000   &       0.006   &       0.003   &       0.003   &       0.004   \\
                    &     (0.003)   &     (0.006)   &     (0.001)***&     (0.004)   &     (0.006)   &     (0.001)** \\
                    &[-0.008 0.013]   &[-0.018 0.018]    &[0.003 0.009]***    &[-0.003 0.015]    &[-0.010 0.025]  &[0.002 0.009]**   \\
R-sq                &       0.010   &       0.016   &       0.026   &       0.010   &       0.016   &       0.023   \\
Observations        &        7205   &        3839   &        3366   &        8807   &        4721   &        4086   \\
         
\addlinespace[0.15cm]\hline\hline\addlinespace[0.15cm]
\multicolumn{7}{p{18cm}}{\scriptsize{\textbf{Notes:} All columns report estimates from Equation (\ref{baselineDiD}),  but instead of using the indicator for being at most 16 years old in 1994, we replace it with an indicator for being that age in the specific year indicated in each row. Specifications in columns (1)–(3) use an age window of $[-4, 4]$, while those in columns (4)–(6) use $[-5, 5]$. The regression samples are drawn from the 2019 and 2021 waves of the Political Culture Survey conducted by DANE. The dependent variable in all columns is the response to the question (in Spanish): ``Durante los \'ultimos 12 meses, usted ha sentido que lo han discriminado [...] por alguno de los siguientes motivos: Sexo [...]'' (``In the past 12 months, have you felt discriminated against [...] based on sex?''). The specifications in columns (2) and (5) include fixed effects for cohort, region, and survey year. Columns (1), (3), and (6) additionally include cohort-by-region, high school-by-region, and survey year-by-region fixed effects, as well as a set of ethnic group dummies. Robust standard errors, clustered at the region level, are reported in parentheses. Wild bootstrap confidence intervals are shown in square brackets.  *, **, and *** indicate statistical significance at the 10\%, 5\%, and 1\% levels, respectively.} }
\end{tabular}

\end{center}
\end{table}
\end{landscape}

\begin{landscape}
\begin{table}[H]
\vspace{0cm}
\begin{center}
{
\renewcommand{\arraystretch}{0.6}
\setlength{\tabcolsep}{5pt}
\caption {Effect on Disagreement with ``Women are Better Suited for Domestic Tasks than Men''  (Years Prior to  1994, Windows [-2,2] and [-3,3])}  \label{table_opinion4_placebo23_boot}
\vspace{-0.3cm}
\scriptsize
\centering  \begin{tabular}{lcccccc}
\hline\hline \addlinespace[0.15cm]
    & (1)& (2)& (3)  & (4)& (5)& (6) \\\addlinespace[0.12cm]\cmidrule[0.2pt](l){2-7}\addlinespace[0.12cm]
& \multicolumn{6}{c}{Dependent Variable: Disagreement with ``women are better suited for domestic tasks''}\\\addlinespace[0.15cm]\cmidrule[0.2pt](l){2-7} \addlinespace[0.10cm]   
      & \multicolumn{3}{c}{Age window [-2,2]} & \multicolumn{3}{c}{Age window [-3,3]}\\\addlinespace[0.1cm]\cmidrule[0.2pt](l){2-4}\cmidrule[0.2pt](l){5-7}
                                 &\multicolumn{1}{c}{All}&\multicolumn{1}{c}{Women}&\multicolumn{1}{c}{Men}&\multicolumn{1}{c}{All}&\multicolumn{1}{c}{Women}&\multicolumn{1}{c}{Men}\\\addlinespace[0.1cm]\cmidrule[0.2pt](l){2-2} \cmidrule[0.2pt](l){3-3}\cmidrule[0.2pt](l){4-4}\cmidrule[0.2pt](l){5-5}\cmidrule[0.2pt](l){6-6}\cmidrule[0.2pt](l){7-7}\addlinespace[0.12cm]

(Age $\leq$ 16 in 1992) $\times$ (Highschool)&       0.017         &       0.029         &       0.004         &       0.010         &       0.023         &      -0.003         \\
                    &     (0.013)         &     (0.020)         &     (0.011)         &     (0.008)         &     (0.015)         &     (0.007)         \\
                    &[-0.042 0.041]      &[-0.062 0.063]     &[-0.018 0.038]      &[-0.026 0.030]     &[-0.053 0.060]       &[-0.016 0.018]         \\
R-sq                &       0.050         &       0.052         &       0.054         &       0.052         &       0.052         &       0.056         \\
Observations        &       17301         &        9348         &        7953         &       24518         &       13228         &       11290         \\

    \addlinespace[0.1cm]\cmidrule[0.2pt](l){2-2} \cmidrule[0.2pt](l){3-3}\cmidrule[0.2pt](l){4-4}\cmidrule[0.2pt](l){5-5}\cmidrule[0.2pt](l){6-6}\cmidrule[0.2pt](l){7-7}\addlinespace[0.12cm]    
(Age $\leq$ 16 in 1990) $\times$ (Highschool)&       0.000         &       0.000         &       0.002         &       0.013         &       0.011         &       0.017         \\
                    &     (0.015)         &     (0.022)         &     (0.013)         &     (0.015)         &     (0.022)         &     (0.012)         \\
                    &[-0.027 0.039]   &[-0.056 0.052]       &[-0.025 0.056]      &[-0.014 0.056]    &[-0.036 0.075]    &[-0.009 0.049]         \\
R-sq                &       0.053         &       0.053         &       0.058         &       0.051         &       0.051         &       0.058         \\
Observations        &       16952         &        9232         &        7720         &       24205         &       13179         &       11026         \\

        \addlinespace[0.1cm]\cmidrule[0.2pt](l){2-2} \cmidrule[0.2pt](l){3-3}\cmidrule[0.2pt](l){4-4}\cmidrule[0.2pt](l){5-5}\cmidrule[0.2pt](l){6-6}\cmidrule[0.2pt](l){7-7}\addlinespace[0.12cm]
(Age $\leq$ 16 in 1988) $\times$ (Highschool)&       0.007         &       0.002         &       0.013         &       0.009         &       0.000         &       0.019         \\
                    &     (0.015)         &     (0.026)         &     (0.005)\sym{*}  &     (0.010)         &     (0.026)         &     (0.011)         \\
                    &[-0.037 0.033]         &[-0.072 0.046]         &[0.003 0.040]       &[-0.019 0.032]              &[-0.073 0.053]          &[-0.012 0.046]         \\
R-sq                &       0.052         &       0.052         &       0.059         &       0.052         &       0.052         &       0.056         \\
Observations        &       17089         &        9357         &        7732         &       23960         &       13116         &       10844         \\

\addlinespace[0.1cm]\cmidrule[0.2pt](l){2-2} \cmidrule[0.2pt](l){3-3}\cmidrule[0.2pt](l){4-4}\cmidrule[0.2pt](l){5-5}\cmidrule[0.2pt](l){6-6}\cmidrule[0.2pt](l){7-7}\addlinespace[0.12cm]
(Age $\leq$ 16 in 1986) $\times$ (Highschool)&       0.014         &       0.013         &       0.012         &       0.005         &       0.008         &      -0.001         \\
                    &     (0.026)         &     (0.038)         &     (0.021)         &     (0.021)         &     (0.032)         &     (0.019)         \\
                    &[-0.054 0.104]        &[-0.088 0.101]      &[-0.044 0.054]        &[-0.035 0.064]           &[-0.052 0.096]      &[-0.065 0.057]         \\
R-sq                &       0.049         &       0.048         &       0.055         &       0.051         &       0.049         &       0.056         \\
Observations        &       17204         &        9460         &        7744         &       24172         &       13216         &       10956         \\

\addlinespace[0.1cm]\cmidrule[0.2pt](l){2-2} \cmidrule[0.2pt](l){3-3}\cmidrule[0.2pt](l){4-4}\cmidrule[0.2pt](l){5-5}\cmidrule[0.2pt](l){6-6}\cmidrule[0.2pt](l){7-7}\addlinespace[0.12cm]
(Age $\leq$ 16 in 1984) $\times$ (Highschool)&      -0.021         &      -0.021         &      -0.017         &      -0.017         &      -0.012         &      -0.021         \\
                    &     (0.008)\sym{*}  &     (0.007)\sym{*}  &     (0.011)         &     (0.008)         &     (0.009)         &     (0.007)\sym{*}  \\
                    &[-0.048 -0.004]**         &[-0.037 0.002]*         &[-0.050 0.002]        &[-0.046 -0.004]        &[-0.042 0.004]             &[-0.042 -0.008]*         \\
R-sq                &       0.051         &       0.053         &       0.051         &       0.050         &       0.049         &       0.054         \\
Observations        &       17178         &        9430         &        7748         &       24254         &       13316         &       10938         \\
         
\addlinespace[0.15cm]\hline\hline\addlinespace[0.15cm]
\multicolumn{7}{p{18cm}}{\scriptsize{\textbf{Notes:} All columns report estimates from Equation (\ref{baselineDiD}), but instead of using the indicator for being at most 16 years old in 1994, we replace it with an indicator for being that age in the specific year indicated in each row. Specifications in columns (1)–(3) use an age window of $[-2, 2]$, while those in columns (4)–(6) use $[-3, 3]$.  The regression samples are drawn from the 2017–2016 and 2020-2021 waves of the ENUT Survey, conducted by DANE. The dependent variable in all columns is a binary indicator equal to 1 if the respondent expresses disagreement with the following statement: “¿Las mujeres son mejores para el trabajo doméstico que los hombres?” (“Women are better suited for domestic tasks than men?”). All the specifications include cohort-by-region, high school-by-region, and survey year-by-region fixed effects, as well as a set of ethnic group dummies. Robust standard errors, clustered at the region level, are reported in parentheses. Wild bootstrap confidence intervals are shown in square brackets.  *, **, and *** indicate statistical significance at the 10\%, 5\%, and 1\% levels, respectively.} }
\end{tabular}
}
\end{center}
\end{table}
\end{landscape}

\begin{landscape}
\begin{table}[H]
\vspace{0cm}
\begin{center}
{
\renewcommand{\arraystretch}{0.6}
\setlength{\tabcolsep}{5pt}
\caption {Effect on Disagreement with ``Women are Better Suited for Domestic Tasks than Men'' (Years Prior to  1994, Windows [-4,4] and [-5,5])}  \label{table_opinion4_placebo45_boot}
\vspace{-0.3cm}
\scriptsize
\centering  \begin{tabular}{lcccccc}
\hline\hline \addlinespace[0.15cm]
    & (1)& (2)& (3)  & (4)& (5)& (6) \\\addlinespace[0.12cm]\cmidrule[0.2pt](l){2-7}\addlinespace[0.12cm]
& \multicolumn{6}{c}{Dependent Variable: Disagreement with ``women are better suited for domestic tasks''}\\\addlinespace[0.15cm]\cmidrule[0.2pt](l){2-7} \addlinespace[0.10cm]   
      & \multicolumn{3}{c}{Age window [-4,4]} & \multicolumn{3}{c}{Age window [-5,5]}\\\addlinespace[0.1cm]\cmidrule[0.2pt](l){2-4}\cmidrule[0.2pt](l){5-7}
                                 &\multicolumn{1}{c}{All}&\multicolumn{1}{c}{Women}&\multicolumn{1}{c}{Men}&\multicolumn{1}{c}{All}&\multicolumn{1}{c}{Women}&\multicolumn{1}{c}{Men}\\\addlinespace[0.1cm]\cmidrule[0.2pt](l){2-2} \cmidrule[0.2pt](l){3-3}\cmidrule[0.2pt](l){4-4}\cmidrule[0.2pt](l){5-5}\cmidrule[0.2pt](l){6-6}\cmidrule[0.2pt](l){7-7}\addlinespace[0.12cm]

(Age $\leq$ 16 in 1992) $\times$ (Highschool)&       0.010   &       0.021   &      -0.001   &       0.015   &       0.021   &       0.009   \\
                    &     (0.010)   &     (0.017)   &     (0.007)   &     (0.006)** &     (0.008)*  &     (0.008)   \\
                    &[-0.026 0.034]    &[-0.033 0.066]      &[-0.020 0.020]   &[-0.000 0.031]**     &[-0.021 0.038]*      &[-0.005 0.032]   \\
R-sq                &       0.052   &       0.051   &       0.056   &       0.052   &       0.050   &       0.057   \\
Observations        &       31837   &       17198   &       14639   &       39700   &       21363   &       18337   \\

    \addlinespace[0.1cm]\cmidrule[0.2pt](l){2-2} \cmidrule[0.2pt](l){3-3}\cmidrule[0.2pt](l){4-4}\cmidrule[0.2pt](l){5-5}\cmidrule[0.2pt](l){6-6}\cmidrule[0.2pt](l){7-7}\addlinespace[0.12cm]    
(Age $\leq$ 16 in 1990) $\times$ (Highschool)&       0.010   &       0.008   &       0.011   &       0.011   &       0.012   &       0.011   \\
                    &     (0.019)   &     (0.025)   &     (0.013)   &     (0.012)   &     (0.016)   &     (0.013)   \\
                    &[-0.023 0.065]      &[-0.044 0.090]     &[-0.018 0.043]     &[-0.011 0.042]   &[-0.019 0.056]           &[-0.032 0.047]   \\
R-sq                &       0.051   &       0.050   &       0.057   &       0.052   &       0.052   &       0.057   \\
Observations        &       31182   &       16980   &       14202   &       38425   &       20880   &       17545   \\

        \addlinespace[0.1cm]\cmidrule[0.2pt](l){2-2} \cmidrule[0.2pt](l){3-3}\cmidrule[0.2pt](l){4-4}\cmidrule[0.2pt](l){5-5}\cmidrule[0.2pt](l){6-6}\cmidrule[0.2pt](l){7-7}\addlinespace[0.12cm]
(Age $\leq$ 16 in 1988) $\times$ (Highschool)&       0.014   &       0.010   &       0.019   &       0.013   &       0.012   &       0.013   \\
                    &     (0.012)   &     (0.022)   &     (0.011)   &     (0.012)   &     (0.022)   &     (0.012)   \\
                    &[-0.017 0.058]     &[-0.052 0.054]    &[-0.026 0.063]   &[-0.019 0.040]    &[-0.049 0.050]      &[-0.033 0.065]   \\
R-sq                &       0.051   &       0.050   &       0.055   &       0.051   &       0.050   &       0.056   \\
Observations        &       30819   &       16829   &       13990   &       37946   &       20690   &       17256   \\

\addlinespace[0.1cm]\cmidrule[0.2pt](l){2-2} \cmidrule[0.2pt](l){3-3}\cmidrule[0.2pt](l){4-4}\cmidrule[0.2pt](l){5-5}\cmidrule[0.2pt](l){6-6}\cmidrule[0.2pt](l){7-7}\addlinespace[0.12cm]
(Age $\leq$ 16 in 1986) $\times$ (Highschool)&       0.003   &       0.002   &       0.004   &       0.002   &       0.001   &       0.003   \\
                    &     (0.014)   &     (0.025)   &     (0.018)   &     (0.008)   &     (0.017)   &     (0.014)   \\
                    &[-0.028 0.043]   &[-0.046 0.065]    &[-0.056 0.048]   &[-0.016 0.030]    &[-0.037 0.041]       &[-0.044 0.036]   \\
R-sq                &       0.051   &       0.052   &       0.054   &       0.051   &       0.050   &       0.055   \\
Observations        &       30817   &       16860   &       13957   &       37644   &       20643   &       17001   \\

\addlinespace[0.1cm]\cmidrule[0.2pt](l){2-2} \cmidrule[0.2pt](l){3-3}\cmidrule[0.2pt](l){4-4}\cmidrule[0.2pt](l){5-5}\cmidrule[0.2pt](l){6-6}\cmidrule[0.2pt](l){7-7}\addlinespace[0.12cm]
(Age $\leq$ 16 in 1984) $\times$ (Highschool)&      -0.012   &      -0.010   &      -0.012   &      -0.007   &      -0.006   &      -0.008   \\
                    &     (0.006)   &     (0.005)   &     (0.008)   &     (0.006)   &     (0.005)   &     (0.009)   \\
                    &[-0.032 0.002] &[-0.024 0.003]    &[-0.037 0.006]    &[-0.028 0.006]     &[-0.022 0.003]    &[-0.029 0.014]   \\
R-sq                &       0.050   &       0.048   &       0.056   &       0.050   &       0.048   &       0.055   \\
Observations        &       30939   &       16978   &       13961   &       37641   &       20659   &       16982   \\
         
\addlinespace[0.15cm]\hline\hline\addlinespace[0.15cm]
\multicolumn{7}{p{18cm}}{\scriptsize{\textbf{Notes:} All columns report estimates from Equation (\ref{baselineDiD}), but instead of using the indicator for being at most 16 years old in 1994, we replace it with an indicator for being that age in the specific year indicated in each row. Specifications in columns (1)–(3) use an age window of $[-4, 4]$, while those in columns (4)–(6) use $[-5, 5]$.  The regression samples are drawn from the 2017–2016 and 2020-2021 waves of the ENUT Survey, conducted by DANE. The dependent variable in all columns is a binary indicator equal to 1 if the respondent expresses disagreement with the following statement: “¿Las mujeres son mejores para el trabajo doméstico que los hombres?” (“Women are better suited for domestic tasks than men?”). All the specifications include cohort-by-region, high school-by-region, and survey year-by-region fixed effects, as well as a set of ethnic group dummies. Robust standard errors, clustered at the region level, are reported in parentheses. Wild bootstrap confidence intervals are shown in square brackets.  *, **, and *** indicate statistical significance at the 10\%, 5\%, and 1\% levels, respectively.} }
\end{tabular}
}
\end{center}
\end{table}
\end{landscape}

\begin{landscape}
\begin{table}[H]
\vspace{0cm}
\begin{center}
{
\renewcommand{\arraystretch}{0.6}
\setlength{\tabcolsep}{5pt}
\caption {Effect on Disagreement with ``the Husband should Make Decisions Concerning his Wife’s Life'' (Years Prior to  1994, Windows [-2,2] and [-3,3]}  \label{table_opinion5_placebo23_boot}
\vspace{-0.3cm}
\scriptsize
\centering  \begin{tabular}{lcccccc}
\hline\hline \addlinespace[0.15cm]
    & (1)& (2)& (3)  & (4)& (5)& (6) \\\addlinespace[0.12cm]\cmidrule[0.2pt](l){2-7}\addlinespace[0.12cm]
& \multicolumn{6}{c}{Dep. Var.: Disagreement ``the husband should make decisions concerning his wife’s life''}\\\addlinespace[0.15cm]\cmidrule[0.2pt](l){2-7} \addlinespace[0.10cm]   
      & \multicolumn{3}{c}{Age window [-2,2]} & \multicolumn{3}{c}{Age window [-3,3]}\\\addlinespace[0.1cm]\cmidrule[0.2pt](l){2-4}\cmidrule[0.2pt](l){5-7}
                                 &\multicolumn{1}{c}{All}&\multicolumn{1}{c}{Women}&\multicolumn{1}{c}{Men}&\multicolumn{1}{c}{All}&\multicolumn{1}{c}{Women}&\multicolumn{1}{c}{Men}\\\addlinespace[0.1cm]\cmidrule[0.2pt](l){2-2} \cmidrule[0.2pt](l){3-3}\cmidrule[0.2pt](l){4-4}\cmidrule[0.2pt](l){5-5}\cmidrule[0.2pt](l){6-6}\cmidrule[0.2pt](l){7-7}\addlinespace[0.12cm]

(Age $\leq$ 16 in 1992) $\times$ (Highschool)&      -0.022         &      -0.031         &      -0.014         &      -0.014         &      -0.027         &      -0.002         \\
                    &     (0.011)         &     (0.018)         &     (0.012)         &     (0.005)\sym{*}  &     (0.012)         &     (0.008)         \\
                    &[-0.057 -0.003]            &[-0.085 0.004]         &[-0.053 0.018]       &[-0.026 -0.002]*      &[-0.058 0.001]      &[-0.021 0.015]         \\
R-sq                &       0.527         &       0.594         &       0.457         &       0.524         &       0.591         &       0.455         \\
Observations        &       17336         &        9349         &        7987         &       24580         &       13228         &       11352         \\

    \addlinespace[0.1cm]\cmidrule[0.2pt](l){2-2} \cmidrule[0.2pt](l){3-3}\cmidrule[0.2pt](l){4-4}\cmidrule[0.2pt](l){5-5}\cmidrule[0.2pt](l){6-6}\cmidrule[0.2pt](l){7-7}\addlinespace[0.12cm]    
(Age $\leq$ 16 in 1990) $\times$ (Highschool)&       0.000         &      -0.002         &       0.002         &       0.012         &       0.006         &       0.016         \\
                    &     (0.017)         &     (0.029)         &     (0.013)         &     (0.015)         &     (0.024)         &     (0.007)         \\
                    &[-0.035 0.060]           &[-0.062 0.096]         &[-0.031 0.030]          &[-0.021 0.057]          &[-0.045 0.086]         &[-0.004 0.031]         \\
R-sq                &       0.514         &       0.573         &       0.451         &       0.516         &       0.573         &       0.454         \\
Observations        &       17039         &        9251         &        7788         &       24305         &       13190         &       11115         \\

        \addlinespace[0.1cm]\cmidrule[0.2pt](l){2-2} \cmidrule[0.2pt](l){3-3}\cmidrule[0.2pt](l){4-4}\cmidrule[0.2pt](l){5-5}\cmidrule[0.2pt](l){6-6}\cmidrule[0.2pt](l){7-7}\addlinespace[0.12cm]
(Age $\leq$ 16 in 1988) $\times$ (Highschool)&       0.019         &       0.024         &       0.014         &       0.024         &       0.027         &       0.019         \\
                    &     (0.008)         &     (0.017)         &     (0.022)         &     (0.011)         &     (0.008)\sym{*}  &     (0.021)         \\
                    &[0.001 0.038]        &[-0.013 0.083]         &[-0.030 0.072]       &[-0.006 0.047]       &[0.014 0.055]**           &[-0.037 0.069]         \\
R-sq                &       0.507         &       0.562         &       0.447         &       0.507         &       0.566         &       0.443         \\
Observations        &       17182         &        9363         &        7819         &       24085         &       13136         &       10949         \\

\addlinespace[0.1cm]\cmidrule[0.2pt](l){2-2} \cmidrule[0.2pt](l){3-3}\cmidrule[0.2pt](l){4-4}\cmidrule[0.2pt](l){5-5}\cmidrule[0.2pt](l){6-6}\cmidrule[0.2pt](l){7-7}\addlinespace[0.12cm]
(Age $\leq$ 16 in 1986) $\times$ (Highschool)&       0.006         &      -0.000         &       0.012         &       0.004         &      -0.001         &       0.008         \\
                    &     (0.012)         &     (0.013)         &     (0.021)         &     (0.010)         &     (0.009)         &     (0.014)         \\
                    &[-0.028 0.051]    &[-0.041 0.025]      &[-0.029 0.089]      &[-0.029 0.025]    &[-0.029 0.017]     &[-0.038 0.048]         \\
R-sq                &       0.502         &       0.556         &       0.443         &       0.494         &       0.549         &       0.434         \\
Observations        &       17306         &        9482         &        7824         &       24337         &       13265         &       11072         \\

\addlinespace[0.1cm]\cmidrule[0.2pt](l){2-2} \cmidrule[0.2pt](l){3-3}\cmidrule[0.2pt](l){4-4}\cmidrule[0.2pt](l){5-5}\cmidrule[0.2pt](l){6-6}\cmidrule[0.2pt](l){7-7}\addlinespace[0.12cm]
(Age $\leq$ 16 in 1984) $\times$ (Highschool)&      -0.007         &      -0.034         &       0.027         &      -0.023         &      -0.047         &       0.006         \\
                    &     (0.017)         &     (0.018)         &     (0.021)         &     (0.011)         &     (0.015)\sym{*}  &     (0.014)         \\
                    &[-0.043 0.049]       &[-0.094 0.020]        &[-0.022 0.094]         &[-0.056 0.013]       &[-0.099 -0.004]**       &[-0.049 0.043]         \\
R-sq                &       0.477         &       0.534         &       0.415         &       0.476         &       0.537         &       0.410         \\
Observations        &       17315         &        9493         &        7822         &       24432         &       13379         &       11053         \\
         
\addlinespace[0.15cm]\hline\hline\addlinespace[0.15cm]
\multicolumn{7}{p{18cm}}{\scriptsize{\textbf{Notes:} All columns report estimates from Equation (\ref{baselineDiD}), but instead of using the indicator for being at most 16 years old in 1994, we replace it with an indicator for being that age in the specific year indicated in each row. Specifications in columns (1)–(3) use an age window of $[-2, 2]$, while those in columns (4)–(6) use $[-3, 3]$.  The regression samples are drawn from the 2017–2016 and 2020-2021 waves of the ENUT Survey, conducted by DANE. The dependent variable in all columns is a binary indicator equal to 1 if the respondent expresses disagreement with the following statement:  “El esposo debe tomar las decisiones relacionadas con la vida de la esposa?” (“The husband should make decisions concerning his wife’s life?”). All the specifications include cohort-by-region, high school-by-region, and survey year-by-region fixed effects, as well as a set of ethnic group dummies. Robust standard errors, clustered at the region level, are reported in parentheses. Wild bootstrap confidence intervals are shown in square brackets.  *, **, and *** indicate statistical significance at the 10\%, 5\%, and 1\% levels, respectively.} }
\end{tabular}
}
\end{center}
\end{table}
\end{landscape}

\begin{landscape}
\begin{table}[H]
\vspace{0cm}
\begin{center}
{
\renewcommand{\arraystretch}{0.6}
\setlength{\tabcolsep}{4pt}
\caption {Effect on Disagreement with ``The Husband should Make Decisions Concerning his Wife’s Life'' (Years Prior to  1994, Windows [-4,4] and [-5,5])}  \label{table_opinion5_placebo45_boot}
\vspace{-0.3cm}
\scriptsize
\centering  \begin{tabular}{lcccccc}
\hline\hline \addlinespace[0.15cm]
    & (1)& (2)& (3)  & (4)& (5)& (6) \\\addlinespace[0.12cm]\cmidrule[0.2pt](l){2-7}\addlinespace[0.12cm]
& \multicolumn{6}{c}{Dep. Var.: Disagreement ``the husband should make decisions concerning his wife’s life''}\\\addlinespace[0.15cm]\cmidrule[0.2pt](l){2-7} \addlinespace[0.10cm]   
      & \multicolumn{3}{c}{Age window [-4,4]} & \multicolumn{3}{c}{Age window [-5,5]}\\\addlinespace[0.1cm]\cmidrule[0.2pt](l){2-4}\cmidrule[0.2pt](l){5-7}
                                 &\multicolumn{1}{c}{All}&\multicolumn{1}{c}{Women}&\multicolumn{1}{c}{Men}&\multicolumn{1}{c}{All}&\multicolumn{1}{c}{Women}&\multicolumn{1}{c}{Men}\\\addlinespace[0.1cm]\cmidrule[0.2pt](l){2-2} \cmidrule[0.2pt](l){3-3}\cmidrule[0.2pt](l){4-4}\cmidrule[0.2pt](l){5-5}\cmidrule[0.2pt](l){6-6}\cmidrule[0.2pt](l){7-7}\addlinespace[0.12cm]

(Age $\leq$ 16 in 1992) $\times$ (Highschool)&      -0.011   &      -0.009   &      -0.017   &       0.001   &       0.008   &      -0.008   \\
                    &     (0.006)   &     (0.008)   &     (0.008)   &     (0.006)   &     (0.010)   &     (0.006)   \\
                    &[-0.033 -0.001]    &[-0.026 0.012]    &[-0.043 -0.002]     &[-0.019 0.012]   &[-0.017 0.031]    &[-0.032 0.003]   \\
R-sq                &       0.528   &       0.589   &       0.464   &       0.530   &       0.591   &       0.467   \\
Observations        &       31946   &       17221   &       14725   &       39853   &       21396   &       18457   \\

    \addlinespace[0.1cm]\cmidrule[0.2pt](l){2-2} \cmidrule[0.2pt](l){3-3}\cmidrule[0.2pt](l){4-4}\cmidrule[0.2pt](l){5-5}\cmidrule[0.2pt](l){6-6}\cmidrule[0.2pt](l){7-7}\addlinespace[0.12cm]    
(Age $\leq$ 16 in 1990) $\times$ (Highschool)&       0.008   &       0.003   &       0.011   &       0.009   &       0.004   &       0.012   \\
                    &     (0.010)   &     (0.012)   &     (0.010)   &     (0.010)   &     (0.012)   &     (0.011)   \\
                    &[-0.015 0.039]   &[-0.019 0.045]    &[-0.016 0.031]     &[-0.016 0.036]    &[-0.026 0.039]     &[-0.039 0.034]   \\
R-sq                &       0.518   &       0.579   &       0.452   &       0.517   &       0.578   &       0.452   \\
Observations        &       31300   &       16985   &       14315   &       38571   &       20896   &       17675   \\

        \addlinespace[0.1cm]\cmidrule[0.2pt](l){2-2} \cmidrule[0.2pt](l){3-3}\cmidrule[0.2pt](l){4-4}\cmidrule[0.2pt](l){5-5}\cmidrule[0.2pt](l){6-6}\cmidrule[0.2pt](l){7-7}\addlinespace[0.12cm]
(Age $\leq$ 16 in 1988) $\times$ (Highschool)&       0.017   &       0.020   &       0.013   &       0.016   &       0.012   &       0.017   \\
                    &     (0.013)   &     (0.011)   &     (0.017)   &     (0.013)   &     (0.010)   &     (0.019)   \\
                    &[-0.019 0.036]     &[-0.008 0.045]     &[-0.033 0.046]      &[-0.021 0.040]     &[-0.024 0.038]    &[-0.035 0.060]   \\
R-sq                &       0.510   &       0.568   &       0.446   &       0.505   &       0.563   &       0.444   \\
Observations        &       30982   &       16864   &       14118   &       38157   &       20749   &       17408   \\

\addlinespace[0.1cm]\cmidrule[0.2pt](l){2-2} \cmidrule[0.2pt](l){3-3}\cmidrule[0.2pt](l){4-4}\cmidrule[0.2pt](l){5-5}\cmidrule[0.2pt](l){6-6}\cmidrule[0.2pt](l){7-7}\addlinespace[0.12cm]
(Age $\leq$ 16 in 1986) $\times$ (Highschool)&       0.003   &      -0.013   &       0.019   &      -0.001   &      -0.015   &       0.015   \\
                    &     (0.013)   &     (0.010)   &     (0.019)   &     (0.013)   &     (0.010)   &     (0.018)   \\
                    &[-0.036 0.036]    &[-0.034 0.009]     &[-0.040 0.062]    &[-0.039 0.033]    &[-0.045 0.016]        &[-0.044 0.048]   \\
R-sq                &       0.488   &       0.543   &       0.428   &       0.486   &       0.545   &       0.421   \\
Observations        &       31018   &       16926   &       14092   &       37889   &       20716   &       17173   \\

\addlinespace[0.1cm]\cmidrule[0.2pt](l){2-2} \cmidrule[0.2pt](l){3-3}\cmidrule[0.2pt](l){4-4}\cmidrule[0.2pt](l){5-5}\cmidrule[0.2pt](l){6-6}\cmidrule[0.2pt](l){7-7}\addlinespace[0.12cm]
(Age $\leq$ 16 in 1984) $\times$ (Highschool)&      -0.017   &      -0.047   &       0.020   &      -0.013   &      -0.037   &       0.017   \\
                    &     (0.008)*  &     (0.009)***&     (0.017)   &     (0.008)   &     (0.010)** &     (0.018)   \\
                    &[-0.032 0.011]*     &[-0.079 -0.009]***     &[-0.022 0.069]    &[-0.033 0.010]     &[-0.069 0.008]**       &[-0.039 0.063]   \\
R-sq                &       0.476   &       0.532   &       0.415   &       0.475   &       0.531   &       0.413   \\
Observations        &       31148   &       17043   &       14105   &       37896   &       20738   &       17158   \\
         
\addlinespace[0.15cm]\hline\hline\addlinespace[0.15cm]
\multicolumn{7}{p{18cm}}{\scriptsize{\textbf{Notes:} All columns report estimates from Equation (\ref{baselineDiD}), but instead of using the indicator for being at most 16 years old in 1994, we replace it with an indicator for being that age in the specific year indicated in each row. Specifications in columns (1)–(3) use an age window of $[-4, 4]$, while those in columns (4)–(6) use $[-5, 5]$.  The regression samples are drawn from the 2017–2016 and 2020-2021 waves of the ENUT Survey, conducted by DANE. The dependent variable in all columns is a binary indicator equal to 1 if the respondent expresses disagreement with the following statement:  “El esposo debe tomar las decisiones relacionadas con la vida de la esposa?” (“The husband should make decisions concerning his wife’s life?”). All the specifications include cohort-by-region, high school-by-region, and survey year-by-region fixed effects, as well as a set of ethnic group dummies. Robust standard errors, clustered at the region level, are reported in parentheses. Wild bootstrap confidence intervals are shown in square brackets.  *, **, and *** indicate statistical significance at the 10\%, 5\%, and 1\% levels, respectively.} }
\end{tabular}
}
\end{center}
\end{table}
\end{landscape}

\begin{landscape}
\begin{table}[H]
\vspace{0cm}
\begin{center}
{
\renewcommand{\arraystretch}{0.6}
\setlength{\tabcolsep}{5pt}
\caption {Effect on Disagreement with ``the Head of the Household Should be the  Man'' (Years Prior to  1994, Windows [-2,2] and [-3,3])}  \label{table_opinion6_placebo23_boot}
\vspace{-0.3cm}
\scriptsize
\centering  \begin{tabular}{lcccccc}
\hline\hline \addlinespace[0.15cm]
    & (1)& (2)& (3)  & (4)& (5)& (6) \\\addlinespace[0.12cm]\cmidrule[0.2pt](l){2-7}\addlinespace[0.12cm]
& \multicolumn{6}{c}{Dependent Variable: Disagreement with ``the head of the household should be the man''}\\\addlinespace[0.15cm]\cmidrule[0.2pt](l){2-7} \addlinespace[0.10cm]   
      & \multicolumn{3}{c}{Age window [-2,2]} & \multicolumn{3}{c}{Age window [-3,3]}\\\addlinespace[0.1cm]\cmidrule[0.2pt](l){2-4}\cmidrule[0.2pt](l){5-7}
                                 &\multicolumn{1}{c}{All}&\multicolumn{1}{c}{Women}&\multicolumn{1}{c}{Men}&\multicolumn{1}{c}{All}&\multicolumn{1}{c}{Women}&\multicolumn{1}{c}{Men}\\\addlinespace[0.1cm]\cmidrule[0.2pt](l){2-2} \cmidrule[0.2pt](l){3-3}\cmidrule[0.2pt](l){4-4}\cmidrule[0.2pt](l){5-5}\cmidrule[0.2pt](l){6-6}\cmidrule[0.2pt](l){7-7}\addlinespace[0.12cm]

    (Age $\leq$ 16 in 1992) $\times$ (Highschool)&       0.002         &       0.040         &      -0.043         &      -0.008         &       0.030         &      -0.053         \\
                    &     (0.026)         &     (0.024)         &     (0.034)         &     (0.022)         &     (0.022)         &     (0.026)         \\
                    &[-0.070 0.062]         &[-0.007 0.104]         &[-0.171 0.024]              &[-0.075 0.043]         &[-0.041 0.087]           &[-0.156 0.003]         \\
R-sq                &       0.096         &       0.065         &       0.102         &       0.099         &       0.065         &       0.104         \\
Observations        &       17484         &        9465         &        8019         &       24776         &       13389         &       11387         \\

    \addlinespace[0.1cm]\cmidrule[0.2pt](l){2-2} \cmidrule[0.2pt](l){3-3}\cmidrule[0.2pt](l){4-4}\cmidrule[0.2pt](l){5-5}\cmidrule[0.2pt](l){6-6}\cmidrule[0.2pt](l){7-7}\addlinespace[0.12cm]    
(Age $\leq$ 16 in 1990) $\times$ (Highschool)&      -0.015         &      -0.025         &      -0.005         &      -0.004         &      -0.014         &       0.005         \\
                    &     (0.022)         &     (0.027)         &     (0.017)         &     (0.023)         &     (0.029)         &     (0.021)         \\
                    &[-0.058 0.038]         &[-0.079 0.037]             &[-0.037 0.042]           &[-0.050 0.054]            &[-0.072 0.045]           &[-0.030 0.068]         \\
R-sq                &       0.094         &       0.064         &       0.100         &       0.091         &       0.064         &       0.092         \\
Observations        &       17186         &        9396         &        7790         &       24533         &       13398         &       11135         \\

        \addlinespace[0.1cm]\cmidrule[0.2pt](l){2-2} \cmidrule[0.2pt](l){3-3}\cmidrule[0.2pt](l){4-4}\cmidrule[0.2pt](l){5-5}\cmidrule[0.2pt](l){6-6}\cmidrule[0.2pt](l){7-7}\addlinespace[0.12cm]
(Age $\leq$ 16 in 1988) $\times$ (Highschool)&       0.027         &       0.013         &       0.045         &       0.019         &      -0.005         &       0.045         \\
                    &     (0.013)         &     (0.013)         &     (0.025)         &     (0.013)         &     (0.015)         &     (0.020)         \\
                    &[0.003 0.060]          &[-0.023 0.057]         &[-0.025 0.114]       &[-0.010 0.061]      &[-0.037 0.048]   &[-0.010 0.106]         \\
R-sq                &       0.090         &       0.064         &       0.092         &       0.090         &       0.065         &       0.092         \\
Observations        &       17337         &        9522         &        7815         &       24284         &       13347         &       10937         \\

\addlinespace[0.1cm]\cmidrule[0.2pt](l){2-2} \cmidrule[0.2pt](l){3-3}\cmidrule[0.2pt](l){4-4}\cmidrule[0.2pt](l){5-5}\cmidrule[0.2pt](l){6-6}\cmidrule[0.2pt](l){7-7}\addlinespace[0.12cm]
(Age $\leq$ 16 in 1986) $\times$ (Highschool)&      -0.000         &       0.008         &      -0.014         &       0.010         &       0.011         &       0.005         \\
                    &     (0.014)         &     (0.025)         &     (0.015)         &     (0.016)         &     (0.026)         &     (0.011)         \\
                    &[-0.031 0.032]     &[-0.057 0.078]       &[-0.053 0.019]      &[-0.035 0.043]        &[-0.056 0.076]        &[-0.021 0.036]         \\
R-sq                &       0.082         &       0.065         &       0.075         &       0.085         &       0.063         &       0.083         \\
Observations        &       17443         &        9618         &        7825         &       24507         &       13437         &       11070         \\

\addlinespace[0.1cm]\cmidrule[0.2pt](l){2-2} \cmidrule[0.2pt](l){3-3}\cmidrule[0.2pt](l){4-4}\cmidrule[0.2pt](l){5-5}\cmidrule[0.2pt](l){6-6}\cmidrule[0.2pt](l){7-7}\addlinespace[0.12cm]
(Age $\leq$ 16 in 1984) $\times$ (Highschool)&      -0.003         &      -0.008         &       0.006         &      -0.016         &      -0.018         &      -0.013         \\
                    &     (0.022)         &     (0.027)         &     (0.032)         &     (0.018)         &     (0.014)         &     (0.030)         \\
                    &[-0.055 0.057]     &[-0.083 0.074]        &[-0.115 0.057]      &[-0.075 0.035]     &[-0.058 0.023]      &[-0.121 0.050]         \\
R-sq                &       0.086         &       0.063         &       0.084         &       0.088         &       0.066         &       0.085         \\
Observations        &       17395         &        9580         &        7815         &       24578         &       13539         &       11039         \\
         
\addlinespace[0.15cm]\hline\hline\addlinespace[0.15cm]
\multicolumn{7}{p{18cm}}{\scriptsize{\textbf{Notes:} All columns report estimates from Equation (\ref{baselineDiD}), but instead of using the indicator for being at most 16 years old in 1994, we replace it with an indicator for being that age in the specific year indicated in each row. Specifications in columns (1)–(3) use an age window of $[-2, 2]$, while those in columns (4)–(6) use $[-3, 3]$.  The regression samples are drawn from the 2017–2016 and 2020-2021 waves of the ENUT Survey, conducted by DANE. The dependent variable in all columns is a binary indicator equal to 1 if the respondent expresses disagreement with the following statement:   “¿La cabeza del hogar debe ser el hombre?” (“The head of the household should be the man?”). All the specifications include cohort-by-region, high school-by-region, and survey year-by-region fixed effects, as well as a set of ethnic group dummies. Robust standard errors, clustered at the region level, are reported in parentheses. Wild bootstrap confidence intervals are shown in square brackets.  *, **, and *** indicate statistical significance at the 10\%, 5\%, and 1\% levels, respectively.} }
\end{tabular}
}
\end{center}
\end{table}
\end{landscape}

\begin{landscape}
\begin{table}[H]
\vspace{0cm}
\begin{center}
{
\renewcommand{\arraystretch}{0.6}
\setlength{\tabcolsep}{5pt}
\caption {Effect on Disagreement with ``the Head of the Household Should be the Man'' (Years Prior to  1994, Windows [-4,4] and [-5,5])}  \label{table_opinion6_placebo45_boot}
\vspace{-0.3cm}
\scriptsize
\centering  \begin{tabular}{lcccccc}
\hline\hline \addlinespace[0.15cm]
    & (1)& (2)& (3)  & (4)& (5)& (6) \\\addlinespace[0.12cm]\cmidrule[0.2pt](l){2-7}\addlinespace[0.12cm]
& \multicolumn{6}{c}{Dependent Variable: Disagreement with ``the head of the household should be the man''}\\\addlinespace[0.15cm]\cmidrule[0.2pt](l){2-7} \addlinespace[0.10cm]   
      & \multicolumn{3}{c}{Age window [-4,4]} & \multicolumn{3}{c}{Age window [-5,5]}\\\addlinespace[0.1cm]\cmidrule[0.2pt](l){2-4}\cmidrule[0.2pt](l){5-7}
                                 &\multicolumn{1}{c}{All}&\multicolumn{1}{c}{Women}&\multicolumn{1}{c}{Men}&\multicolumn{1}{c}{All}&\multicolumn{1}{c}{Women}&\multicolumn{1}{c}{Men}\\\addlinespace[0.1cm]\cmidrule[0.2pt](l){2-2} \cmidrule[0.2pt](l){3-3}\cmidrule[0.2pt](l){4-4}\cmidrule[0.2pt](l){5-5}\cmidrule[0.2pt](l){6-6}\cmidrule[0.2pt](l){7-7}\addlinespace[0.12cm]

(Age $\leq$ 16 in 1992) $\times$ (Highschool)&      -0.005   &       0.025   &      -0.040   &       0.006   &       0.033   &      -0.025   \\
                    &     (0.014)   &     (0.014)   &     (0.017)*  &     (0.014)   &     (0.017)   &     (0.015)   \\
                    &[-0.046 0.032]    &[-0.029 0.063]      &[-0.105 0.003]     &[-0.042 0.035]   &[-0.040 0.079]      &[-0.085 -0.003]   \\
R-sq                &       0.097   &       0.067   &       0.101   &       0.093   &       0.066   &       0.095   \\
Observations        &       32205   &       17436   &       14769   &       40229   &       21702   &       18527   \\

    \addlinespace[0.1cm]\cmidrule[0.2pt](l){2-2} \cmidrule[0.2pt](l){3-3}\cmidrule[0.2pt](l){4-4}\cmidrule[0.2pt](l){5-5}\cmidrule[0.2pt](l){6-6}\cmidrule[0.2pt](l){7-7}\addlinespace[0.12cm]    
(Age $\leq$ 16 in 1990) $\times$ (Highschool)&      -0.004   &      -0.009   &       0.000   &      -0.000   &      -0.003   &       0.001   \\
                    &     (0.014)   &     (0.022)   &     (0.010)   &     (0.008)   &     (0.013)   &     (0.009)   \\
                    &[-0.043 0.026]    &[-0.056 0.052]    &[-0.025 0.026]     &[-0.021 0.020]    &[-0.033 0.041]     &[-0.019 0.023]   \\
R-sq                &       0.093   &       0.065   &       0.093   &       0.093   &       0.065   &       0.093   \\
Observations        &       31578   &       17237   &       14341   &       38898   &       21186   &       17712   \\

        \addlinespace[0.1cm]\cmidrule[0.2pt](l){2-2} \cmidrule[0.2pt](l){3-3}\cmidrule[0.2pt](l){4-4}\cmidrule[0.2pt](l){5-5}\cmidrule[0.2pt](l){6-6}\cmidrule[0.2pt](l){7-7}\addlinespace[0.12cm]
(Age $\leq$ 16 in 1988) $\times$ (Highschool)&       0.020   &       0.008   &       0.032   &       0.013   &       0.006   &       0.019   \\
                    &     (0.011)   &     (0.012)   &     (0.016)*  &     (0.008)   &     (0.014)   &     (0.007)** \\
                    &[-0.005 0.050]    &[-0.026 0.051]       &[-0.003 0.081]*    &[-0.008 0.032]    &[-0.033 0.044]    &[0.004 0.045]**   \\
R-sq                &       0.088   &       0.063   &       0.089   &       0.087   &       0.061   &       0.088   \\
Observations        &       31231   &       17105   &       14126   &       38446   &       21014   &       17432   \\

\addlinespace[0.1cm]\cmidrule[0.2pt](l){2-2} \cmidrule[0.2pt](l){3-3}\cmidrule[0.2pt](l){4-4}\cmidrule[0.2pt](l){5-5}\cmidrule[0.2pt](l){6-6}\cmidrule[0.2pt](l){7-7}\addlinespace[0.12cm]
(Age $\leq$ 16 in 1986) $\times$ (Highschool)&       0.006   &       0.002   &       0.010   &       0.000   &      -0.010   &       0.010   \\
                    &     (0.013)   &     (0.022)   &     (0.013)   &     (0.012)   &     (0.015)   &     (0.015)   \\
                    &[. 0.043]    &[-0.045 0.110]    &[-0.034 0.044]    &[-0.034 0.032]     &[-0.035 0.068]      &[-0.036 0.052]   \\
R-sq                &       0.088   &       0.063   &       0.087   &       0.090   &       0.065   &       0.091   \\
Observations        &       31232   &       17151   &       14081   &       38138   &       20998   &       17140   \\

\addlinespace[0.1cm]\cmidrule[0.2pt](l){2-2} \cmidrule[0.2pt](l){3-3}\cmidrule[0.2pt](l){4-4}\cmidrule[0.2pt](l){5-5}\cmidrule[0.2pt](l){6-6}\cmidrule[0.2pt](l){7-7}\addlinespace[0.12cm]
(Age $\leq$ 16 in 1984) $\times$ (Highschool)&      -0.006   &      -0.008   &      -0.004   &      -0.000   &      -0.007   &       0.009   \\
                    &     (0.012)   &     (0.011)   &     (0.023)   &     (0.010)   &     (0.009)   &     (0.018)   \\
                    &[-0.042 0.025]     &[-0.039 0.025]   &[-0.085 0.031]     &[-0.023 0.032]     &[-0.021 0.023]     &[-0.056 0.040]   \\
R-sq                &       0.087   &       0.066   &       0.082   &       0.088   &       0.067   &       0.084   \\
Observations        &       31359   &       17262   &       14097   &       38148   &       21003   &       17145   \\
         
\addlinespace[0.15cm]\hline\hline\addlinespace[0.15cm]
\multicolumn{7}{p{18cm}}{\scriptsize{\textbf{Notes:} All columns report estimates from Equation (\ref{baselineDiD}), but instead of using the indicator for being at most 16 years old in 1994, we replace it with an indicator for being that age in the specific year indicated in each row. Specifications in columns (1)–(3) use an age window of $[-4, 4]$, while those in columns (4)–(6) use $[-5, 5]$.  The regression samples are drawn from the 2017–2016 and 2020-2021 waves of the ENUT Survey, conducted by DANE. The dependent variable in all columns is a binary indicator equal to 1 if the respondent expresses disagreement with the following statement:   “¿La cabeza del hogar debe ser el hombre?” (“The head of the household should be the man?”). All the specifications include cohort-by-region, high school-by-region, and survey year-by-region fixed effects, as well as a set of ethnic group dummies. Robust standard errors, clustered at the region level, are reported in parentheses. Wild bootstrap confidence intervals are shown in square brackets.  *, **, and *** indicate statistical significance at the 10\%, 5\%, and 1\% levels, respectively.} }
\end{tabular}
}
\end{center}
\end{table}
\end{landscape}


\begin{figure}[H]
             \caption{Effect of Exposure to Courses on the 1991 Constitution on Support for Gender Equality: Robustness to Alternative Age Windows}
        \label{fig_didgenderequalityagewindows_gender}
\begin{subfigure}{0.5\textwidth}
\caption{Women's Support for Gender Equality} \label{}
\includegraphics[width=\linewidth]{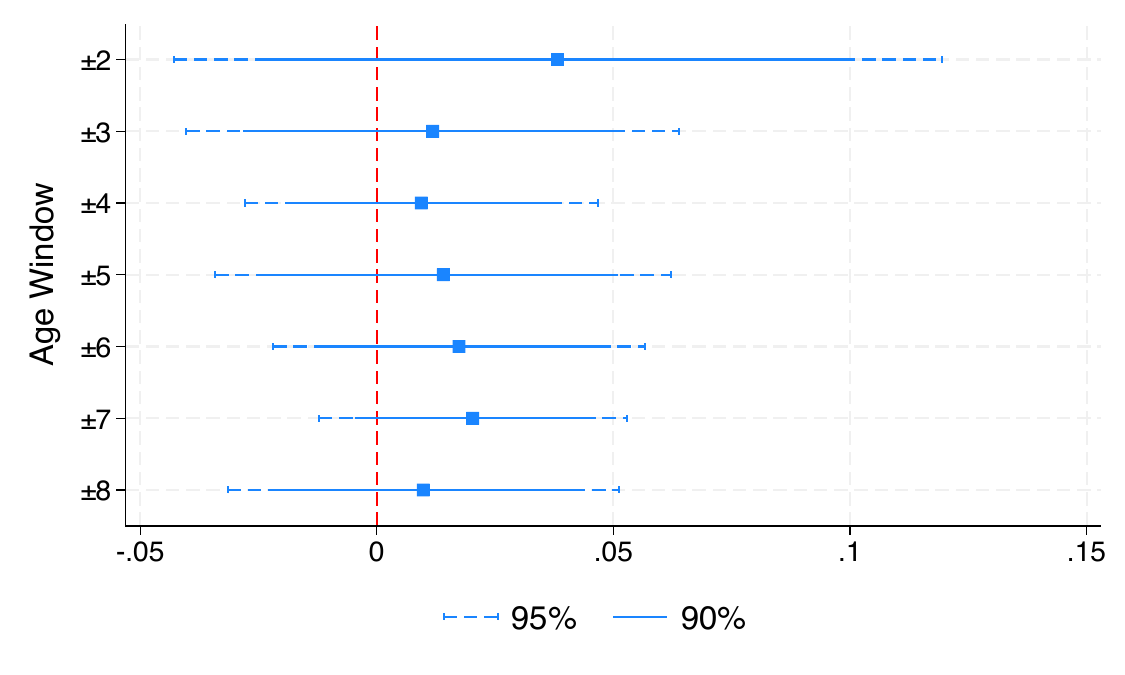}
\end{subfigure}\hspace*{\fill}
\begin{subfigure}{0.5\textwidth}
\caption{Men's Support for Gender Equality} \label{}
\includegraphics[width=\linewidth]{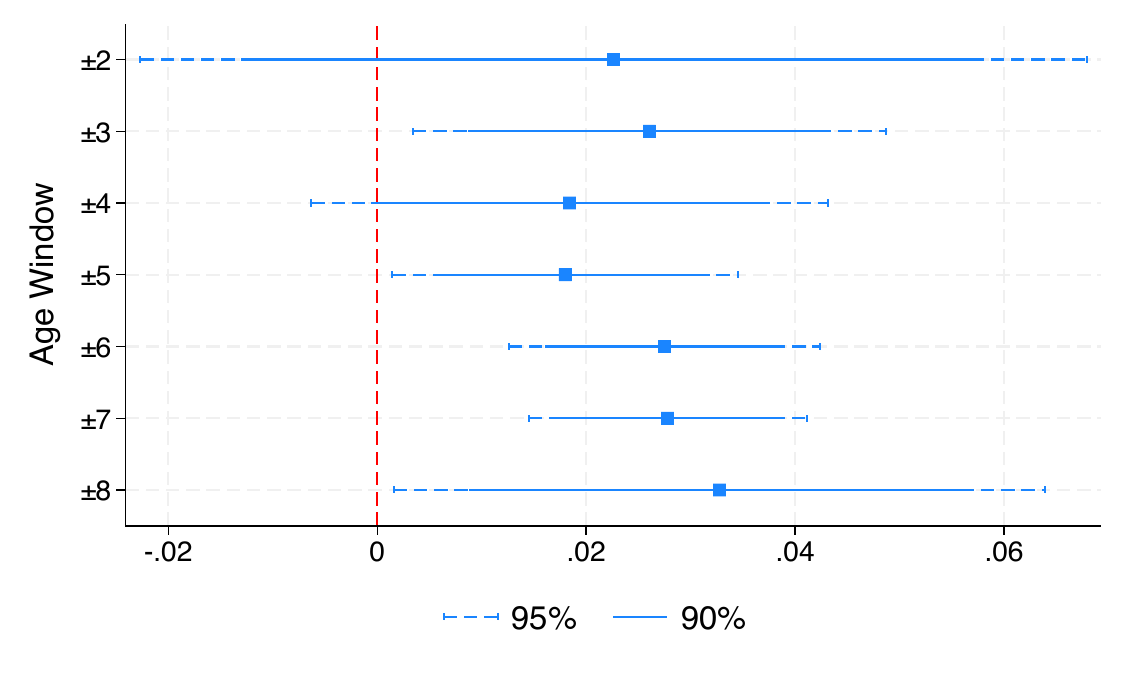}
\end{subfigure}
     \begin{minipage}{16cm} \footnotesizes These figures show the estimates from Eq. (\ref{baselineDiD}), in a specification that includes cohort $\times$ region, high school $\times$ region, and survey year $\times$ region fixed effects, along with a set of ethnic group dummies. 
\end{minipage}
\end{figure}

\begin{table}[H]
\begin{center}
{
\renewcommand{\arraystretch}{0.5}
\setlength{\tabcolsep}{15pt}
\caption {Effect on Support for Gender
Equality: Robustness to Different Age Windows}  \label{table_baselineresults_agewindows}
\vspace{-0.3cm}
\footnotesize
\centering  \begin{tabular}{lccc}
\hline\hline \addlinespace[0.05cm]
    & (1)& (2)& (3)\\\addlinespace[0.05cm]\cmidrule[0.2pt](l){2-4}\addlinespace[0.05cm]
      & \multicolumn{3}{c}{Dependent Variable: Support for Gender Equality}\\\addlinespace[0.05cm]\cmidrule[0.2pt](l){2-4}\addlinespace[0.05cm]    
                      &\multicolumn{1}{c}{All}&\multicolumn{1}{c}{Women}&\multicolumn{1}{c}{Men}\\\addlinespace[0.05cm]\hline
                \addlinespace[0.05cm]
      \multicolumn{1}{l}{\emph{\underline{Panel A}:  }}      & \multicolumn{3}{c}{Age Window: [-2,2]}\\\addlinespace[0.05cm]\cmidrule[0.2pt](l){2-4}\addlinespace[0.05cm]               
(Age $\leq$ 16 in 1994) $\times$ (Highschool)&       0.034   &       0.038   &       0.023   \\
                    &     (0.020)   &     (0.029)   &     (0.016)   \\
                    &[-0.051 0.068]      &[-0.131 0.108]      &[-0.024 0.051]   \\
R-sq                &       0.024   &       0.021   &       0.037   \\
Observations        &        8350   &        4415   &        3935   \\

\addlinespace[0.05cm]\hline\addlinespace[0.05cm]
      \multicolumn{1}{l}{\emph{\underline{Panel B}:  }}      & \multicolumn{3}{c}{Age Window: [-3,3]}\\\addlinespace[0.05cm]\cmidrule[0.2pt](l){2-4}\addlinespace[0.05cm]               
(Age $\leq$ 16 in 1994) $\times$ (Highschool)&       0.021   &       0.012   &       0.026   \\
                    &     (0.013)   &     (0.019)   &     (0.008)** \\
                    &[-0.008 0.059]       &[-0.036 0.052]       &[0.010 0.063]***   \\
R-sq                &       0.021   &       0.019   &       0.033   \\
Observations        &       11816   &        6275   &        5541   \\

\addlinespace[0.05cm]\hline\addlinespace[0.05cm]
      \multicolumn{1}{l}{\emph{\underline{Panel C}:  }}      & \multicolumn{3}{c}{Age Window: [-4,4]}\\\addlinespace[0.05cm]\cmidrule[0.2pt](l){2-4}\addlinespace[0.05cm]               
(Age $\leq$ 16 in 1994) $\times$ (Highschool)&       0.015   &       0.009   &       0.018   \\
                    &     (0.009)   &     (0.013)   &     (0.009)*   \\
                    &[-0.007 0.105]     &[-0.036 0.143]       &[0.000 0.059]***   \\
R-sq                &       0.019   &       0.020   &       0.028   \\
Observations        &       15120   &        8076   &        7044   \\

\addlinespace[0.05cm]\hline\addlinespace[0.05cm]
      \multicolumn{1}{l}{\emph{\underline{Panel D}:  }}      & \multicolumn{3}{c}{Age Window: [-5,5]}\\\addlinespace[0.05cm]\cmidrule[0.2pt](l){2-4}\addlinespace[0.05cm]               
(Age $\leq$ 16 in 1994) $\times$ (Highschool)&       0.017   &       0.014   &       0.018   \\
                    &     (0.008)*  &     (0.017)   &     (0.006)** \\
                    &[-0.006 0.088]*       &[-0.032 0.121]      &[-0.004 0.056]***   \\
R-sq                &       0.021   &       0.023   &       0.028   \\
Observations        &       18429   &        9856   &        8573   \\

\addlinespace[0.05cm]\hline\addlinespace[0.05cm]
      \multicolumn{1}{l}{\emph{\underline{Panel E}:  }}      & \multicolumn{3}{c}{Age Window: [-6,6]}\\\addlinespace[0.05cm]\cmidrule[0.2pt](l){2-4}\addlinespace[0.05cm]               
(Age $\leq$ 16 in 1994) $\times$ (Highschool)&       0.023   &       0.017   &       0.028   \\
                    &     (0.008)** &     (0.014)   &     (0.005)***\\
                    &[0.001 0.095]**       &[-0.018 0.096]      &[0.017 0.103]***   \\
R-sq                &       0.020   &       0.022   &       0.026   \\
Observations        &       21698   &       11617   &       10081   \\

\addlinespace[0.05cm]\hline\addlinespace[0.05cm]
      \multicolumn{1}{l}{\emph{\underline{Panel F}:  }}      & \multicolumn{3}{c}{Age Window: [-7,7]}\\\addlinespace[0.05cm]\cmidrule[0.2pt](l){2-4}\addlinespace[0.05cm]               
(Age $\leq$ 16 in 1994) $\times$ (Highschool)&       0.024   &       0.020   &       0.028   \\
                    &     (0.008)** &     (0.012)   &     (0.005)***\\
                    &[0.006 0.080]**       &[-0.001 0.105]       &[0.011 0.061]***   \\
R-sq                &       0.020   &       0.023   &       0.024   \\
Observations        &       25105   &       13432   &       11673   \\

\addlinespace[0.05cm]\hline\addlinespace[0.05cm]
      \multicolumn{1}{l}{\emph{\underline{Panel G}:  }}      & \multicolumn{3}{c}{Age Window: [-8,8]}\\\addlinespace[0.05cm]\cmidrule[0.2pt](l){2-4}\addlinespace[0.05cm]               
(Age $\leq$ 16 in 1994) $\times$ (Highschool)&       0.021   &       0.010   &       0.033   \\
                    &     (0.012)   &     (0.015)   &     (0.011)** \\
                    &[-0.004 0.090]     &[-0.015 0.120]      &[0.000 0.065]***   \\
R-sq                &       0.019   &       0.021   &       0.023   \\
Observations        &       28431   &       15270   &       13161   \\

\addlinespace[0.05cm]\hline\hline\addlinespace[0.05cm]
\multicolumn{4}{p{15.5cm}}{\scriptsize{\textbf{Notes:} All columns report estimates from Equation (\ref{baselineDiD}), using the age window indicated in each panel. The regression samples are drawn from the 2019a and 2021 waves of the Political Culture Survey conducted by DANE. The dependent variable in all columns is the response to the question (in Spanish): ``¿Usted est\'a de acuerdo con que las mujeres participen en igualdad de condiciones y oportunidades?'' (``Do you agree that women should participate on equal terms and have the same opportunities?''). All the specifications include cohort-by-region, high school-by-region, and survey year-by-region fixed effects, as well as a set of ethnic group dummies. Robust standard errors, clustered at the region level, are reported in parentheses. Wild bootstrap confidence intervals are shown in square brackets.  *, **, and *** indicate statistical significance at the 10\%, 5\%, and 1\% levels, respectively.} }
\end{tabular}
}
\end{center}
\end{table}

\begin{table}[H]
\begin{center}
\renewcommand{\arraystretch}{0.5}
\setlength{\tabcolsep}{15pt}
\caption {Effect on Perception of Gender Discrimination: Robustness to Different Age Windows}  \label{table_sexdiscrimination_agewindows}
\vspace{-0.3cm}
\footnotesize
\centering  \begin{tabular}{lccc}
\hline\hline \addlinespace[0.05cm]
    & (1)& (2)& (3)\\\addlinespace[0.05cm]\cmidrule[0.2pt](l){2-4}\addlinespace[0.05cm]
      & \multicolumn{3}{c}{Dependent Variable: Perception of Gender Discrimination}\\\addlinespace[0.05cm]\cmidrule[0.2pt](l){2-4}\addlinespace[0.05cm]    
                      &\multicolumn{1}{c}{All}&\multicolumn{1}{c}{Women}&\multicolumn{1}{c}{Men}\\\addlinespace[0.05cm]\hline
                \addlinespace[0.05cm]
      \multicolumn{1}{l}{\emph{\underline{Panel A}:  }}      & \multicolumn{3}{c}{Age Window: [-2,2]}\\\addlinespace[0.05cm]\cmidrule[0.2pt](l){2-4}\addlinespace[0.05cm]               
(Age $\leq$ 16 in 1994) $\times$ (Highschool)&      -0.003   &      -0.018   &       0.013   \\
                    &     (0.006)   &     (0.013)   &     (0.010)   \\
                    &[-0.027 0.007]         &[-0.049 0.009]        &[-0.000 0.045]   \\
R-sq                &       0.015   &       0.025   &       0.014   \\
Observations        &        4568   &        2432   &        2136   \\

\addlinespace[0.05cm]\hline\addlinespace[0.05cm]
      \multicolumn{1}{l}{\emph{\underline{Panel B}:  }}      & \multicolumn{3}{c}{Age Window: [-3,3]}\\\addlinespace[0.05cm]\cmidrule[0.2pt](l){2-4}\addlinespace[0.05cm]               
(Age $\leq$ 16 in 1994) $\times$ (Highschool)&      -0.005   &      -0.020   &       0.012   \\
                    &     (0.003)   &     (0.007)** &     (0.010)   \\
                    &[-0.033 0.003]     &[-0.059 -0.005]***      &[-0.000 0.044]   \\
R-sq                &       0.013   &       0.020   &       0.013   \\
Observations        &        6555   &        3495   &        3060   \\

\addlinespace[0.05cm]\hline\addlinespace[0.05cm]
      \multicolumn{1}{l}{\emph{\underline{Panel C}:  }}      & \multicolumn{3}{c}{Age Window: [-4,4]}\\\addlinespace[0.05cm]\cmidrule[0.2pt](l){2-4}\addlinespace[0.05cm]               
(Age $\leq$ 16 in 1994) $\times$ (Highschool)&      -0.006   &      -0.020   &       0.010   \\
                    &     (0.003)   &     (0.003)***&     (0.006)   \\
                    &[-0.032 0.001]*      &[-0.063 -0.011]***     &[-0.001 0.026]   \\
R-sq                &       0.016   &       0.019   &       0.039   \\
Observations        &        8327   &        4467   &        3860   \\

\addlinespace[0.05cm]\hline\addlinespace[0.05cm]
      \multicolumn{1}{l}{\emph{\underline{Panel D}:  }}      & \multicolumn{3}{c}{Age Window: [-5,5]}\\\addlinespace[0.05cm]\cmidrule[0.2pt](l){2-4}\addlinespace[0.05cm]               
(Age $\leq$ 16 in 1994) $\times$ (Highschool)&      -0.005   &      -0.016   &       0.007   \\
                    &     (0.003)   &     (0.003)***&     (0.005)   \\
                    &[-0.028 0.001]*        &[-0.047 -0.008]***        &[-0.002 0.021]   \\
R-sq                &       0.015   &       0.020   &       0.029   \\
Observations        &       10139   &        5450   &        4689   \\

\addlinespace[0.05cm]\hline\addlinespace[0.05cm]
      \multicolumn{1}{l}{\emph{\underline{Panel E}:  }}      & \multicolumn{3}{c}{Age Window: [-6,6]}\\\addlinespace[0.05cm]\cmidrule[0.2pt](l){2-4}\addlinespace[0.05cm]               
(Age $\leq$ 16 in 1994) $\times$ (Highschool)&      -0.004   &      -0.015   &       0.008   \\
                    &     (0.002)*  &     (0.005)** &     (0.006)   \\
                    &[-0.009 -0.001]*       &[-0.023 0.001]        &[-0.004 0.026]   \\
R-sq                &       0.013   &       0.018   &       0.024   \\
Observations        &       11923   &        6407   &        5516   \\

\addlinespace[0.05cm]\hline\addlinespace[0.05cm]
      \multicolumn{1}{l}{\emph{\underline{Panel F}:  }}      & \multicolumn{3}{c}{Age Window: [-7,7]}\\\addlinespace[0.05cm]\cmidrule[0.2pt](l){2-4}\addlinespace[0.05cm]               
(Age $\leq$ 16 in 1994) $\times$ (Highschool)&      -0.006   &      -0.015   &       0.005   \\
                    &     (0.002)*  &     (0.005)** &     (0.006)   \\
                    &[-0.011 -0.002]*        &[-0.026 -0.000]***          &[-0.004 0.024]   \\
R-sq                &       0.013   &       0.017   &       0.022   \\
Observations        &       13881   &        7463   &        6418   \\

\addlinespace[0.05cm]\hline\addlinespace[0.05cm]
      \multicolumn{1}{l}{\emph{\underline{Panel G}:  }}      & \multicolumn{3}{c}{Age Window: [-8,8]}\\\addlinespace[0.05cm]\cmidrule[0.2pt](l){2-4}\addlinespace[0.05cm]               
(Age $\leq$ 16 in 1994) $\times$ (Highschool)&      -0.005   &      -0.014   &       0.005   \\
                    &     (0.002)*  &     (0.005)*  &     (0.005)   \\
                    &[-0.011 -0.001]        &[-0.022 0.003]      &[-0.003 0.018]   \\
R-sq                &       0.012   &       0.016   &       0.021   \\
Observations        &       15665   &        8424   &        7241   \\

\addlinespace[0.05cm]\hline\hline\addlinespace[0.05cm]
\multicolumn{4}{p{16cm}}{\scriptsize{\textbf{Notes:} All columns report estimates from Equation (\ref{baselineDiD}), using the age window indicated in each panel. The regression samples are drawn from the 2019 and 2021 waves of the Political Culture Survey conducted by DANE. The dependent variable in all columns is the response to the question (in Spanish): ``Durantelos \'ultimos 12 meses, usted ha sentido que lo han discriminado [...] por alguno de los siguientes motivos: Sexo [...]'' (``In the past 12 months, have you felt discriminated against [...] based on sex?''). All the include cohort-by-region, high school-by-region, and survey year-by-region fixed effects, as well as a set of ethnic group dummies. Robust standard errors, clustered at the region level, are reported in parentheses. Wild bootstrap confidence intervals are shown in square brackets.  *, **, and *** indicate statistical significance at the 10\%, 5\%, and 1\% levels, respectively.}}
\end{tabular}
\end{center}
\end{table}

\begin{table}[H]
\begin{center}
{
\renewcommand{\arraystretch}{0.5}
\setlength{\tabcolsep}{15pt}
\caption {Effect on Disagreement with ``Women are better suited for domestic tasks than men'': Robustness to Different Age Windows}  \label{table_domesticopinion4_agewindows}
\vspace{-0.3cm}
\footnotesize
\centering  \begin{tabular}{lccc}
\hline\hline \addlinespace[0.05cm]
    & (1)& (2)& (3)\\\addlinespace[0.05cm]\cmidrule[0.2pt](l){2-4}\addlinespace[0.05cm]
          & \multicolumn{3}{c}{Dep. Var.: Disagree with the statement}\\\addlinespace[0.05cm]
      & \multicolumn{3}{c}{``Women are better suited for domestic tasks''}\\\addlinespace[0.05cm]\cmidrule[0.2pt](l){2-4}\addlinespace[0.05cm]    
                      &\multicolumn{1}{c}{All}&\multicolumn{1}{c}{Women}&\multicolumn{1}{c}{Men}\\\addlinespace[0.05cm]\hline
                \addlinespace[0.05cm]
      \multicolumn{1}{l}{\emph{\underline{Panel A}:  }}      & \multicolumn{3}{c}{Age Window: [-2,2]}\\\addlinespace[0.05cm]\cmidrule[0.2pt](l){2-4}\addlinespace[0.05cm]               
(Age $\leq$ 16 in 1994) $\times$ (Highschool)&      -0.020   &      -0.020   &      -0.021   \\
                    &     (0.010)  &     (0.017)   &     (0.018)   \\
                    &[-0.047 0.001]          &[-0.065 0.075]      &[-0.075 0.046]   \\
R-sq                &       0.051   &       0.051   &       0.055   \\
Observations        &       18447   &        9871   &        8576   \\

\addlinespace[0.05cm]\hline\addlinespace[0.05cm]
      \multicolumn{1}{l}{\emph{\underline{Panel B}:  }}      & \multicolumn{3}{c}{Age Window: [-3,3]}\\\addlinespace[0.05cm]\cmidrule[0.2pt](l){2-4}\addlinespace[0.05cm]               
(Age $\leq$ 16 in 1994) $\times$ (Highschool)&      -0.006   &      -0.003   &      -0.010   \\
                    &     (0.009)   &     (0.016)   &     (0.009)   \\
                    &[-0.036 0.020]       &[-0.046 0.041]        &[-0.030 0.019]   \\
R-sq                &       0.052   &       0.050   &       0.056   \\
Observations        &       26061   &       13933   &       12128   \\

\addlinespace[0.05cm]\hline\addlinespace[0.05cm]
      \multicolumn{1}{l}{\emph{\underline{Panel C}:  }}      & \multicolumn{3}{c}{Age Window: [-4,4]}\\\addlinespace[0.05cm]\cmidrule[0.2pt](l){2-4}\addlinespace[0.05cm]               
(Age $\leq$ 16 in 1994) $\times$ (Highschool)&      -0.003   &      -0.001   &      -0.007   \\
                    &     (0.011)   &     (0.017)   &     (0.010)   \\
                    &[-0.025 0.021]         &[-0.054 0.047]        &[-0.050 0.014]   \\
R-sq                &       0.050   &       0.049   &       0.055   \\
Observations        &       33242   &       17840   &       15402   \\

\addlinespace[0.05cm]\hline\addlinespace[0.05cm]
      \multicolumn{1}{l}{\emph{\underline{Panel D}:  }}      & \multicolumn{3}{c}{Age Window: [-5,5]}\\\addlinespace[0.05cm]\cmidrule[0.2pt](l){2-4}\addlinespace[0.05cm]               
(Age $\leq$ 16 in 1994) $\times$ (Highschool)&      -0.001   &       0.002   &      -0.007   \\
                    &     (0.010)   &     (0.014)   &     (0.009)   \\
                    &[-0.027 0.020]      &[-0.032 0.033]       &[-0.041 0.015]   \\
R-sq                &       0.049   &       0.048   &       0.054   \\
Observations        &       40748   &       21904   &       18844   \\

\addlinespace[0.05cm]\hline\addlinespace[0.05cm]
      \multicolumn{1}{l}{\emph{\underline{Panel E}:  }}      & \multicolumn{3}{c}{Age Window: [-6,6]}\\\addlinespace[0.05cm]\cmidrule[0.2pt](l){2-4}\addlinespace[0.05cm]               
(Age $\leq$ 16 in 1994) $\times$ (Highschool)&       0.004   &       0.007   &      -0.001   \\
                    &     (0.008)   &     (0.012)   &     (0.006)   \\
                    &[-0.007 0.025]      &[-0.026 0.037]        &[-0.019 0.017]   \\
R-sq                &       0.049   &       0.048   &       0.055   \\
Observations        &       48173   &       26041   &       22132   \\

\addlinespace[0.05cm]\hline\addlinespace[0.05cm]
      \multicolumn{1}{l}{\emph{\underline{Panel F}:  }}      & \multicolumn{3}{c}{Age Window: [-7,7]}\\\addlinespace[0.05cm]\cmidrule[0.2pt](l){2-4}\addlinespace[0.05cm]               
(Age $\leq$ 16 in 1994) $\times$ (Highschool)&       0.010   &       0.012   &       0.008   \\
                    &     (0.006)   &     (0.009)   &     (0.007)   \\
                    &[-0.000 0.027]       &[-0.018 0.046]       &[-0.021 0.022]   \\
R-sq                &       0.050   &       0.048   &       0.056   \\
Observations        &       55949   &       30272   &       25677   \\

\addlinespace[0.05cm]\hline\addlinespace[0.05cm]
      \multicolumn{1}{l}{\emph{\underline{Panel G}:  }}      & \multicolumn{3}{c}{Age Window: [-8,8]}\\\addlinespace[0.05cm]\cmidrule[0.2pt](l){2-4}\addlinespace[0.05cm]              
(Age $\leq$ 16 in 1994) $\times$ (Highschool)&       0.007   &       0.010   &       0.003   \\
                    &     (0.004)   &     (0.010)   &     (0.008)   \\
                    &[-0.002 0.020]        &[-0.021 0.029]       &[-0.031 0.031]   \\
R-sq                &       0.049   &       0.048   &       0.054   \\
Observations        &       63534   &       34406   &       29128   \\

\addlinespace[0.05cm]\hline\hline\addlinespace[0.05cm]
\multicolumn{4}{p{15cm}}{\scriptsize{\textbf{Notes:} All columns report estimates from Equation (\ref{baselineDiD}), using the age window indicated in each panel. The regression samples are drawn from the 2017–2016 and 2020-2021 waves of the ENUT Survey, conducted by DANE. The dependent variable in all columns is a binary indicator equal to 1 if the respondent expresses disagreement with the following statement: “¿Las mujeres son mejores para el trabajo doméstico que los hombres?” (“Women are better suited for domestic tasks than men?”). All the include cohort-by-region, high school-by-region, and survey year-by-region fixed effects, as well as a set of ethnic group dummies. Robust standard errors, clustered at the region level, are reported in parentheses. Wild bootstrap confidence intervals are shown in square brackets.  *, **, and *** indicate statistical significance at the 10\%, 5\%, and 1\% levels, respectively.} }
\end{tabular}
}
\end{center}
\end{table}

\begin{table}[H]
\begin{center}
{
\renewcommand{\arraystretch}{0.5}
\setlength{\tabcolsep}{15pt}
\caption {Effect on Disagreement with the statement ``The husband should make decisions concerning his wife’s life'': Robustness to Different Age Windows}  \label{table_domesticopinion5_agewindows}
\vspace{-0.3cm}
\footnotesize
\centering  \begin{tabular}{lccc}
\hline\hline \addlinespace[0.05cm]
    & (1)& (2)& (3)\\\addlinespace[0.05cm]\cmidrule[0.2pt](l){2-4}\addlinespace[0.05cm]
          & \multicolumn{3}{c}{Dep. Var.: Disagreement with:  the husband}\\\addlinespace[0.05cm]
      & \multicolumn{3}{c}{should make decisions concerning his wife's life}\\\addlinespace[0.05cm]\cmidrule[0.2pt](l){2-4}\addlinespace[0.05cm]    
                      &\multicolumn{1}{c}{All}&\multicolumn{1}{c}{Women}&\multicolumn{1}{c}{Men}\\\addlinespace[0.05cm]\hline
                \addlinespace[0.05cm]
      \multicolumn{1}{l}{\emph{\underline{Panel A}:  }}      & \multicolumn{3}{c}{Age Window: [-2,2]}\\\addlinespace[0.05cm]\cmidrule[0.2pt](l){2-4}\addlinespace[0.05cm]               
(Age $\leq$ 16 in 1994) $\times$ (Highschool)&       0.010   &       0.037   &      -0.016   \\
                    &     (0.010)   &     (0.022)   &     (0.016)   \\
                    &[-0.018 0.035]      &[-0.015 0.095]     &[-0.077 0.027]   \\
R-sq                &       0.543   &       0.608   &       0.479   \\
Observations        &       18489   &        9884   &        8605   \\

\addlinespace[0.05cm]\hline\addlinespace[0.05cm]
      \multicolumn{1}{l}{\emph{\underline{Panel B}:  }}      & \multicolumn{3}{c}{Age Window: [-3,3]}\\\addlinespace[0.05cm]\cmidrule[0.2pt](l){2-4}\addlinespace[0.05cm]               
(Age $\leq$ 16 in 1994) $\times$ (Highschool)&       0.005   &       0.028   &      -0.018   \\
                    &     (0.010)   &     (0.016)   &     (0.012)   \\
                    &[-0.021 0.026]        &[-0.019 0.067]         &[-0.053 0.004]   \\
R-sq                &       0.545   &       0.611   &       0.478   \\
Observations        &       26150   &       13963   &       12187   \\

\addlinespace[0.05cm]\hline\addlinespace[0.05cm]
      \multicolumn{1}{l}{\emph{\underline{Panel C}:  }}      & \multicolumn{3}{c}{Age Window: [-4,4]}\\\addlinespace[0.05cm]\cmidrule[0.2pt](l){2-4}\addlinespace[0.05cm]               
(Age $\leq$ 16 in 1994) $\times$ (Highschool)&      -0.002   &       0.023   &      -0.026   \\
                    &     (0.009)   &     (0.014)   &     (0.007)***\\
                    &[-0.027 0.018]    &[-0.024 0.047]       &[-0.043 -0.012]**   \\
R-sq                &       0.547   &       0.610   &       0.483   \\
Observations        &       33344   &       17867   &       15477   \\

\addlinespace[0.05cm]\hline\addlinespace[0.05cm]
      \multicolumn{1}{l}{\emph{\underline{Panel D}:  }}      & \multicolumn{3}{c}{Age Window: [-5,5]}\\\addlinespace[0.05cm]\cmidrule[0.2pt](l){2-4}\addlinespace[0.05cm]               
(Age $\leq$ 16 in 1994) $\times$ (Highschool)&      -0.002   &       0.018   &      -0.020   \\
                    &     (0.009)   &     (0.012)   &     (0.013)   \\
                    &[-0.022 0.024]         &[-0.027 0.039]      &[-0.056 0.012]   \\
R-sq                &       0.545   &       0.612   &       0.476   \\
Observations        &       40868   &       21928   &       18940   \\

\addlinespace[0.05cm]\hline\addlinespace[0.05cm]
      \multicolumn{1}{l}{\emph{\underline{Panel E}:  }}      & \multicolumn{3}{c}{Age Window: [-6,6]}\\\addlinespace[0.05cm]\cmidrule[0.2pt](l){2-4}\addlinespace[0.05cm]               
(Age $\leq$ 16 in 1994) $\times$ (Highschool)&      -0.003   &       0.017   &      -0.023   \\
                    &     (0.005)   &     (0.005)** &     (0.008)** \\
                    &[-0.014 0.010]         &[-0.001 0.033]*      &[-0.045 -0.002]***   \\
R-sq                &       0.546   &       0.608   &       0.480   \\
Observations        &       48303   &       26057   &       22246   \\

\addlinespace[0.05cm]\hline\addlinespace[0.05cm]
      \multicolumn{1}{l}{\emph{\underline{Panel F}:  }}      & \multicolumn{3}{c}{Age Window: [-7,7]}\\\addlinespace[0.05cm]\cmidrule[0.2pt](l){2-4}\addlinespace[0.05cm]               
(Age $\leq$ 16 in 1994) $\times$ (Highschool)&       0.000   &       0.020   &      -0.020   \\
                    &     (0.004)   &     (0.004)***&     (0.007)** \\
                    &[-0.009 0.011]     &[0.007 0.031]***        &[-0.040 -0.004]***   \\
R-sq                &       0.547   &       0.610   &       0.481   \\
Observations        &       56115   &       30286   &       25829   \\

\addlinespace[0.05cm]\hline\addlinespace[0.05cm]
      \multicolumn{1}{l}{\emph{\underline{Panel G}:  }}      & \multicolumn{3}{c}{Age Window: [-8,8]}\\\addlinespace[0.05cm]\cmidrule[0.2pt](l){2-4}\addlinespace[0.05cm]               
(Age $\leq$ 16 in 1994) $\times$ (Highschool)&      -0.003   &       0.018   &      -0.024   \\
                    &     (0.003)   &     (0.005)** &     (0.004)***\\
                    &[-0.010 0.008]    &[0.001 0.032]**    &[-0.032 -0.016]***   \\
R-sq                &       0.549   &       0.612   &       0.483   \\
Observations        &       63751   &       34434   &       29317   \\

\addlinespace[0.05cm]\hline\hline\addlinespace[0.05cm]
\multicolumn{4}{p{16cm}}{\scriptsize{\textbf{Notes:} All columns report estimates from Equation (\ref{baselineDiD}), using the age window indicated in each panel. The regression samples are drawn from the 2017–2016 and 2020-2021 waves of the ENUT Survey, conducted by DANE. The dependent variable in all columns is a binary indicator equal to 1 if the respondent expresses disagreement with the following statement:  “El esposo debe tomar las decisiones relacionadas con la vida de la esposa?” (“The husband should make decisions concerning his wife’s life”). All the include cohort-by-region, high school-by-region, and survey year-by-region fixed effects, as well as a set of ethnic group dummies. Robust standard errors, clustered at the region level, are reported in parentheses. Wild bootstrap confidence intervals are shown in square brackets.  *, **, and *** indicate statistical significance at the 10\%, 5\%, and 1\% levels, respectively.} }
\end{tabular}
}
\end{center}
\end{table}

\begin{table}[H]
\begin{center}
{
\renewcommand{\arraystretch}{0.5}
\setlength{\tabcolsep}{15pt}
\caption {Effect on Disagreement with ``The head of the household should be the man'': Robustness to Different Age Windows}  \label{table_domesticopinion6_agewindows}
\vspace{-0.3cm}
\footnotesize
\centering  \begin{tabular}{lccc}
\hline\hline \addlinespace[0.05cm]
    & (1)& (2)& (3)\\\addlinespace[0.05cm]\cmidrule[0.2pt](l){2-4}\addlinespace[0.05cm]
          & \multicolumn{3}{c}{Dep. Var.: Disagreement with: ``the head of the}\\\addlinespace[0.05cm]
      & \multicolumn{3}{c}{household should be the man''}\\\addlinespace[0.05cm]\cmidrule[0.2pt](l){2-4}\addlinespace[0.05cm]    
                      &\multicolumn{1}{c}{All}&\multicolumn{1}{c}{Women}&\multicolumn{1}{c}{Men}\\\addlinespace[0.05cm]\hline
                \addlinespace[0.05cm]
      \multicolumn{1}{l}{\emph{\underline{Panel A}:  }}      & \multicolumn{3}{c}{Age Window: [-2,2]}\\\addlinespace[0.05cm]\cmidrule[0.2pt](l){2-4}\addlinespace[0.05cm]               
(Age $\leq$ 16 in 1994) $\times$ (Highschool)&       0.014   &       0.017   &       0.013   \\
                    &     (0.020)   &     (0.017)   &     (0.029)   \\
                    &[-0.028 0.077]       &[-0.028 0.055]       &[-0.042 0.130]   \\
R-sq                &       0.099   &       0.071   &       0.100   \\
Observations        &       18645   &        9977   &        8668   \\

\addlinespace[0.05cm]\hline\addlinespace[0.05cm]
      \multicolumn{1}{l}{\emph{\underline{Panel B}:  }}      & \multicolumn{3}{c}{Age Window: [-3,3]}\\\addlinespace[0.05cm]\cmidrule[0.2pt](l){2-4}\addlinespace[0.05cm]               
(Age $\leq$ 16 in 1994) $\times$ (Highschool)&       0.017   &       0.042   &      -0.007   \\
                    &     (0.012)   &     (0.011)** &     (0.013)   \\
                    &[-0.006 0.051]       &[0.011 0.076]***       &[-0.033 0.026]   \\
R-sq                &       0.096   &       0.067   &       0.098   \\
Observations        &       26392   &       14131   &       12261   \\

\addlinespace[0.05cm]\hline\addlinespace[0.05cm]
      \multicolumn{1}{l}{\emph{\underline{Panel C}:  }}      & \multicolumn{3}{c}{Age Window: [-4,4]}\\\addlinespace[0.05cm]\cmidrule[0.2pt](l){2-4}\addlinespace[0.05cm]               
(Age $\leq$ 16 in 1994) $\times$ (Highschool)&       0.007   &       0.039   &      -0.023   \\
                    &     (0.010)   &     (0.014)** &     (0.007)** \\
                    &[-0.024 0.042]       &[-0.017 0.061]*       &[-0.040 -0.000]**   \\
R-sq                &       0.094   &       0.064   &       0.099   \\
Observations        &       33648   &       18093   &       15555   \\

\addlinespace[0.05cm]\hline\addlinespace[0.05cm]
      \multicolumn{1}{l}{\emph{\underline{Panel D}:  }}      & \multicolumn{3}{c}{Age Window: [-5,5]}\\\addlinespace[0.05cm]\cmidrule[0.2pt](l){2-4}\addlinespace[0.05cm]               
(Age $\leq$ 16 in 1994) $\times$ (Highschool)&       0.001   &       0.037   &      -0.034   \\
                    &     (0.011)   &     (0.015)*  &     (0.007)***\\
                    &[-0.050 0.033]       &[-0.027 0.063]*      &[-0.062 -0.015]*   \\
R-sq                &       0.095   &       0.064   &       0.101   \\
Observations        &       41224   &       22208   &       19016   \\

\addlinespace[0.05cm]\hline\addlinespace[0.05cm]
      \multicolumn{1}{l}{\emph{\underline{Panel E}:  }}      & \multicolumn{3}{c}{Age Window: [-6,6]}\\\addlinespace[0.05cm]\cmidrule[0.2pt](l){2-4}\addlinespace[0.05cm]               
(Age $\leq$ 16 in 1994) $\times$ (Highschool)&       0.001   &       0.028   &      -0.026   \\
                    &     (0.010)   &     (0.013)*  &     (0.008)** \\
                    &[-0.051 0.021]    &[-0.058 0.052]*      &[-0.047 -0.010]**   \\
R-sq                &       0.094   &       0.065   &       0.098   \\
Observations        &       48723   &       26404   &       22319   \\

\addlinespace[0.05cm]\hline\addlinespace[0.05cm]
      \multicolumn{1}{l}{\emph{\underline{Panel F}:  }}      & \multicolumn{3}{c}{Age Window: [-7,7]}\\\addlinespace[0.05cm]\cmidrule[0.2pt](l){2-4}\addlinespace[0.05cm]               
(Age $\leq$ 16 in 1994) $\times$ (Highschool)&       0.012   &       0.037   &      -0.011   \\
                    &     (0.010)   &     (0.013)** &     (0.012)   \\
                    &[-0.040 0.035]    &[-0.061 0.058]**     &[-0.041 0.015]   \\
R-sq                &       0.094   &       0.066   &       0.096   \\
Observations        &       56629   &       30719   &       25910   \\

\addlinespace[0.05cm]\hline\addlinespace[0.05cm]
      \multicolumn{1}{l}{\emph{\underline{Panel G}:  }}      & \multicolumn{3}{c}{Age Window: [-8,8]}\\\addlinespace[0.05cm]\cmidrule[0.2pt](l){2-4}\addlinespace[0.05cm]               
(Age $\leq$ 16 in 1994) $\times$ (Highschool)&       0.009   &       0.035   &      -0.017   \\
                    &     (0.007)   &     (0.012)** &     (0.008)*  \\
                    &[-0.037 0.024]    &[-0.061 0.050]**        &[-0.038 -0.001]*   \\
R-sq                &       0.094   &       0.067   &       0.094   \\
Observations        &       64327   &       34914   &       29413   \\

\addlinespace[0.05cm]\hline\hline\addlinespace[0.05cm]
\multicolumn{4}{p{16cm}}{\scriptsize{\textbf{Notes:} All columns report estimates from Equation (\ref{baselineDiD}), using the age window indicated in eah panel. The regression samples are drawn from the 2017–2016 and 2020-2021 waves of the ENUT Survey, conducted by DANE. The regression samples are drawn from the 2017–2016 and 2020-2021 waves of the ENUT Survey, conducted by DANE. The dependent variable in all columns is a binary indicator equal to 1 if the respondent expresses disagreement with the following statement:   “La cabeza del hogar debe ser el hombre?” (“The head of the household should be the man”). All the include cohort-by-region, high school-by-region, and survey year-by-region fixed effects, as well as a set of ethnic group dummies. Robust standard errors, clustered at the region level, are reported in parentheses. Wild bootstrap confidence intervals are shown in square brackets.  *, **, and *** indicate statistical significance at the 10\%, 5\%, and 1\% levels, respectively.} }
\end{tabular}
}
\end{center}
\end{table}

\newpage


\begin{table}[H]
\begin{center}
{
\renewcommand{\arraystretch}{0.5}
\setlength{\tabcolsep}{10pt}
\caption {Effect of Treatment on Support for Gender Equality: Robustness to Controlling for Number of Children in the Household}  \label{table_supportgenderequality_robustnchildren}
\vspace{-0.3cm}
\footnotesize
\centering  \begin{tabular}{lccc}
\hline\hline \addlinespace[0.15cm]
    & (1)& (2)& (3)\\\addlinespace[0.12cm]\cmidrule[0.2pt](l){2-4}\addlinespace[0.12cm]
      & \multicolumn{3}{c}{Dep. Var.: Support for Gender Equality}\\\addlinespace[0.1cm]\cmidrule[0.2pt](l){2-4}\addlinespace[0.1cm]
                      &\multicolumn{1}{c}{All}&\multicolumn{1}{c}{Women}&\multicolumn{1}{c}{Men}\\
                \addlinespace[0.1cm]\cmidrule[0.2pt](l){1-1} \cmidrule[0.2pt](l){2-2} \cmidrule[0.2pt](l){3-3}\cmidrule[0.2pt](l){4-4}\addlinespace[0.05cm]
(Age $\leq$ 16 in 1994) $\times$ (Highschool)&       0.020         &       0.010         &       0.026         \\
                    &     (0.013)         &     (0.019)         &     (0.008)\sym{**}  \\
                    &[-0.009 0.053]       &[-0.037 0.049]       &[0.009 0.053]**         \\
(Highschool)        &       0.084         &       0.178         &       0.002         \\
                    &     (0.006)\sym{***}&     (0.009)\sym{***}&     (0.004)         \\
                    &[-0.235 0.391]       &[-0.272 0.694]         &[-0.222 0.222]         \\
(Age $\leq$ 16 in 1994) $\times$ (Children)&      -0.010         &      -0.019         &       0.000         \\
                    &     (0.012)         &     (0.017)         &     (0.014)         \\
                    &[-0.056 0.020]       &[-0.086 0.016]         &[-0.104 0.025]         \\
(Children)          &       0.007         &       0.007         &       0.006         \\
                    &     (0.010)         &     (0.012)         &     (0.013)         \\
                    &[-0.016 0.042]     &[-0.007 0.065]         &[-0.025 0.063]         \\
R-sq                &       0.021         &       0.020         &       0.033         \\
Observations        &       11816         &        6275         &        5541         \\

\addlinespace[0.15cm]\hline\hline\addlinespace[0.15cm]
\multicolumn{4}{p{14.4cm}}{\scriptsize{\textbf{Notes:} All columns report estimates from Equation (\ref{baselineDiD}) using an age window of [-3, 3].  The regression samples are drawn from the 2019 and 2021 waves of the Political Culture Survey conducted by DANE. The dependent variable in all columns is the response to the question (in Spanish): ``¿Usted est\'a de acuerdo con que las mujeres participen en igualdad de condiciones y oportunidades?'' (``Do you agree that women should participate on equal terms and have the same opportunities?''). All the specifications include cohort-by-region, high school-by-region, and survey year-by-region fixed effects, as well as a set of ethnic group dummies. Robust standard errors, clustered at the region level, are reported in parentheses. Wild bootstrap confidence intervals are shown in square brackets.  *, **, and *** indicate statistical significance at the 10\%, 5\%, and 1\% levels, respectively.} }
\end{tabular}
}
\end{center}
\end{table}


\begin{table}[H]
\begin{center}
{
\renewcommand{\arraystretch}{0.5}
\setlength{\tabcolsep}{10pt}
\caption {Effect of Treatment on Perception of Gender Discrimination: Robustness to Controlling for Number of Children  in the Household}  \label{table_didsexdiscrim_robustnchildren}
\vspace{-0.3cm}
\footnotesize
\centering  \begin{tabular}{lccc}
\hline\hline \addlinespace[0.15cm]
    & (1)& (2)& (3)\\\addlinespace[0.12cm]\cmidrule[0.2pt](l){2-4}\addlinespace[0.12cm]
      & \multicolumn{3}{c}{Dep. Var.: Perception of Gender Discrimination}\\\addlinespace[0.1cm]\cmidrule[0.2pt](l){2-4}\addlinespace[0.1cm]
                      &\multicolumn{1}{c}{All}&\multicolumn{1}{c}{Women}&\multicolumn{1}{c}{Men}\\
                \addlinespace[0.1cm]\cmidrule[0.2pt](l){1-1} \cmidrule[0.2pt](l){2-2} \cmidrule[0.2pt](l){3-3}\cmidrule[0.2pt](l){4-4}\addlinespace[0.05cm]
(Age $\leq$ 16 in 1994) $\times$ (Highschool)&      -0.005         &      -0.021         &       0.012         \\
                    &     (0.003)         &     (0.006)\sym{*}  &     (0.010)         \\
                    &[-0.029 0.004]        &[-0.037 -0.007]***         &[0.000 0.046]**         \\
(Highschool)        &      -0.002         &      -0.001         &      -0.004         \\
                    &     (0.002)         &     (0.004)         &     (0.005)         \\
                    &[-0.152 0.291]          &[-0.163 0.564]            &[-0.514 0.174]         \\
(Age $\leq$ 16 in 1994) $\times$ (Children)&       0.001         &      -0.003         &       0.005         \\
                    &     (0.006)         &     (0.010)         &     (0.003)         \\
                    &[-0.023 0.038]          &[-0.043 0.088]          &[-0.007 0.014]         \\
(Children)          &      -0.000         &       0.004         &      -0.005         \\
                    &     (0.005)         &     (0.008)         &     (0.003)         \\
                    &[-0.015 0.017]           &[-0.015 0.034]       &[-0.017 0.000]         \\
R-sq                &       0.013         &       0.020         &       0.013         \\
Observations        &        6555         &        3495         &        3060         \\

\addlinespace[0.15cm]\hline\hline\addlinespace[0.15cm]
\multicolumn{4}{p{14.4cm}}{\scriptsize{\textbf{Notes:} All columns report estimates from Equation (\ref{baselineDiD}) using an age window of [-3, 3].  The regression samples are drawn from the 2021 wave of the Political Culture Survey conducted by DANE. The dependent variable in all columns is the response to the question (in Spanish): ``Durante los \'ultimos 12 meses, usted ha sentido que lo han discriminado [...] por alguno de los siguientes motivos: Sexo [...]'' (``In the past 12 months, have you felt discriminated against [...] based on sex?'').  All the specifications include cohort-by-region, high school-by-region, and survey year-by-region fixed effects, as well as a set of ethnic group dummies. Robust standard errors, clustered at the region level, are reported in parentheses. Wild bootstrap confidence intervals are shown in square brackets.  *, **, and *** indicate statistical significance at the 10\%, 5\%, and 1\% levels, respectively.} }
\end{tabular}
}
\end{center}
\end{table}


\begin{table}[H]
\begin{center}
{
\renewcommand{\arraystretch}{0.5}
\setlength{\tabcolsep}{10pt}
\caption {Effect of Treatment on Disagreement with ``Women are Better Suited for Domestic Tasks than Men'': Robustness to Controlling for Number of Children  in the Household}  \label{table_opinion4_robustnchildren}
\vspace{-0.3cm}
\footnotesize
\centering  \begin{tabular}{lccc}
\hline\hline \addlinespace[0.15cm]
    & (1)& (2)& (3)\\\addlinespace[0.12cm]\cmidrule[0.2pt](l){2-4}\addlinespace[0.12cm]
      & \multicolumn{3}{c}{Dep. Var.: Disagreement with ``Women}\\
      & \multicolumn{3}{c}{are better suited for domestic tasks''}\\
      \addlinespace[0.1cm]\cmidrule[0.2pt](l){2-4}\addlinespace[0.1cm]
                      &\multicolumn{1}{c}{All}&\multicolumn{1}{c}{Women}&\multicolumn{1}{c}{Men}\\
                \addlinespace[0.1cm]\cmidrule[0.2pt](l){1-1} \cmidrule[0.2pt](l){2-2} \cmidrule[0.2pt](l){3-3}\cmidrule[0.2pt](l){4-4}\addlinespace[0.05cm]
(Age $\leq$ 16 in 1994) $\times$ (Highschool)&      -0.005         &      -0.002         &      -0.010         \\
                    &     (0.009)         &     (0.016)         &     (0.010)         \\
                    &[-0.037 0.022]          &[-0.045 0.043]                &[-0.029 0.022]         \\
(Highschool)        &       0.168         &       0.173         &       0.164         \\
                    &     (0.004)\sym{***}&     (0.008)\sym{***}&     (0.005)\sym{***}\\
                    &[0.100 0.209]***           &[0.000 0.229]**          &[0.105 0.213]**         \\
(Age $\leq$ 16 in 1994) $\times$ (Children)&       0.003         &       0.010         &      -0.005         \\
                    &     (0.007)         &     (0.007)         &     (0.017)         \\
                    &[-0.011 0.025]        &[-0.017 0.028]              &[-0.036 0.051]         \\
(Children)          &      -0.006         &      -0.005         &      -0.007         \\
                    &     (0.009)         &     (0.007)         &     (0.013)         \\
                    &[-0.031 0.013]    &[-0.024 0.011]       &[-0.041 0.020]         \\
R-sq                &       0.052         &       0.050         &       0.056         \\
Observations        &       26061         &       13933         &       12128         \\

\addlinespace[0.15cm]\hline\hline\addlinespace[0.15cm]
\multicolumn{4}{p{14.4cm}}{\scriptsize{\textbf{Notes:} All columns report estimates from Equation (\ref{baselineDiD}) using an age window of [-3, 3]. The regression samples are drawn from the 2017–2016 and 2020-2021 waves of the ENUT Survey, conducted by DANE. The dependent variable in all columns is a binary indicator equal to 1 if the respondent expresses disagreement with the following statement: “¿Las mujeres son mejores para el trabajo doméstico que los hombres?” (“Women are better suited for domestic tasks than men?”)..  All the specifications include cohort-by-region, high school-by-region, and survey year-by-region fixed effects, as well as a set of ethnic group dummies. Robust standard errors, clustered at the region level, are reported in parentheses. Wild bootstrap confidence intervals are shown in square brackets.  *, **, and *** indicate statistical significance at the 10\%, 5\%, and 1\% levels, respectively.} }
\end{tabular}
}
\end{center}
\end{table}


\begin{table}[H]
\begin{center}
{
\renewcommand{\arraystretch}{0.5}
\setlength{\tabcolsep}{10pt}
\caption {Effect of Treatment on Disagreement with ``The Husband should make Decisions Concerning his Wife’s Life'': Robustness to Controlling for Number of Children  in the Household}  \label{table_opinion5_robustnchildren}
\vspace{-0.3cm}
\footnotesize
\centering  \begin{tabular}{lccc}
\hline\hline \addlinespace[0.15cm]
    & (1)& (2)& (3)\\\addlinespace[0.12cm]\cmidrule[0.2pt](l){2-4}\addlinespace[0.12cm]
      & \multicolumn{3}{c}{Dep. Var.: Disagreement with ``The husband}\\
    & \multicolumn{3}{c}{should make decisions concerning his wife’s life''}\\
      \addlinespace[0.1cm]\cmidrule[0.2pt](l){2-4}\addlinespace[0.1cm]
                      &\multicolumn{1}{c}{All}&\multicolumn{1}{c}{Women}&\multicolumn{1}{c}{Men}\\
                \addlinespace[0.1cm]\cmidrule[0.2pt](l){1-1} \cmidrule[0.2pt](l){2-2} \cmidrule[0.2pt](l){3-3}\cmidrule[0.2pt](l){4-4}\addlinespace[0.05cm]
(Age $\leq$ 16 in 1994) $\times$ (Highschool)&       0.005         &       0.029         &      -0.018         \\
                    &     (0.010)         &     (0.016)         &     (0.012)         \\
                    &[-0.021 0.027]          &[-0.020 0.070]            &[-0.052 0.004]         \\
(Highschool)        &       0.099         &       0.073         &       0.126         \\
                    &     (0.006)\sym{***}&     (0.008)\sym{***}&     (0.007)\sym{***}\\
                    &[0.045 0.164]***           &[-0.011 0.086]*              &[0.099 0.199]***         \\
(Age $\leq$ 16 in 1994) $\times$ (Children)&       0.003         &       0.002         &       0.007         \\
                    &     (0.006)         &     (0.009)         &     (0.005)         \\
                    &[-0.013 0.022]          &[-0.028 0.023]           &[-0.005 0.023]         \\
(Children)          &      -0.001         &      -0.005         &       0.000         \\
                    &     (0.003)         &     (0.004)         &     (0.004)         \\
                    &[-0.008 0.006]        &[-0.015 0.011]         &[-0.013 0.008]         \\
R-sq                &       0.545         &       0.611         &       0.478         \\
Observations        &       26150         &       13963         &       12187         \\

\addlinespace[0.15cm]\hline\hline\addlinespace[0.15cm]
\multicolumn{4}{p{14.4cm}}{\scriptsize{\textbf{Notes:} All columns report estimates from Equation (\ref{baselineDiD}) using an age window of [-3, 3]. The regression samples are drawn from the 2017–2016 and 2020-2021 waves of the ENUT Survey, conducted by DANE. The dependent variable in all columns is a binary indicator equal to 1 if the respondent expresses disagreement with the following statement:  “El esposo debe tomar las decisiones relacionadas con la vida de la esposa?” (“The husband should make decisions concerning his wife’s life”).  All the specifications include cohort-by-region, high school-by-region, and survey year-by-region fixed effects, as well as a set of ethnic group dummies. Robust standard errors, clustered at the region level, are reported in parentheses. Wild bootstrap confidence intervals are shown in square brackets.  *, **, and *** indicate statistical significance at the 10\%, 5\%, and 1\% levels, respectively.} }
\end{tabular}
}
\end{center}
\end{table}


\begin{table}[H]
\begin{center}
{
\renewcommand{\arraystretch}{0.5}
\setlength{\tabcolsep}{10pt}
\caption {Effect of Treatment on Disagreement with ``the Head of the Household Should be the Man'': Robustness to Controlling for Number of Children  in the Household}  \label{table_opinion6_robustnchildren}
\vspace{-0.3cm}
\footnotesize
\centering  \begin{tabular}{lccc}
\hline\hline \addlinespace[0.15cm]
    & (1)& (2)& (3)\\\addlinespace[0.12cm]\cmidrule[0.2pt](l){2-4}\addlinespace[0.12cm]
      & \multicolumn{3}{c}{Dep. Var.: Disagreement with ``the head of}\\
            & \multicolumn{3}{c}{the household should be the man''}\\
      \addlinespace[0.1cm]\cmidrule[0.2pt](l){2-4}\addlinespace[0.1cm]
                      &\multicolumn{1}{c}{All}&\multicolumn{1}{c}{Women}&\multicolumn{1}{c}{Men}\\
                \addlinespace[0.1cm]\cmidrule[0.2pt](l){1-1} \cmidrule[0.2pt](l){2-2} \cmidrule[0.2pt](l){3-3}\cmidrule[0.2pt](l){4-4}\addlinespace[0.05cm]
(Age $\leq$ 16 in 1994) $\times$ (Highschool)&       0.016         &       0.041         &      -0.007         \\
                    &     (0.012)         &     (0.010)\sym{**} &     (0.014)         \\
                    &[-0.006 0.049]       &[0.016 0.073]***          &[-0.037 0.029]         \\
(Highschool)        &       0.135         &       0.132         &       0.131         \\
                    &     (0.008)\sym{***}&     (0.006)\sym{***}&     (0.009)\sym{***}\\
                    &[0.051 0.250]**             &[0.073 0.223]***           &[0.015 0.211]**         \\
(Age $\leq$ 16 in 1994) $\times$ (Children)&      -0.008         &      -0.014         &      -0.001         \\
                    &     (0.009)         &     (0.015)         &     (0.015)         \\
                    &[-0.034 0.025]        &[-0.079 0.009]           &[-0.034 0.034]         \\
(Children)          &      -0.006         &       0.004         &      -0.017         \\
                    &     (0.007)         &     (0.009)         &     (0.009)         \\
                    &[-0.028 0.009]         &[-0.026 0.028]        &[-0.039 0.005]         \\
R-sq                &       0.096         &       0.067         &       0.098         \\
Observations        &       26392         &       14131         &       12261         \\

\addlinespace[0.15cm]\hline\hline\addlinespace[0.15cm]
\multicolumn{4}{p{14.4cm}}{\scriptsize{\textbf{Notes:} All columns report estimates from Equation (\ref{baselineDiD}) using an age window of [-3, 3].  The regression samples are drawn from the 2017–2016 and 2020-2021 waves of the ENUT Survey, conducted by DANE. The dependent variable in all columns is a binary indicator equal to 1 if the respondent expresses disagreement with the following statement:   “La cabeza del hogar debe ser el hombre?” (“The head of the household should be the man”).  All the specifications include cohort-by-region, high school-by-region, and survey year-by-region fixed effects, as well as a set of ethnic group dummies. Robust standard errors, clustered at the region level, are reported in parentheses. Wild bootstrap confidence intervals are shown in square brackets.  *, **, and *** indicate statistical significance at the 10\%, 5\%, and 1\% levels, respectively.} }
\end{tabular}
}
\end{center}
\end{table}




\end{document}